\documentclass[prb,aps,reprint,amsmath,amssymb]{revtex4-2}
\usepackage{graphicx} 
\usepackage{hyperref,braket,graphicx,color,physics,mathtools,subcaption,caption}

\begin{document}

\title{Analysis of exciton-polariton condensation under different pumping schemes for 1D and 2D microcavities including the effect of strong correlation between polaritons}
\author{Varad R.~Pande$^1$}
\affiliation{$^1$Walton Institute for Information and Communication Systems Science, South East Technological University (SETU), Waterford, Ireland.
}
\date{\today}

\begin{abstract}
Strongly correlated polaritons are necessary for entering the quantum photonic regime with many applications. We simulate exciton-polariton condensation using the finite-difference and 4th order Runge-Kutta methods with the strongly correlated polariton condition incorporated in the mean-field equations and analyze the polariton dynamics. This is done for coherent, near-resonant pumping as well as homogeneous, incoherent, non-resonant pumping. We find conditions akin to the polariton blockade in the dynamics. 
\end{abstract}
		
\maketitle

\section{Motivation}

Exciton-polaritons are hybrid light-matter quasiparticles resulting from the strong coupling between excitons in quantum wells and photons confined in distributed Bragg reflectors (DBR)~\cite{byrnes2014exciton}. They inherit strong interactions from their excitonic component and small mass from their photonic component. Strong coupling occurs when the quantum well is placed at the antinodes of the photonic field confined inside the microcavity and the energy of the photonic field is near the exciton energy~\cite{ghosh2022microcavity}. Phenomena such as Bose-Einstein condensation, superfluidity, quantum vortices, stimulated parametric scattering and long-range spatial coherence have been observed in exciton-polaritons~\cite{timofeev2012exciton}. Exciton-polaritons are observed in microcavities containing quantum wells such as GaAs~\cite{weisbuch1992observation}, CdTe~\cite{kasprzak2006bose}, carbon nanotubes~\cite{graf2016near}, GaN~\cite{christopoulos2007room}, halide perovskites~\cite{su2021perovskite}, organic semiconductors~\cite{lidzey1998strong}, transition metal dichalcogenides~\cite{liu2015strong}, and ZnO~\cite{zhang2015weak}. The temperature at which exciton-polaritons are observed depends on the exciton binding energies of the quantum wells: the observation occurs when the binding energy is greater than room temperature thermal fluctuations~\cite{ghosh2022microcavity}. 

Exciton-polariton condensation has been shown at a variety of temperatures, including room temperature~\cite{su2020observation,su2021perovskite}. Different laser pumping schemes have also been demonstrated to cause exciton-polariton condensation in 1-dimensional (1D) and 2-dimensional (2D) microcavities~\cite{carusotto2013quantum}. Coherent, near-resonant pumping is when the laser frequency is resonant with the energy of the lower polariton branch. Homogeneous, incoherent, non-resonant pumping is due to a high energy continuous wave illumination resulting in the creation of hot free carriers in the quantum well. For a graphical representation of these pumping methods, see the middle panel of figure 1 in Ref.~\cite{carusotto2013quantum} for homogeneous, incoherent, non-resonant pumping and figure 1b in Ref.~\cite{byrnes2014exciton} for a representation of both schemes. Equations governing the dynamics of exciton-polaritons are in the mean-field regime and they are a modification of the Gross-Pitaevskii equation~\cite{wouters2007excitations,carusotto2004probing,carusotto2013quantum}.

However, strongly correlated polaritons have not been explored in the mean-field regime which can lead to polariton blockade~\cite{verger2006polariton} like effects. A polariton blockade is when a single polariton excited within a photonic lattice site prevents the addition of another polariton to the same lattice site by an external laser~\cite{delteil2019towards,munoz2019emergence,liew2023future}. This effect is manifested when the polariton-polariton interaction strength is greater than the polariton linewidth.  Polariton nonlinearity due to their excitonic components results in applications such as all-optical chips and circuits, ultrafast transistors, switches~\cite{ghosh2022microcavity}, and parametric amplifiers~\cite{savvidis2000angle,saba2001high}. Furthermore, due to the polariton blockade, prospects for controlling and manipulating single photonic qubits are enhanced potentially resulting in quantum computing~\cite{ghosh2020quantum} and quantum reservoir computing~\cite{ghosh2021realising}. 

In this work, we simulate exciton-polariton condensation under distinct laser pumping schemes (corresponding to three different coupled equations) with the strongly correlated polariton condition incorporated in the mean-field equations. To the best of our knowledge, this is also the first time that all these pumping schemes are considered in a unified framework. This framework is based on a finite-difference method combined with 4th order Runge-Kutta method (RK4). These methods have been used previously to study exciton-polaritons under near-resonant~\cite{voronych2017numerical} and non-resonant~\cite{gargoubi2016polariton} pumping schemes. They have also been used to simulate the Gross-Pitaevskii equation (GPE) for studying the Bose-Einstein condensate (BEC) of ultracold atoms.
Exciton-polariton condensate phase dynamics have been mapped to the Kardar-Parisi-Zhang equation under some assumptions while using the 4th order Runge-Kutta method for time evolution of the coupled equations~\cite{vercesi2023phase}. The RK4 method has also been used to analyze the generalized dissipative GPE on a one-dimensional torus~\cite{antonelli2019dissipative}. The second and fourth order finite-difference method has been used for finding the wave functions and energy levels of the GPE in geometries such as a cylinder and cubic box~\cite{chang2008finite,chang2009computing}. Adaptive 3rd and 4th order Runge-Kutta scheme has been used to estimate the local error in the Interaction Picture method at very little extra cost~\cite{balac2013embedded}.

This paper is organized as follows: in section~\ref{MEPD}, we describe the mathematical equations governing the dynamics of exciton-polaritons; specifically, equations of the first~\ref{CNRP1} (CNRP1) and second~\ref{CNRP2} (CNRP2) type corresponding to coherent, near-resonant pumping, and coupled equations corresponding to homogeneous, incoherent, non-resonant pumping~\ref{HINRP} (HINRP). Then we move to the numerical methods to simulate these equations in section~\ref{NM} and describe the finite-difference~\ref{FD} and 4th order Runge-Kutta~\ref{RK4} methods. After this, we plot and interpret the results in section~\ref{res}, corresponding to all three types of equations~\ref{CNRP1res}\ref{CNRP2res}\ref{HINRPres}. This is followed by a conclusion of this work in section~\ref{conc}. 

\section{Mathematics of exciton-polariton dynamics}
\label{MEPD}
Depending on the laser pumping method, there are a few ways by which the dynamics of exciton-polaritons can be accessed. 

\subsection{Coherent, near-resonant pumping 1}
\label{CNRP1}

The following model Hamiltonian incorporates the effect of strong coupling between excitons and cavity photons in a planar microcavity~\cite{carusotto2004probing,timofeev2012exciton}:
\begin{eqnarray}
\label{H}
	H &=& \int d\mathbf{x} \sum_{i,j\in \{x,c\}} \hat{\psi}_i^\dagger (\mathbf{x})[\mathbf{h}^0_{ij} (-i\nabla) + V_i(\mathbf{x}) \delta_{ij}] \hat{\psi}_j (\mathbf{x}) \nonumber\\
	&&+ \frac{\hbar g}{2} \int d\mathbf{x} \hat{\psi}_x^\dagger (\mathbf{x}) \hat{\psi}_x^\dagger (\mathbf{x}) \hat{\psi}_x (\mathbf{x}) \hat{\psi}_x (\mathbf{x}) \nonumber\\
	&&+ \int d\mathbf{x} \hbar F_p (\mathbf{x},t) \hat{\psi}^\dagger_c (\mathbf{x}) + h.c.,
\end{eqnarray}
where the field operators $\hat{\psi}_x(\mathbf{x})$ and $\hat{\psi}_c(\mathbf{x})$ correspond to exciton ($x$) and cavity photon ($c$) indexes with $i,j$ running over them respectively and $\mathbf{x}$ is the spatial position within the plane. The single particle Hamiltonian $\mathbf{h}^0$ is k-space reads:
\begin{equation}
\begin{aligned}
	\mathbf{h}^0(\mathbf{k}) = \hbar \begin{bmatrix}
		\omega_x(\mathbf{k}) & \Omega_R \\
		\Omega_R & \omega_c(\mathbf{k})
	\end{bmatrix},\end{aligned}
\end{equation}
where the exciton-photon coupling is incorporated by the Rabi frequency $\Omega_R$. Energy dispersion of the cavity mode as a function of $\mathbf{k}$ (in-plane wavevector) and $k_z$ (quantized photon wavevector along growth direction) is $\omega_c(\mathbf{k}) = \omega_c^0 \sqrt{1 + \mathbf{k}^2/k_z^2}$, and energy dispersion of the exciton is assumed to be flat: $\omega_x(\mathbf{k}) = \omega_x$. In real space, we replace $\mathbf{k}$ with $ -i\nabla $, to get $\omega_c(-i\nabla) = \omega_c^0 - \frac{\hbar^2 \nabla^2}{2m_c}$ and $\omega_x(-i\nabla) = \omega_x^0 - \frac{\hbar^2 \nabla^2}{2m_x}$. The exciton dispersion is flat because the exciton mass $m_x$ is considered infinite relative to the polariton mass $m_c$.

Under the Hamiltonian~\ref{H}, within the mean-field approximation, the time evolution of the mean fields $\psi_{x,c} (\mathbf{x}) = \langle \hat{\psi}_{x,c} (\mathbf{x}) \rangle = \int d\mathbf{x} \hat{\psi}_{x,c}^\dagger (\mathbf{x}) \hat{\psi}_{x,c} (\mathbf{x}) $, can be expressed as,
\begin{widetext}
\begin{equation}
	\begin{aligned}
		i\hbar \frac{d}{dt} \begin{bmatrix}
			\psi_x(\mathbf{x}) \\
			\psi_c(\mathbf{x})
		\end{bmatrix} = \begin{bmatrix}
			0 \\
			F_p(\mathbf{x},t)
		\end{bmatrix} + \bigg(\mathbf{h}^0 + \begin{bmatrix}
			\hbar g |\psi_x(\mathbf{x})|^2 + V_x(\mathbf{x}) - \frac{i \hbar \gamma_x}{2} & 0 \\
			0 & V_c(\mathbf{x}) - \frac{i \hbar \gamma_c}{2}
		\end{bmatrix}\bigg) \begin{bmatrix}
			\psi_x(\mathbf{x}) \\
			\psi_c(\mathbf{x})
	\end{bmatrix}\end{aligned}
\end{equation}	
\end{widetext}
In the above, the photon and exciton modes broaden homogeneously due to the dissipation rates $\gamma_c$ and $\gamma_x$ respectively. These equations are relevant for resonant pumping of the microcavity by the laser where the laser frequency is in tune with the exciton resonance of the quantum well.

The above equation contains several parameters which are unnecessary for practically investigating the system~\cite{voronych2017numerical}. These include the single particle potentials $V_x(\mathbf{x})$ and $V_c(\mathbf{x})$. Moreover, $\omega_p$, $\omega_x^0$, and $\omega_c^0$ can be replaced by a couple of parameters representing polariton detuning $\delta$ and pump field detuning $\delta_\omega$ from the cavity mode frequency $\omega_c^0$. Further, it is reasonable to fix the reference frame such that center of the laser spot $\mathbf{x}_0$ corresponds to $\mathbf{x} = 0$. Thus, we have $\omega_p \rightarrow \omega_p + \omega_c^0 = \delta_\omega$, $ \omega_c^0 \rightarrow \omega_c^0 - \omega_c^0 = 0 $, $ \omega_x^0 \rightarrow \omega_x^0 - \omega_c^0 = \delta $, $\omega_c(-i\nabla) = -\frac{\hbar^2 \nabla^2}{2m_c}$, $ \omega_x(-i\nabla) = \delta $, $ \mathbf{x}_0 = 0 $, and $ V_x(\mathbf{x}) = V_c(\mathbf{x}) = 0 $. After making these approximations and assumptions, the coupled equations, without taking spin into account, can be expressed as:
\begin{eqnarray}
\label{CNRP1eq}
i\hbar\frac{d}{dt} \psi_c(\mathbf{x},t) &=& \Omega_R \psi_x(\mathbf{x},t) + F(\mathbf{x},t) \nonumber\\
&&+ \left(-\frac{\hbar^2 \nabla^2}{2m_c} - i\hbar\frac{\gamma_c}{2}\right) \psi_c(\mathbf{x},t) \nonumber\\
i\hbar\frac{d}{dt} \psi_x(\mathbf{x},t) &=& \Omega_R \psi_c(\mathbf{x},t) \nonumber\\
&&+ \left(g|\psi_x(\mathbf{x},t)|^2 + \delta - i\hbar\frac{\gamma_x}{2}\right) \psi_x(\mathbf{x},t), 
\end{eqnarray}
where, 
\begin{equation}
\begin{aligned}
\label{pump}
	F(\mathbf{x},t) = F_p e^{i(\mathbf{k}_p.\mathbf{x} - \omega_p t)} e^{-\frac{(\mathbf{x} - \mathbf{x}_0)^2}{2w_x^2}}. \end{aligned}
\end{equation}
Here, $w_x$ is the spread of the Gaussian laser spot on the microcavity, $\mathbf{x}_0$ is its center, $\mathbf{k}_p$ is the momentum of the laser pump, $\omega_p$ its frequency, and $F_p$ its field amplitude. Moreover, $\Omega_R$ is the Rabi frequency, $\gamma_c$ and $\gamma_x$ are the decay rates for photons and excitons respectively, $m_c$ is the mass of the polariton which is of the order of $10^{-5} m_e$, where $m_e$ is the mass of an electron, $g$ is the nonlinear interaction strength between excitons, $\delta$ is the detuning of the polaritons from the cavity mode frequency. 

The set of equations we get when spin is accounted~\cite{voronych2017numerical} for is,
\begin{widetext}
\begin{eqnarray}
	\label{CNRP1eqspin}
	i\hbar\frac{d}{dt} \psi_{c,-1}(\mathbf{x},t) &=& \Omega_R \psi_{x,-1}(\mathbf{x},t) + F_{-1}(\mathbf{x},t) + \left(-\frac{\hbar^2 \nabla^2}{2m_c} - i\hbar\frac{\gamma_c}{2}\right) \psi_{c,-1}(\mathbf{x},t) \nonumber\\
	i\hbar\frac{d}{dt} \psi_{c,+1}(\mathbf{x},t) &=& \Omega_R \psi_{x,+1}(\mathbf{x},t) + F_{+1}(\mathbf{x},t) + \left(-\frac{\hbar^2 \nabla^2}{2m_c} - i\hbar\frac{\gamma_c}{2}\right) \psi_{c,+1}(\mathbf{x},t) \nonumber\\
	i\hbar\frac{d}{dt} \psi_{x,-1}(\mathbf{x},t) &=& \Omega_R \psi_{c,-1}(\mathbf{x},t) + \left(g_1|\psi_{x,-1}(\mathbf{x},t)|^2 + g_2|\psi_{x,+1}(\mathbf{x},t)|^2 + \delta - i\hbar\frac{\gamma_x}{2}\right) \psi_{x,-1}(\mathbf{x},t) \nonumber\\
	i\hbar\frac{d}{dt} \psi_{x,+1}(\mathbf{x},t) &=& \Omega_R \psi_{c,+1}(\mathbf{x},t) + \left(g_1|\psi_{x,+1}(\mathbf{x},t)|^2 + g_2|\psi_{x,-1}(\mathbf{x},t)|^2 + \delta - i\hbar\frac{\gamma_x}{2}\right) \psi_{x,+1}(\mathbf{x},t) \nonumber\\
\end{eqnarray}
\end{widetext}

\subsection{Coherent, near-resonant pumping 2}
\label{CNRP2}

For an atomic Bose-Einstein condensate (BEC), the Gross-Pitaevskii equation looks like,
\begin{equation}
\begin{aligned}
	i\hbar\frac{d\psi}{dt} = \bigg(-\frac{\hbar^2}{2m} \nabla^2 + V_{ext} (\mathbf{x}) + g |\psi|^2\bigg) \psi, \end{aligned}
\end{equation}
where $g = 4\pi \hbar^2 a/m$ and $a$ is the atom-atom scattering length.

Incorporating the finite lifetime of excitons and the fact that DBR pairs used to confine photons are imperfect, we introduce a decay or loss term in the above equation. Additionally, we introduce a source term in the equation to account for the injection of polaritons by a coherently applied laser field resonant with the lower polariton branch~\cite{deveaud2007physics,carusotto2013quantum,ghosh2022microcavity}.
 
Therefore, using a near-resonant laser corresponding to a coherent excitation, the above equation is modified as follows:
\begin{eqnarray}
\label{CNRP2eq}
i\hbar \pdv{\psi(\mathbf{r})}{t} &=& \left\{ \frac{\hbar^2}{2m} \nabla_r^2 + V_{ext}(\mathbf{r}) - \frac{i\hbar \gamma_c}{2} + \hbar g |\psi(\mathbf{r})|^2 \right\} \psi(\mathbf{r}) \nonumber\\
&&+ i \eta(\mathbf{r}) F(\mathbf{r},t),
\end{eqnarray}
where the incident optical field represented by $F(\mathbf{r},t)$ is coupled to the lower polariton mode via the function $\eta(\mathbf{r})$. Note that $F(\mathbf{r},t)$ is same as equation~\ref{pump}. 

\subsection{Homogeneous, incoherent, non-resonant pumping}	
\label{HINRP}

This method comprises non-resonant pumping of the microcavity and is described by the driven-dissipative generalized Gross-Pitaevskii equation for the lower polaritons coupled with the reservoir equation responsible for replenishing the condensate with polaritons lost from the cavity~\cite{carusotto2013quantum,gargoubi2016polariton,wouters2007excitations}:
\begin{widetext}
\begin{eqnarray}
	\label{HINRPeq}
	i\hbar \pdv{\psi(\mathbf{r})}{t} &=& \left\{ E_0 - \frac{\hbar^2}{2m} \nabla_r^2 + \frac{i\hbar}{2} (R[n_R(\mathbf{r})] - \gamma_c) + V_{ext}(\mathbf{r}) + \hbar g |\psi(\mathbf{r})|^2 + V_R(\mathbf{r})\right\} \psi(\mathbf{r}), \nonumber\\
	\Dot{n}_R(\mathbf{r}) &=& P(\mathbf{r}) - \gamma_R n_R(\mathbf{r}) - R[n_R(\mathbf{r})]|\psi(\mathbf{r})|^2.
\end{eqnarray}
\end{widetext}

Here, $m$ and $E_0$ correspond to the effective mass and minimum energy of the lower polariton branch respectively (we consider $ E_0 = 0 $ when simulating this equation without loss of generality). $g>0$ is the repulsive polariton-polariton interaction strength. $\gamma_c$ represents the linear loss rate of the condensate polaritons and $V_{ext}(\mathbf{r})$ represents external potential due to cavity and exciton disorder. $R[n_R]$ is the rate of polariton gain. $V_R(\mathbf{r})$ is the mean-field repulsive potential produced by the reservoir which is given by $V_R(\mathbf{r}) \approx \hbar G P(\mathbf{r}) + \hbar g_R n_R(\mathbf{r}) $, where $G>0$ are experimentally extracted phenomenological coefficients and $P(\mathbf{r})$ is the pumping rate which is spatially dependent. $\gamma_R$ is the effective relaxation rate of reservoir polaritons. 
\section{Numerical methods}
\label{NM}
We consider zero wavefunction as the initial state for coherent, near-resonant pumping of type 1~\cite{voronych2017numerical} (CNRP1; see section \ref{CNRP1}). We consider Gaussian distributions as initial states for $\psi$ and $n_R$ for coherent, near-resonant pumping of type 2 (CNRP2; see section \ref{CNRP2}) and homogeneous, incoherent, non-resonant pumping (HINRP; see section \ref{HINRP})~\cite{gargoubi2016polariton}:
\begin{eqnarray}
	\psi(x,y,0) = \frac{\sqrt{N_c}}{\sigma_p \sqrt{\pi}} \exp{-(x^2 + y^2)/2\sigma_p^2} \\
	n_R(x,y,0) = \frac{P_0}{\gamma_R} \exp{-(x^2 + y^2)/2\sigma_p^2}
\end{eqnarray}

The left hand side (LHS) of the equations~\ref{CNRP1eq}, \ref{CNRP1eqspin}, \ref{CNRP2eq}, and \ref{HINRPeq} all have a single time derivative of the respective wave functions. The right hand side (RHS) for the set of equations \ref{CNRP1eq} and \ref{CNRP1eqspin} have an almost identical structure, in that, there are second-order partial derivatives, linear terms involving $\psi_c$ and $\psi_x$ in the equation computing the time derivative of $\psi_c$ (and also $\psi_{c,-1}$, $\psi_{c,+1}$), and non-linear terms involving $\psi_x$ (and also $\psi_{x,-1}$, $\psi_{x,+1}$) in the equation computing the time derivative of $\psi_x$ (and also $\psi_{x,-1}$, $\psi_{x,+1}$). The RHS of equations \ref{CNRP2eq} and \ref{HINRPeq} also contain non-linear terms of $\psi$ and $n_R$, as well as second-order partial derivatives along with linear terms. 

We use the fourth order Runge-Kutta method for the time stepping and the finite-difference method for encoding the second-order partial derivatives. Throughout, we use cylindrical boundary conditions for the 2D microcavity and boundary conditions in which the ends are equated to 0 for the 1D microcavity. These are specified below:
\begin{eqnarray}
	\psi(x_{min}) &=& 0 \nonumber\\
	\psi(x_{max}) &=& 0 \nonumber\\
	\psi(y_{max}) &=& \psi(y_{min})
\end{eqnarray}

\subsection{Finite-difference}
\label{FD}
Since any computer has a limited memory, the wavefunctions $\psi_{x,c}$ and $\psi$ and $n_R$ need to be represented by discretizing the X-Y plane by introducing a finite set of space coordinates. We do this by considering an $N$-node uniform mesh in 1D and an $N \cross N$-node square uniform mesh in 2D~\cite{voronych2017numerical}. We consider the center of the mesh to be $\mathbf{x} = 0$ and distance between consecutive mesh nodes to be $\Delta_x$ and $\Delta_y$. Thus, for a 1D microwire, the size of the microcavty is $N\Delta_x$ and for a 2D microcavity, the size is $N\Delta_x \cross N\Delta_y$. The partial derivatives on the RHS of the equations can be approximated by the central finite difference formula for a uniform mesh in 1D:
\begin{equation}
\begin{aligned}
	\frac{d^2}{dx^2} \psi_c(x,t) = \frac{\psi_c(x-\Delta_x, t) - 2 \psi_c(x,t) + \psi_c(x + \Delta_x, t)}{\Delta_x^2}.\end{aligned}
\end{equation}
And in 2D:
\begin{eqnarray*}
	&&\nabla^2\psi_c(x,y,t) = \bigg(\frac{d^2}{dx^2} + \frac{d^2}{dy^2}\bigg) \psi_c(x,y,t) \nonumber\\
	&&= \frac{\psi_c(x-\Delta_x,y, t) - 2 \psi_c(x,y,t) + \psi_c(x + \Delta_x,y, t)}{\Delta_x^2} \nonumber\\
	&&+ \frac{\psi_c(x, y -\Delta_y, t) - 2 \psi_c(x,y,t) + \psi_c(x,y + \Delta_y, t)}{\Delta_y^2}
\end{eqnarray*}
The errors accumulated due to the finite size meshing in x and y are $\mathcal{O}(\Delta_x^2)$ and $\mathcal{O}(\Delta_x^2 \Delta_y^2)$ for 1D and 2D respectively. 

\subsection{Fourth order Runge-Kutta}
\label{RK4}
The 4th order Runge-Kutta method was developed from the Euler algorithm, where the differential equation $\frac{d}{dt}y(t) = f(t,y)$ is solved by replacing $\frac{d}{dt}y(t) \approx [y(t+h)-y(t)]/h $, with $h$ as the time step, leading to the approximation $y(t+h) \approx y(t) + h f(t+y)$. The 4th order Runge-Kutta method introduces additional steps to enhance the stability and accuracy of computation as follows~\cite{voronych2017numerical}:
\begin{eqnarray}
	k_1 &=& h f(y,t), \nonumber\\
	k_2 &=& h f(y+\frac{1}{2}k_1, t + \frac{1}{2} h), \nonumber\\
	k_3 &=& h f(y+\frac{1}{2}k_2, t + \frac{1}{2} h), \nonumber\\
	k_4 &=& h f(y+ k_3, t + h).
\end{eqnarray}
This leads to 
\begin{equation}
\begin{aligned}
	y(t+h) = y(t) + \frac{1}{6} (k_1 + 2k_2 + 2k_3 + k_4)\end{aligned}
\end{equation}
Applying the same RK4 method to the set of equations \ref{CNRP1eq} involves same steps, but for both excitons ($x$) and polaritons ($c$) respectively. Moreover, the method for the equations with spin effects included is twice as long as the spinless case.

The errors accumulated due to the 4th order Runge-Kutta method are $\mathcal{O}(h^5)$.
The stability of the method -- 4th order Runge-Kutta combined with finite-difference -- is determined by the CFL (Courant-Friedrichs-Lewy) condition which states that,
\begin{equation}
\begin{aligned}
	\frac{\hbar}{m_c} \frac{h}{\Delta_x^2} \leq 1\end{aligned}
\end{equation}
for the 1D case, and 
\begin{equation}
\begin{aligned}
	\frac{\hbar}{m_c} \frac{h}{\Delta_x^2} + \frac{\hbar}{m_c} \frac{h}{\Delta_y^2} \leq 1\end{aligned}
\end{equation}
for the 2D case.

We evaluate the CFL condition for the parameters considered in the simulations; see tables in the next section for respective values in the calculations below. For 1D (table~\ref{table2}):
\begin{equation}
	\frac{\hbar}{m_c} \frac{h}{\Delta_x^2} = 0.0232,
\end{equation}
and for 2D (table~\ref{table1}):
\begin{equation}
\frac{\hbar}{m_c} \frac{h}{\Delta_x^2} + \frac{\hbar}{m_c} \frac{h}{\Delta_y^2} = 0.3117.
\end{equation}

\section{Results}
\label{res}
Results of the polariton condensation in different parameter regimes as well those corresponding to different pumping schemes for both 1D and 2D microcavities will be presented here. We include cases when there are strong correlations between polaritons.

To observe effects such as the polariton blockade, conditions similar to which are necessary to implement purely quantum phenomena in exciton-polaritons, we need to move towards the strongly correlated polariton regime~\cite{liew2023future}. Polariton blockade is when interactions between polaritons are significant to the extent that excitation of one exciton-polariton in a microcavity prevents the excitation of another when the microcavity is impinged by a two-photon pulse~\cite{gerace2019quantum}. In other words, the interaction of one polariton on another results in a $\pi$-phase shift enabling the control-Z gate crucial for realizing universal quantum computation. This regime occurs when the polariton-polariton interaction strength ($g$ or $g_1$ and $g_2$) is greater than the polariton linewidth ($ \Gamma_p = \hbar \gamma_c $) or the dissipation rate ($\gamma_c$; which is inversely proportional to polariton lifetime) resulting in a blueshift greater than the linewidth~\cite{delteil2019towards,munoz2019emergence}. 

We plot the results for three values of this ratio: $g/\gamma_c = 1.132$, $g/\gamma_c = 10$ and $g/\gamma_c = 100$ for both coherent, near-resonant pumping and homogeneous, incoherent, non-resonant pumping. These values are inspired by the proposal to implement SWAP, and square-root SWAP gate (which is necessary to implement the C-NOT) with high fidelity (see Figure 3 in Ref.~\cite{ghosh2020quantum}) using exciton-polaritons. Moreover, experimentally, a value of the ratio slightly greater than 1 has been achieved~\cite{delteil2019towards,munoz2019emergence}. It is found that the number of polaritons added to the system decreases sharply up to single digits with increasing values of the ratio $g/\gamma_c$. This implies that the photons emitted from the microcavity (as the polaritons reconvert to photons) are strongly antibunched indicating the quantum polariton blockade regime.

The following parameters are inspired partly by Refs.~\cite{wouters2008spatial,gargoubi2016polariton,voronych2017numerical}. See the later text for explanation pertaining to the use of specific values for the parameters when relevant.

\begin{table}[h]
	\centering
	\begin{tabular}{lll}
		\hline
		\textbf{Parameter} & \textbf{Value} & \textbf{Unit}\\
		\hline
		$\hbar$ & 0.6582 & meV.ps \\
		$ N_c(t=0) $ & 1 & \\
		$ P (\text{also } P_0 = P) $ & 60.790 (or 25.835, 85.106, 0.6242) & $ \mu m^{-1} ps^{-1} $ \\
		$m_0$ & $5.677 \times 10^3$ & meV$/(\mu\text{m}/\text{ps})^2$ \\
		$m$ & $7.44 \times 10^{-5}\, m_0$ & meV$/(\mu\text{m}/\text{ps})^2$ \\
		$g_R$ & $0$ & meV $ \mu m^{1/2} $ \\
		$g$ & $0.86$ (or 0.015~\cite{gargoubi2016polariton}, 7.596, 75.96) & meV~$\mu$m \\
		$G$ & $0.0175$ & $\mu$m \\
		$\gamma_R$ & $2/\hbar $ & $ \mu m^{-1/2} ps^{-1} $ \\
		$\gamma_c$ & $0.5/\hbar $ & $ps^{-1}$ \\
		$R$ & $0.05/\hbar $ & $\mu$m $ps^{-1}$ \\
		$V_{\text{ext}}$ & $0$ & meV \\
		$\sigma_p$ & $20$ & $\mu$m \\
		cavsizex & 100 & $ \mu $m \\
		cavsizex (for 2D) & 24 & $ \mu $m \\
		cavsizey (for 2D) & 24 & $ \mu $m \\
		$ h $ & 0.001 & ps \\
		$ \eta $ & 1 & \\
		$ \delta_\omega $ & 0 & $ ps^{-1} $ \\
		$ F_p $ & 0.05 (or 0.5, 5) & meV~$\mu m^{-1/2}$ \\
		$ w $ & 10 & $ \mu m $ \\
		$ k_p (\text{also } k_{px} \text{ and } k_{py}) $ & 0 & $ \mu m^{-1} $ \\
		Mesh nodes x (xsize) & 201 (for 1D) and 241 (for 2D) & \\ 
		$ \Delta_x $ & cavsizex/(xsize - 1) & \\
		Mesh nodes y (ysize) & 241 (for 2D) & \\ 
		$ \Delta_y $ & cavsizey/(ysize - 1) & \\
		\hline
	\end{tabular}
	\caption{List of parameters used in the simulation of equations corresponding to CNRP2 and HINRP for both 1D and 2D cases. The units are given for the 1D case, but the 2D units can be derived straightforwardly.}
	\label{table1}
\end{table}

\begin{table}[h]
	\centering
	\begin{tabular}{lll}
		\hline
		\textbf{Parameter} & \textbf{Value} & \textbf{Unit} \\
		\hline
		$d$ & $5$ & $\mu\text{m}$ \\
		$F_p$ & $0.5$ & meV~$(\mu\text{m})^{-1/2}$ \\
		$k_p$ & $1$ & $1/\mu\text{m}$ \\
		$\delta_\omega$ & $5$ & $1/\text{ps}$ \\
		$w$ & $10$ & $\mu\text{m}$ \\
		$\delta$ & $5$ & meV \\
		$g$ & $\Gamma_p*1.132 $ (or 10, 100) & meV~$\mu\text{m}$ \\
		$ \Gamma_p $ & $ \hbar*\gamma_c $ & meV \\
		$\Omega_R$ & $4.4$ & meV \\
		$\gamma_x$ & $0.01$ & $1/\text{ps}$ \\
		$\gamma_c$ & $0.1$ & $1/\text{ps}$ \\
		$\text{cavsizex}$ & $100$ & $\mu\text{m}$ \\
		$h$ & $0.001$ & ps \\
		$m_0$ & $5.677 \times 10^3$ & meV$/(\mu\text{m}/\text{ps})^2$ \\
		$m_c$ & $m_0 \times 2 \times 10^{-5}$ & meV$/(\mu\text{m}/\text{ps})^2$ \\
		Mesh nodes x (xsize) & 201 & \\ 
		$ \Delta_x $ & cavsizex/(xsize - 1) & \\
		\hline
	\end{tabular}
	\caption{List of parameters used in the separate coupled equations for excitons and photons (polaritons), that is, CNRP1.}
	\label{table2}
\end{table}

\begin{table}[h]
	\centering
	\begin{tabular}{lll}
		\hline
		\textbf{Parameter} & \textbf{Value} & \textbf{Unit} \\
		\hline
		$d$ & $5$ & $\mu\text{m}$ \\
		$F_{pA}$ & $0.5$ & meV~$(\mu\text{m})^{-1/2}$ \\
		$k_{pA}$ & $1$ & $1/\mu\text{m}$ \\
		$\delta_{\omega A}$ & $5$ & $1/\text{ps}$ \\
		$w_A$ & $10$ & $\mu\text{m}$ \\
		$F_{pB}$ & $0.5$ & meV~$(\mu\text{m})^{1/2}$ \\
		$k_{pB}$ & $1$ & $1/\mu\text{m}$ \\
		$w_B$ & $10$ & $\mu\text{m}$ \\
		$\delta$ & $5$ & meV \\
		$g_1$ & $\Gamma_p*1.132 $ (or 10, 100) & meV~$\mu\text{m}$ \\
		$g_2$ & $\Gamma_p*0.1132 $ (or 1, 10) & meV~$\mu\text{m}$ \\
		$ \Gamma_p $ & $ \hbar*\gamma_c $ & meV \\
		$\gamma_x$ & $0.01$ & $1/\text{ps}$ \\
		$\gamma_c$ & $0.1$ & $1/\text{ps}$ \\
		\hline
	\end{tabular}
	\caption{List of parameters used in the separate coupled equations for excitons and photons (polaritons) (CNRP1) including the effect of spin. The parameters common to the earlier table are not shown.}
	\label{table3}
\end{table}

\subsection{Coherent, near-resonant pumping 1}
\label{CNRP1res}
\begin{figure}[htbp!]
	\centering
	\begin{subfigure}{0.48\textwidth}
		\centering
		\centering
\includegraphics[width=0.8\columnwidth]{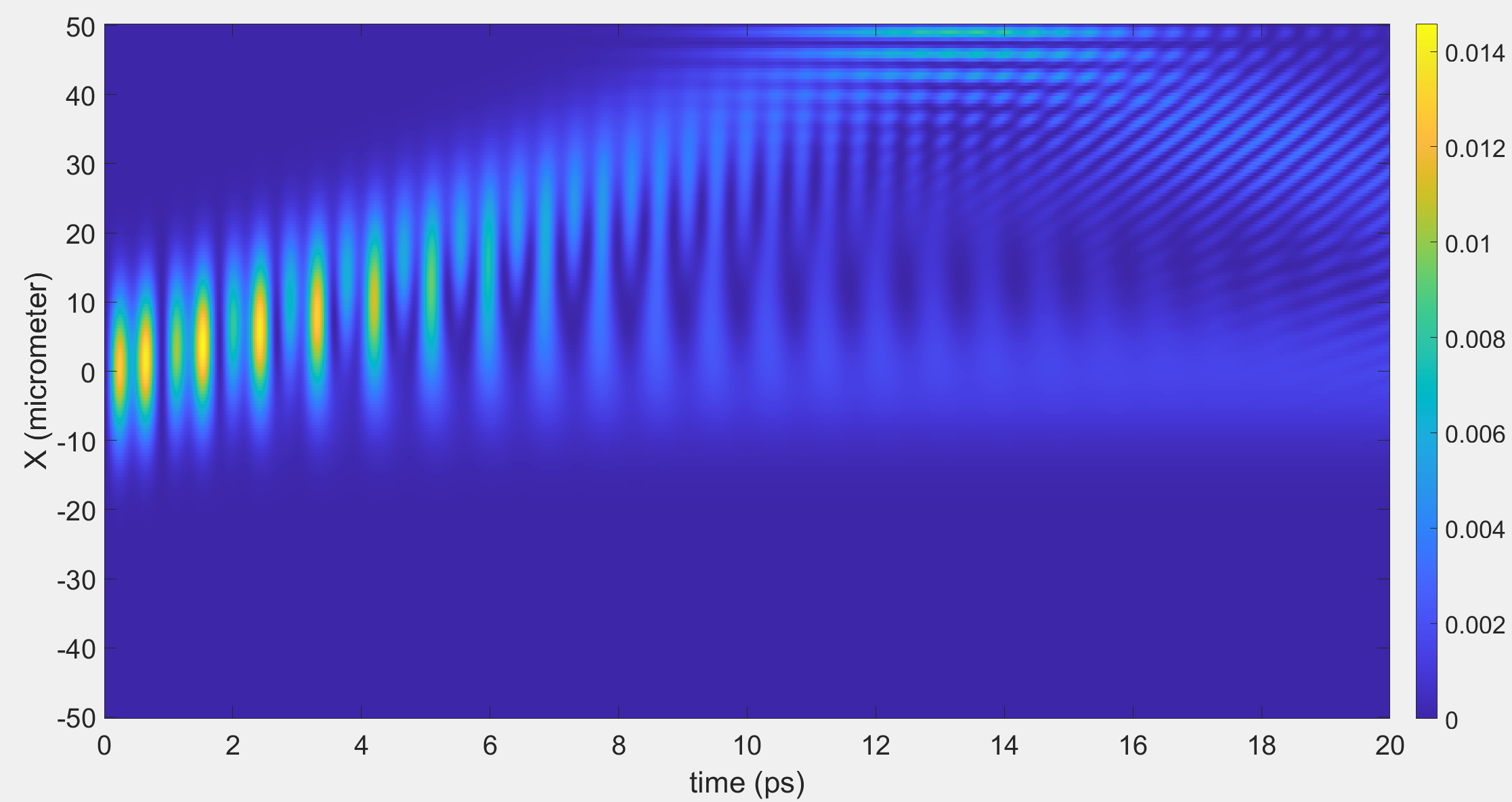}
		\caption{Photons}
		\label{psic_str_corr1}
	\end{subfigure}
	\hfill
	\begin{subfigure}{0.48\textwidth}
		\centering
		\includegraphics[width=0.8\columnwidth]{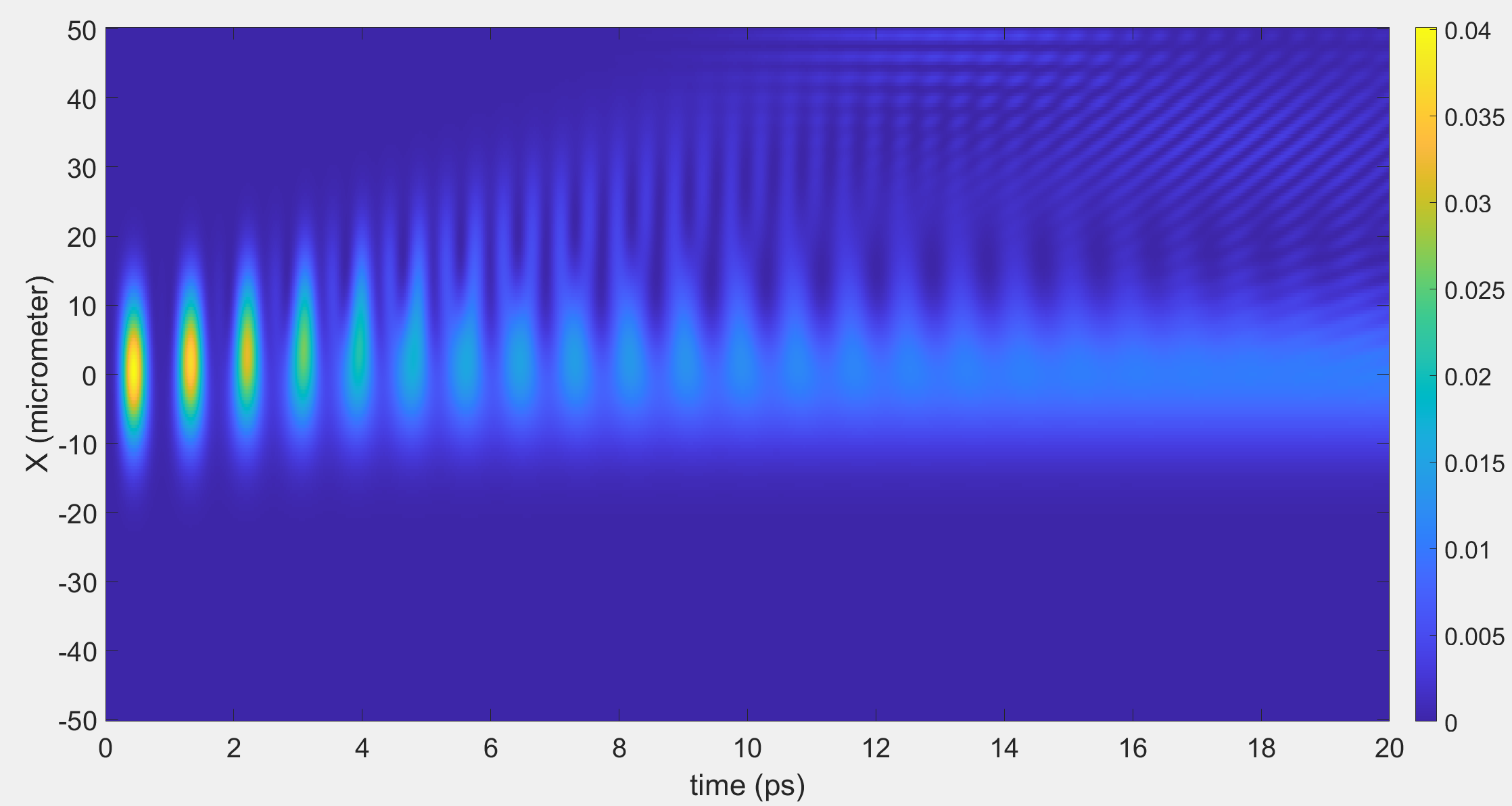}
		\caption{Excitons}
		\label{psix_str_corr1}
	\end{subfigure}
	\caption{Evolution of the density of exciton-polaritons against time with the ratio $g/\Gamma_p = 1.132$ for coherent, near-resonant pumping.}
	\label{CNRP1_str_corr1}
\centering
\end{figure}

\begin{figure}[htbp!]
	\centering
	\begin{subfigure}{0.48\textwidth}
		\centering
		\centering
\includegraphics[width=0.8\columnwidth]{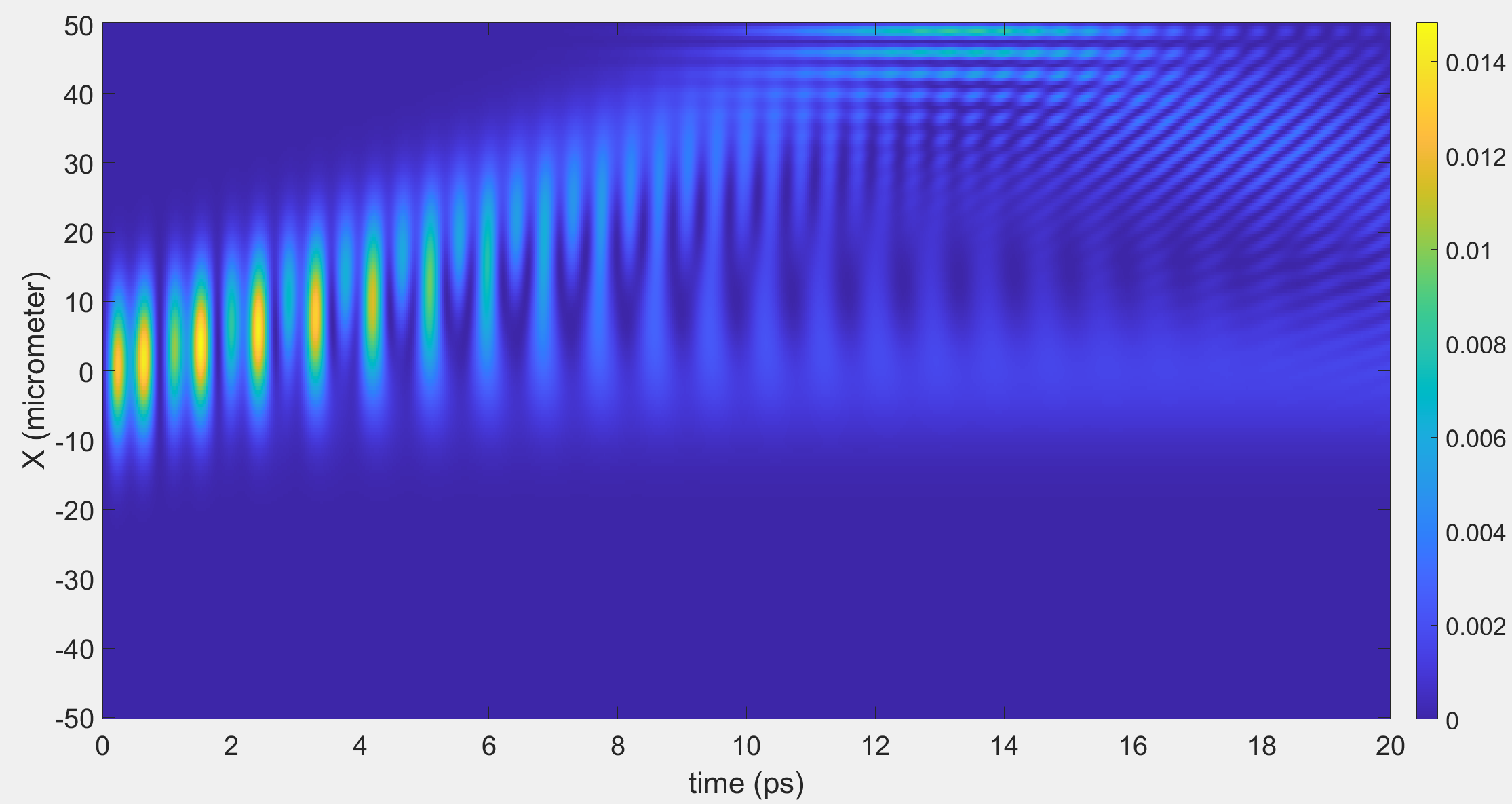}
		\caption{Photons}
		\label{psic_str_corr2}
	\end{subfigure}
	\hfill
	\begin{subfigure}{0.48\textwidth}
		\centering
		\includegraphics[width=0.8\columnwidth]{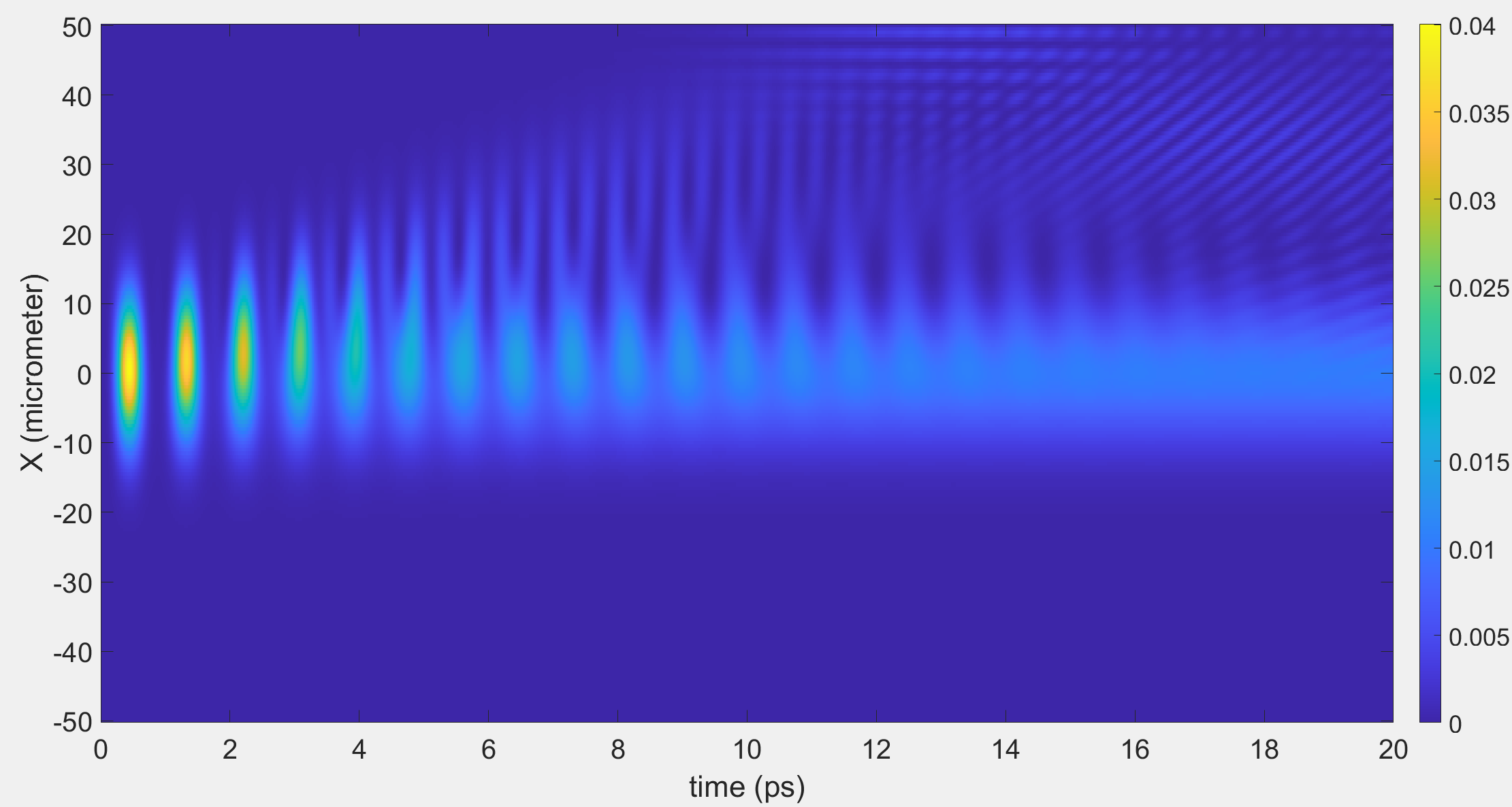}
		\caption{Excitons}
		\label{psix_str_corr2}
	\end{subfigure}
	\caption{Evolution of the density of exciton-polaritons against time with the ratio $g/\Gamma_p = 10$ for coherent, near-resonant pumping.}
	\label{CNRP1_str_corr2}
\centering
\end{figure}

\begin{figure}[htbp!]
	\centering
	\begin{subfigure}{0.48\textwidth}
		\centering
		\centering
\includegraphics[width=0.8\columnwidth]{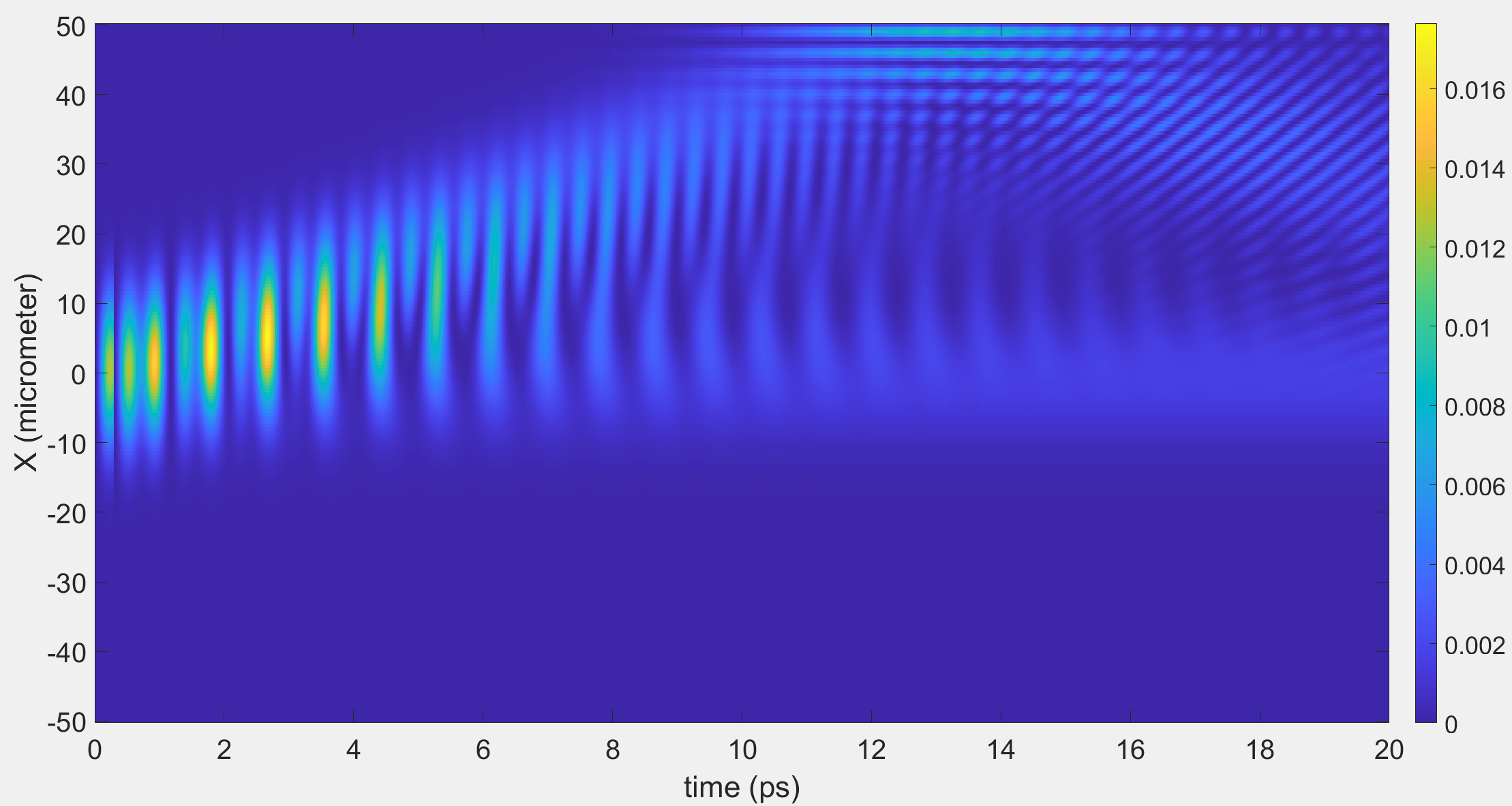}
		\caption{Photons}
		\label{psic_str_corr3}
	\end{subfigure}
	\hfill
	\begin{subfigure}{0.48\textwidth}
		\centering
		\includegraphics[width=0.8\columnwidth]{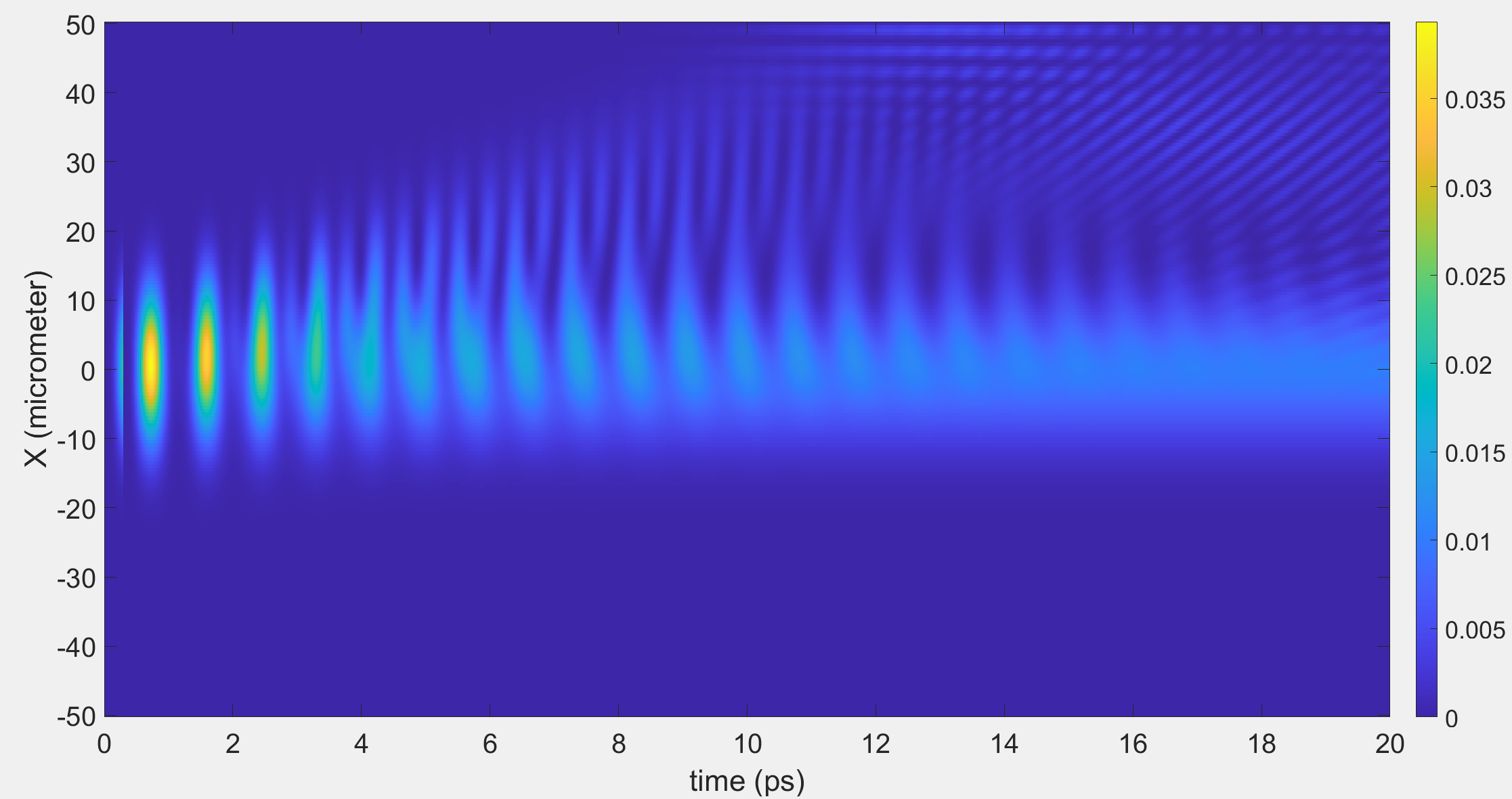}
		\caption{Excitons}
		\label{psix_str_corr3}
	\end{subfigure}
	\caption{Evolution of the density of exciton-polaritons against time with the ratio $g/\Gamma_p = 100$ for coherent, near-resonant pumping.}
	\label{CNRP1_str_corr3}
\centering
\end{figure}

For accessing the strongly correlated polariton regime, we have to consider the condition $g/\Gamma_p$, where $g$ is the polariton-polariton interaction strength and $\Gamma_p$ is the polariton linewidth. The linewidth is related to the polariton decay rate by the condition $ \Gamma_p = \hbar \gamma_c $ and we equate $g$ with the polariton-polariton interaction strength appearing in the mean-field equations~\ref{CNRP1eq}. We show the results for a 1D microcavity in Figures \ref{CNRP1_str_corr1}, \ref{CNRP1_str_corr2}, and \ref{CNRP1_str_corr3}. The density of photons is $ |\psi_c|^2 $ and the density of excitons is $ |\psi_x|^2 $.

We observe that there is a certain periodicity in the density of both excitons and photons and there is a spread in their distribution in both time and space. 

With time, the density of photons is spatially confined  between -15 $\mu m$ to 15 $\mu m$ initially and goes up to 50 $\mu m$ inside the cavity as time moves forward. It is also observed that as time passes, the density of photons gradually decreases. Beyond about 10 ps, the density of photons is delocalized in space, that is, several photon clusters are visible at the same time points. Between about 11 to 15 ps, the density peaks a little between 40 to 50 $\mu m$. There is almost no difference between the plots corresponding to $ U/\Gamma = 1.132 $ (figure \ref{psic_str_corr1}) and $ U/\Gamma = 10 $ (figure \ref{psic_str_corr2}) for photons. The plot corresponding to $ U/\Gamma = 100 $ (figure \ref{psic_str_corr3}) shows a slight increase in probability density as well as a rightward shift in time. Otherwise, the trend remains unchanged. 

The density of excitons is initially confined between -15 to 15 $\mu m$ and goes up to 50 $ \mu m $ inside the cavity as time moves forward. It is also observed that as time passes, the density of excitons gradually decreases. Beyond about 10 ps, the density of excitons is delocalized in space, that is, several exciton clusters are visible at the same time points. There is almost no difference between the plots corresponding to $ U/\Gamma = 1.132 $ ((figure \ref{psix_str_corr1})) and $ U/\Gamma = 10 $ ((figure \ref{psix_str_corr2})) for excitons. The plot corresponding to $ U/\Gamma = 100 $ ((figure \ref{psix_str_corr3})) shows a very slight (relative to the photon plots difference between $ U/\Gamma = 1.132 $ or $ U/\Gamma = 10 $ and $ U/\Gamma = 100 $) increase in probability density as well as a rightward shift in time. Otherwise, the trend remains unchanged.

In general, for the spinless case (figures \ref{CNRP1_str_corr1}, \ref{CNRP1_str_corr2}, and \ref{CNRP1_str_corr3}), there are more high density photon clusters relative to exciton clusters at the initial time. The photon density is higher than the exciton density. The exciton clusters, at initial times, are more spread out in space compared to the photon clusters.

\begin{figure}[htbp!]
	\centering
	\begin{subfigure}{0.48\textwidth}
		\centering
		\centering
\includegraphics[width=0.8\columnwidth]{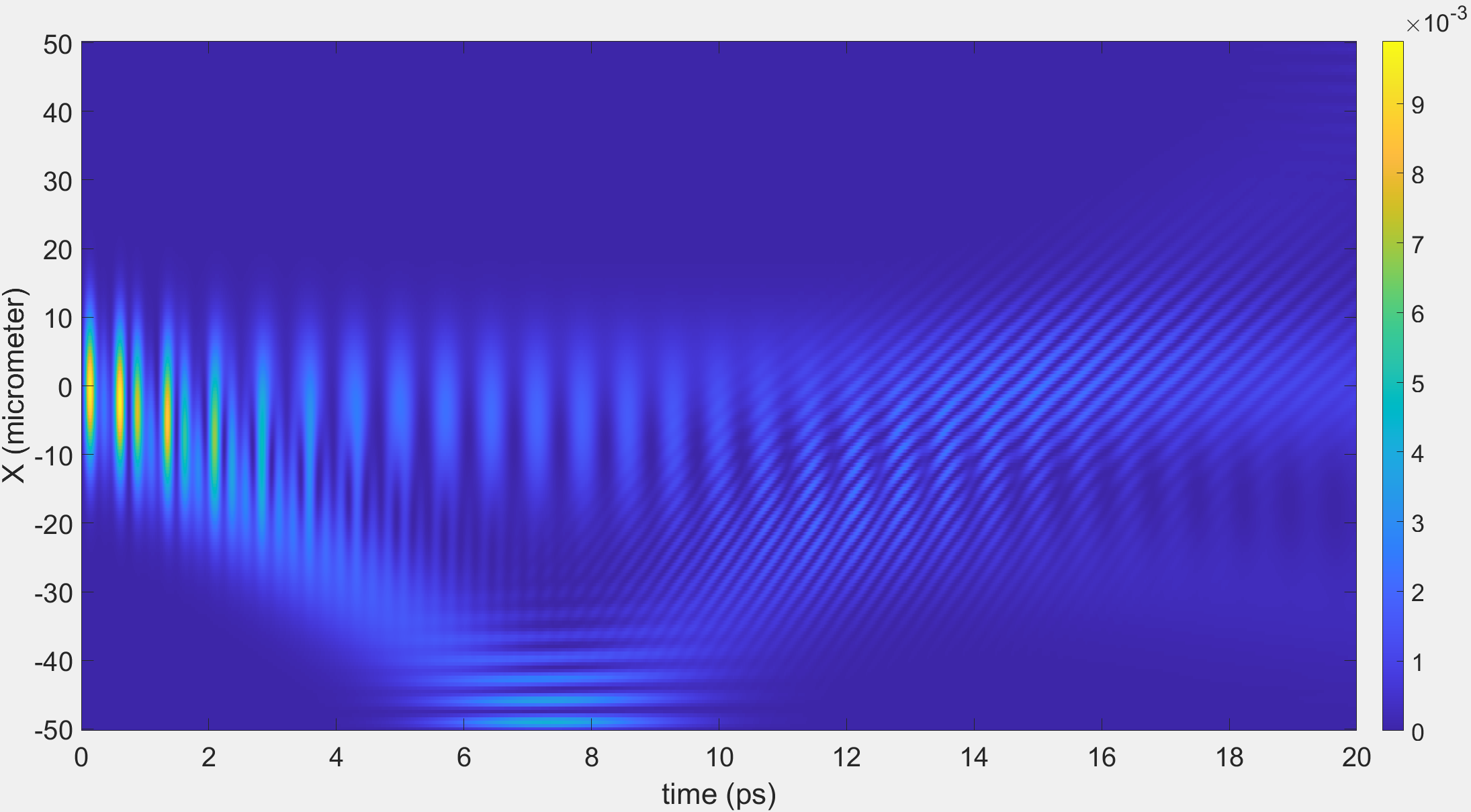}
		\caption{Photons}
		\label{psicA1_str_corr1_30sec}
	\end{subfigure}
	\hfill
	\begin{subfigure}{0.48\textwidth}
		\centering
		\includegraphics[width=0.8\columnwidth]{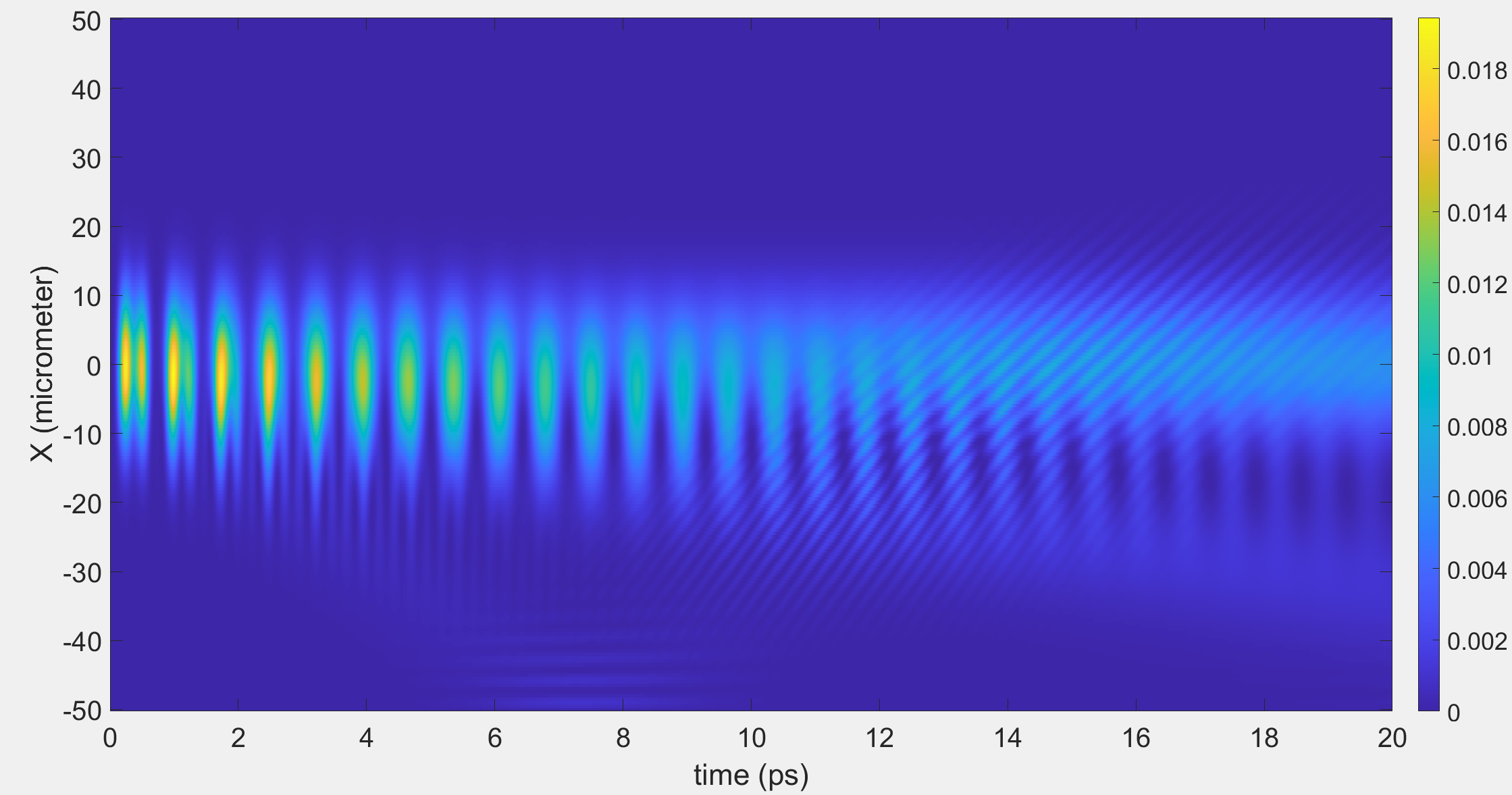}
		\caption{Excitons}
		\label{psixA1_str_corr1_30sec}
	\end{subfigure}
	\caption{Density of exciton-polaritons with the ratio $g/\Gamma_p = 1.132$ for coherent, near-resonant pumping for spin $\sigma = +1$.}
	\label{CNRP1A_str_corr1}
\centering
\end{figure}

\begin{figure}[htbp!]
	\centering
	\begin{subfigure}{0.48\textwidth}
		\centering
		\centering
\includegraphics[width=0.8\columnwidth]{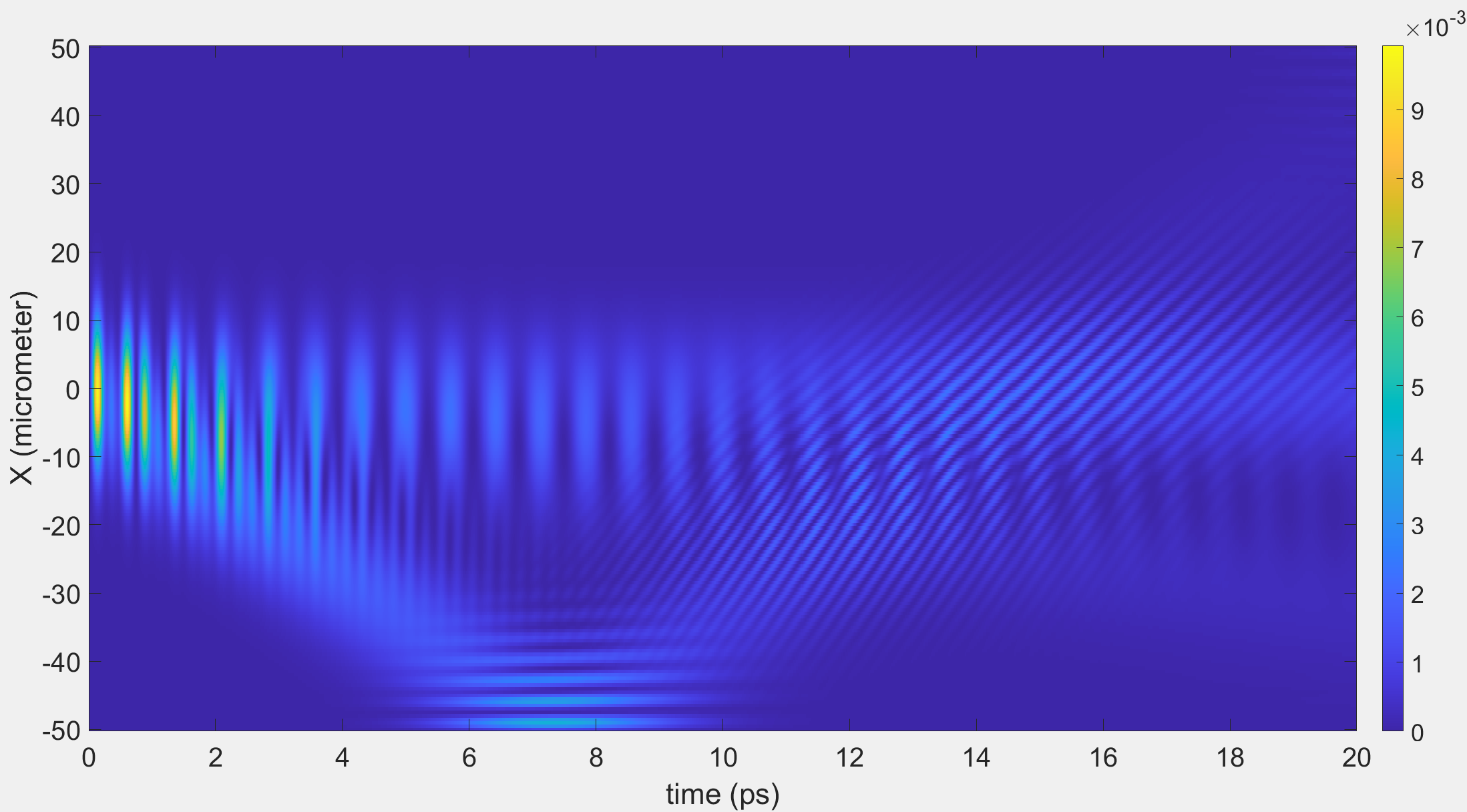}
		\caption{Photons}
		\label{psicA1_str_corr2_30sec}
	\end{subfigure}
	\hfill
	\begin{subfigure}{0.48\textwidth}
		\centering
		\includegraphics[width=0.8\columnwidth]{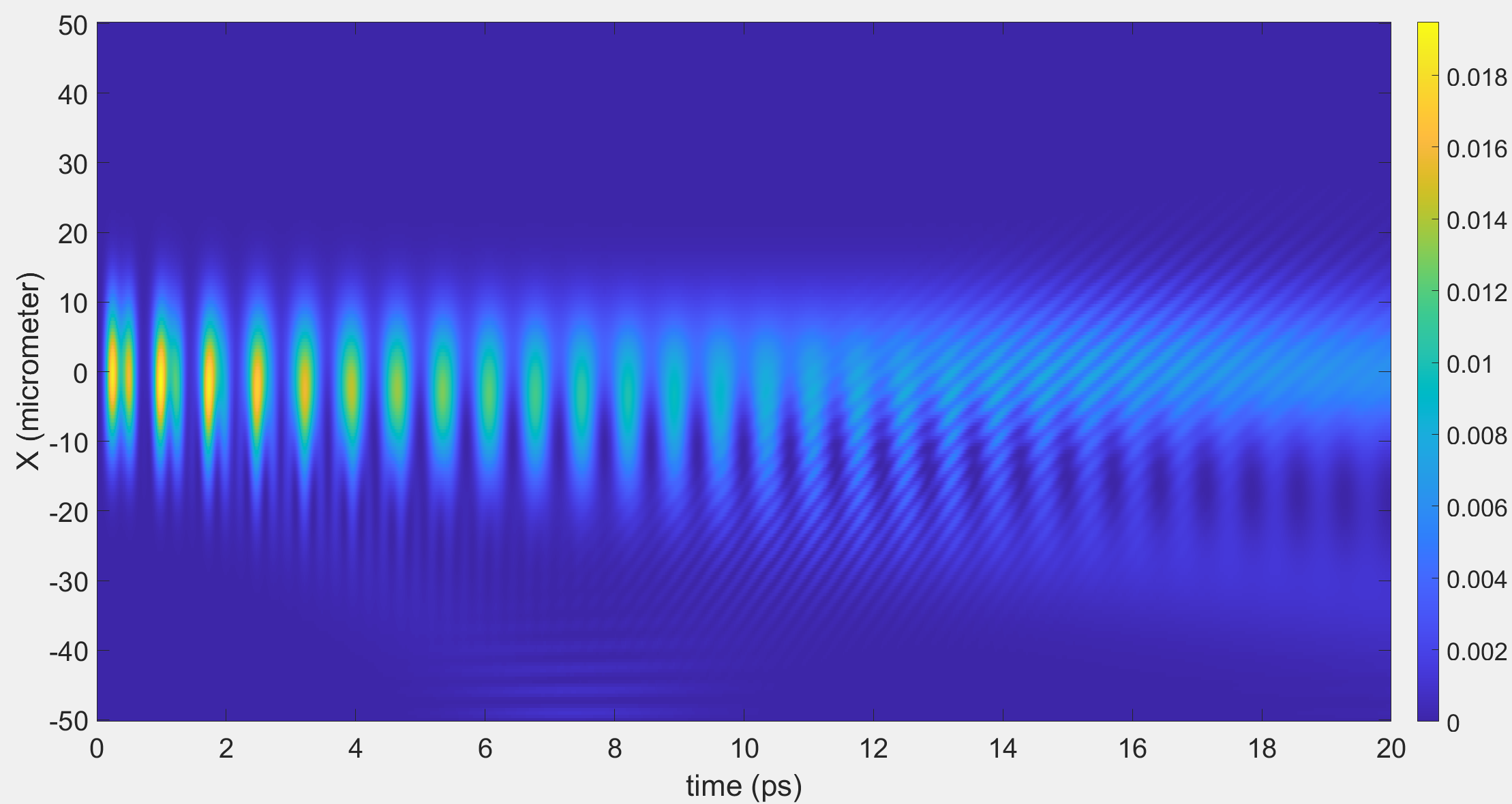}
		\caption{Excitons}
		\label{psixA1_str_corr2_30sec}
	\end{subfigure}
	\caption{Density of exciton-polaritons with the ratio $g/\Gamma_p = 10$ for coherent, near-resonant pumping for spin $\sigma = +1$.}
	\label{CNRP1A_str_corr2}
\centering
\end{figure}

\begin{figure}[htbp!]
	\centering
	\begin{subfigure}{0.48\textwidth}
		\centering
		\centering
\includegraphics[width=0.8\columnwidth]{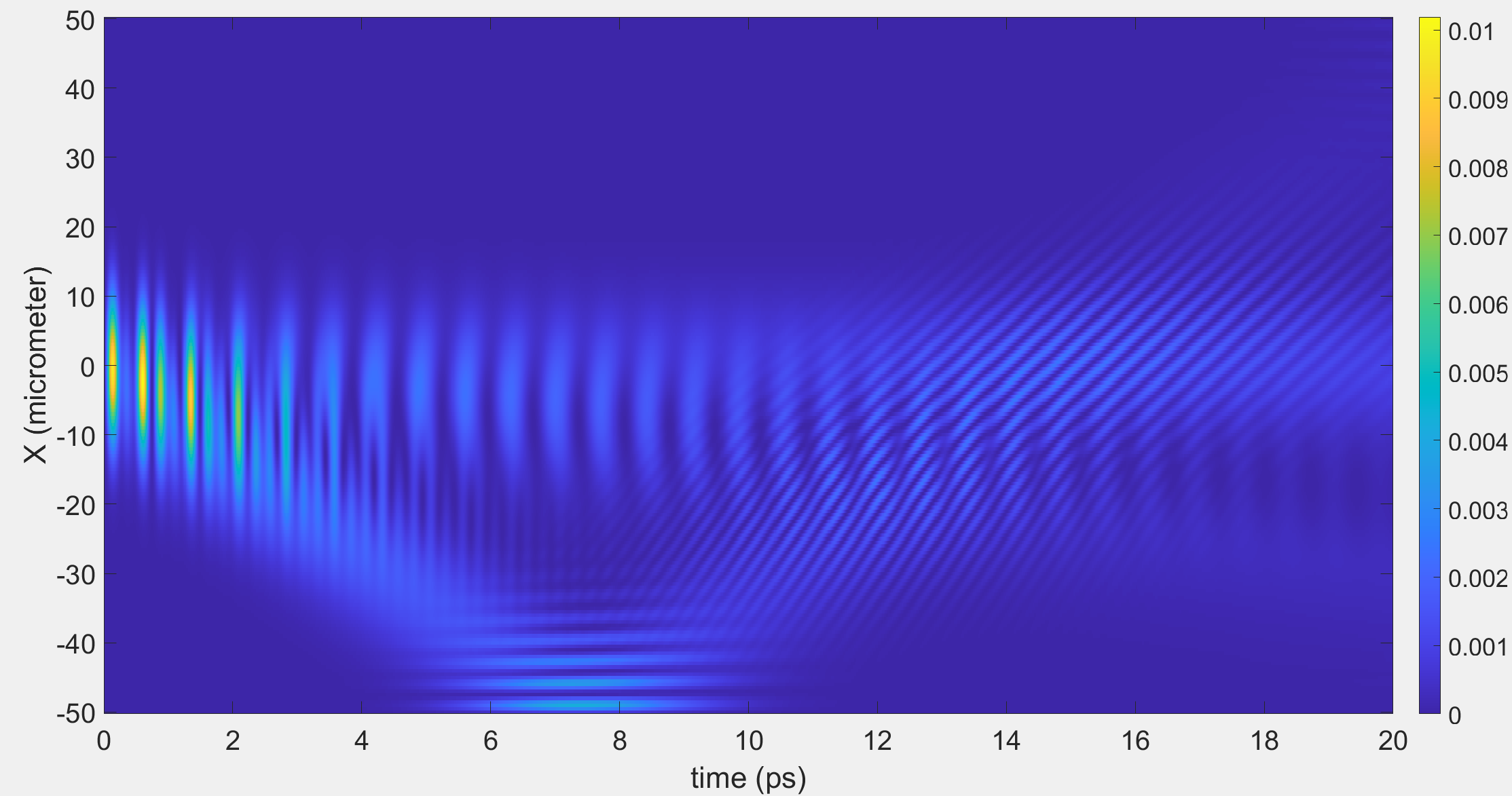}
		\caption{Photons}
		\label{psicA1_str_corr3_30sec}
	\end{subfigure}
	\hfill
	\begin{subfigure}{0.48\textwidth}
		\centering
		\includegraphics[width=0.8\columnwidth]{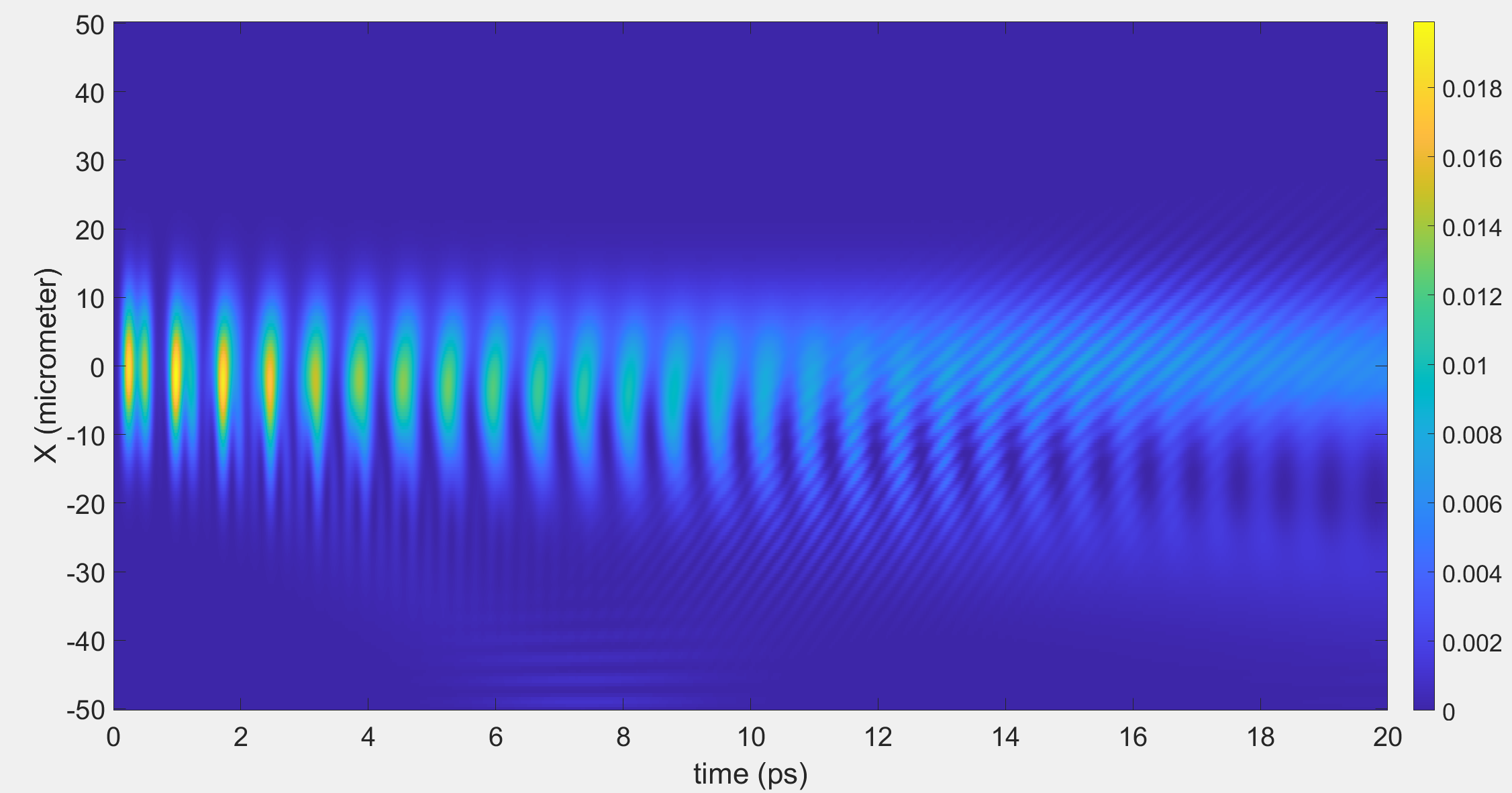}
		\caption{Excitons}
		\label{psixA1_str_corr3_30sec}
	\end{subfigure}
	\caption{Density of exciton-polaritons with the ratio $g/\Gamma_p = 100$ for coherent, near-resonant pumping for spin $\sigma = +1$.}
	\label{CNRP1A_str_corr3}
\centering
\end{figure}

\begin{figure}[htbp!]
	\centering
	\begin{subfigure}{0.48\textwidth}
		\centering
		\centering
\includegraphics[width=0.8\columnwidth]{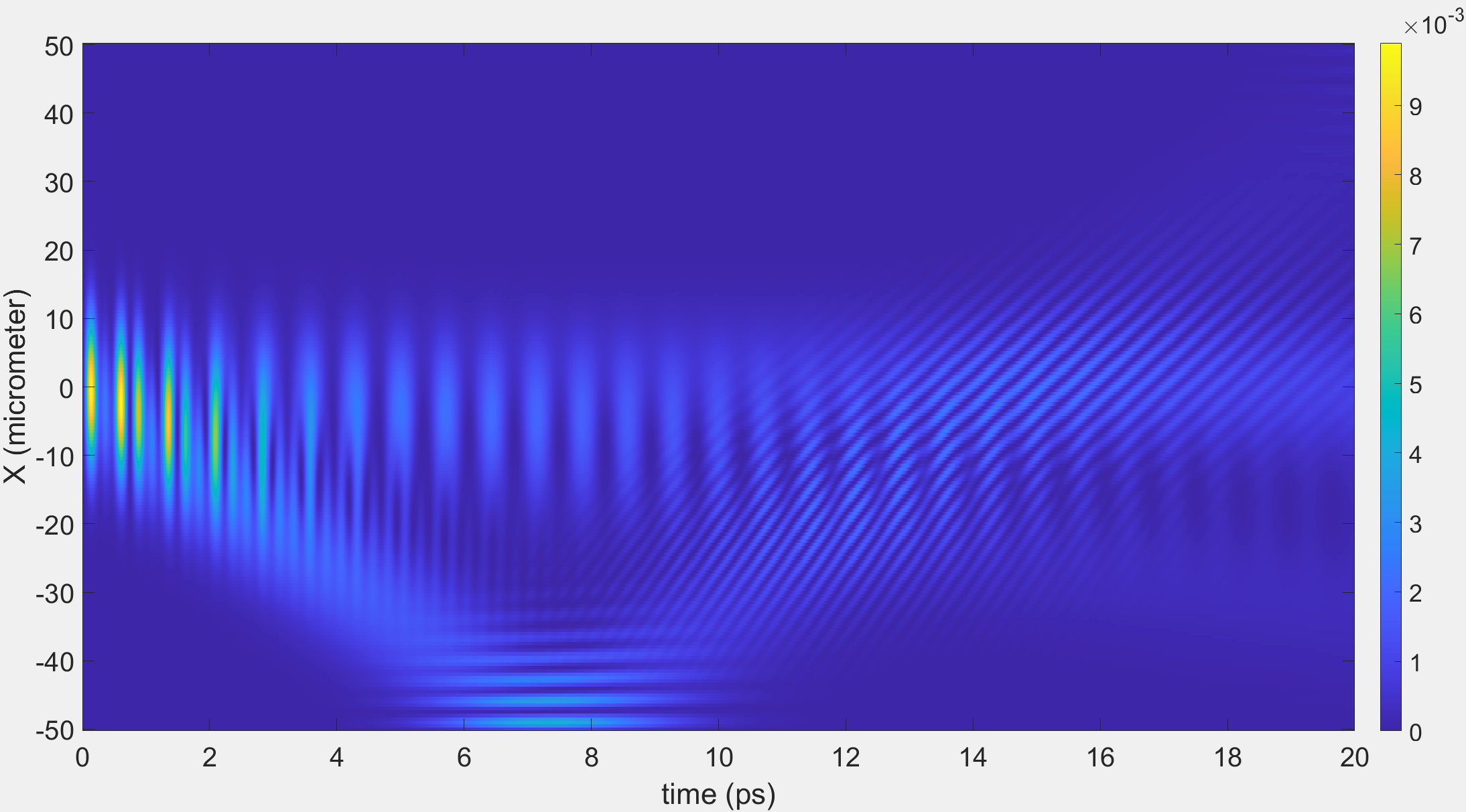}
		\caption{Photons}
		\label{psicB1_str_corr1_30sec}
	\end{subfigure}
	\hfill
	\begin{subfigure}{0.48\textwidth}
		\centering
		\includegraphics[width=0.8\columnwidth]{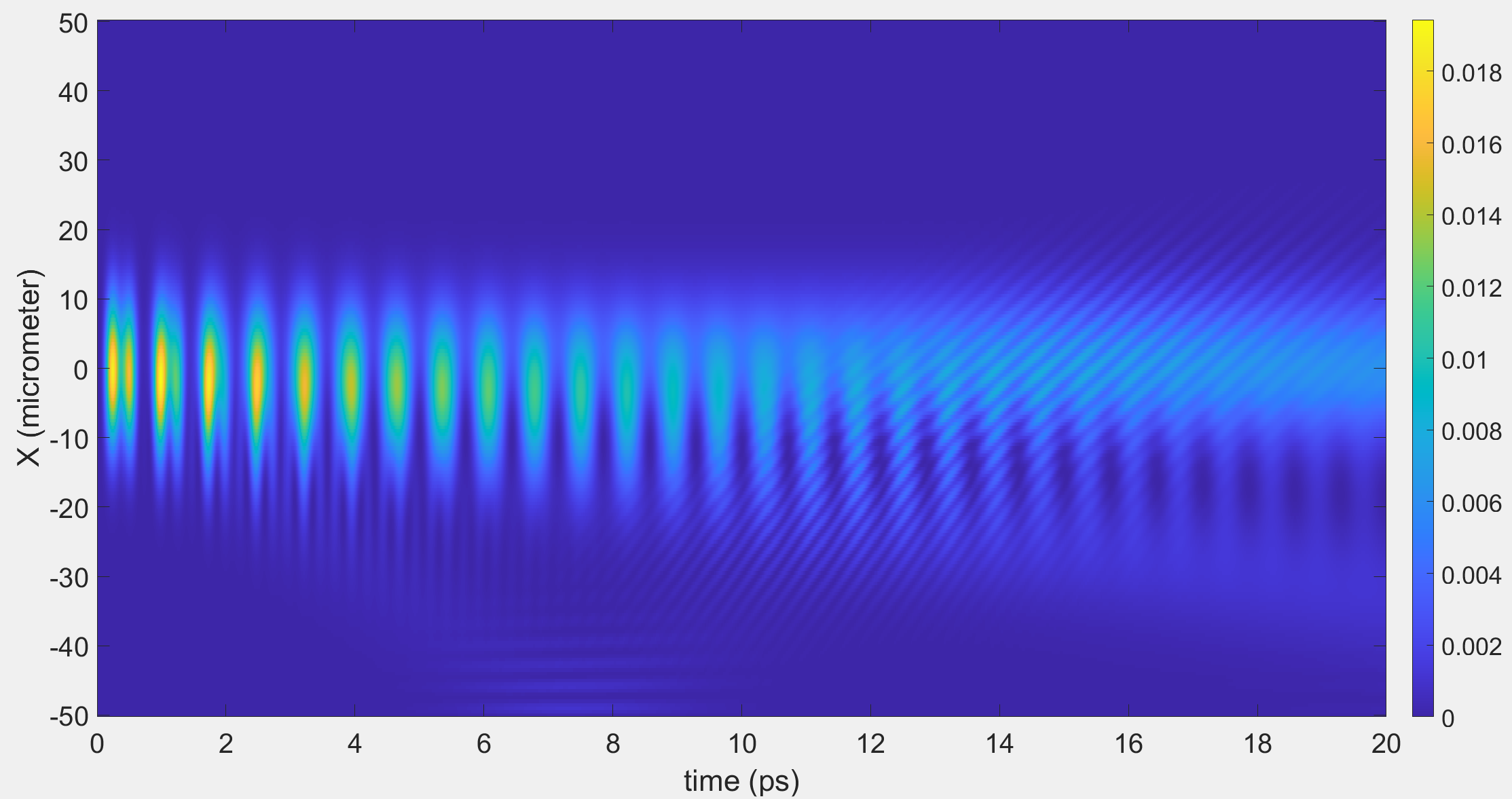}
		\caption{Excitons}
		\label{psixB1_str_corr1_30sec}
	\end{subfigure}
	\caption{Density of exciton-polaritons with the ratio $g/\Gamma_p = 1.132$ for coherent, near-resonant pumping for spin $\sigma = -1$.}
	\label{CNRP1B_str_corr1}
\centering
\end{figure}

\begin{figure}[htbp!]
	\centering
	\begin{subfigure}{0.48\textwidth}
		\centering
		\centering
\includegraphics[width=0.8\columnwidth]{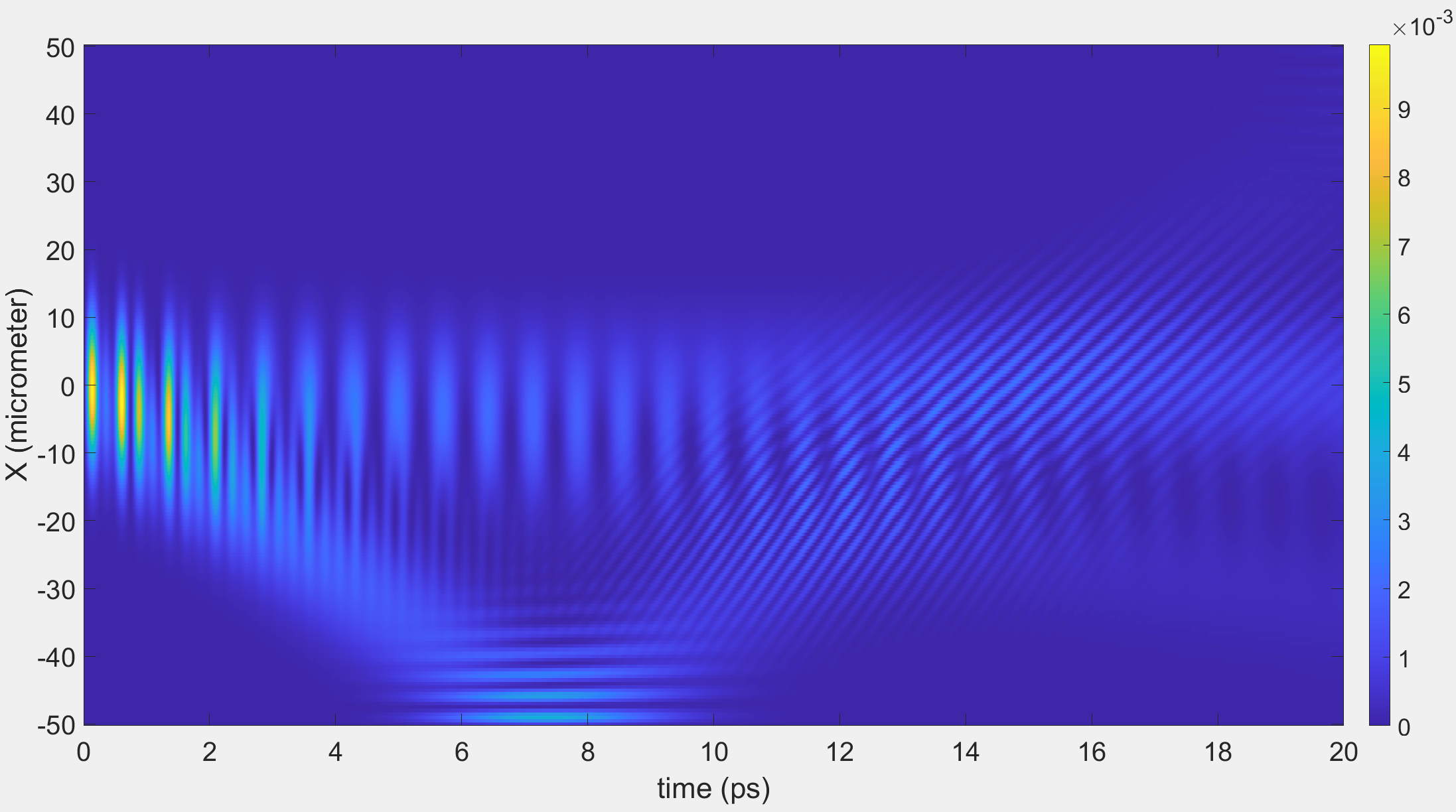}
		\caption{Photons}
		\label{psicB1_str_corr2_30sec}
	\end{subfigure}
	\hfill
	\begin{subfigure}{0.48\textwidth}
		\centering
		\includegraphics[width=0.8\columnwidth]{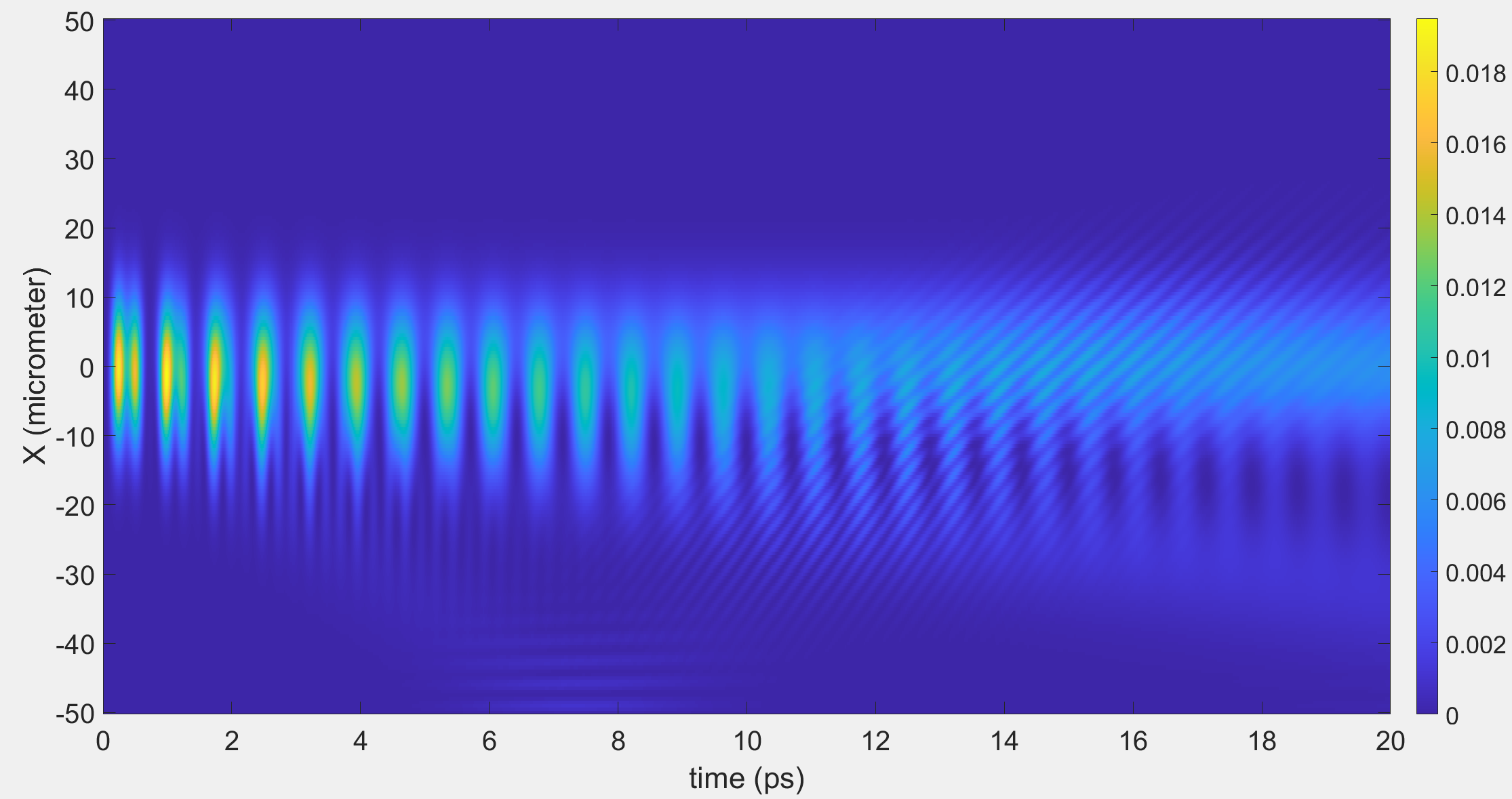}
		\caption{Excitons}
		\label{psixB1_str_corr2_30sec}
	\end{subfigure}
	\caption{Density of exciton-polaritons with the ratio $g/\Gamma_p = 10$ for coherent, near-resonant pumping for spin $\sigma = -1$.}
	\label{CNRP1B_str_corr2}
\centering
\end{figure}

\begin{figure}[htbp!]
	\centering
	\begin{subfigure}{0.48\textwidth}
		\centering
		\centering
\includegraphics[width=0.8\columnwidth]{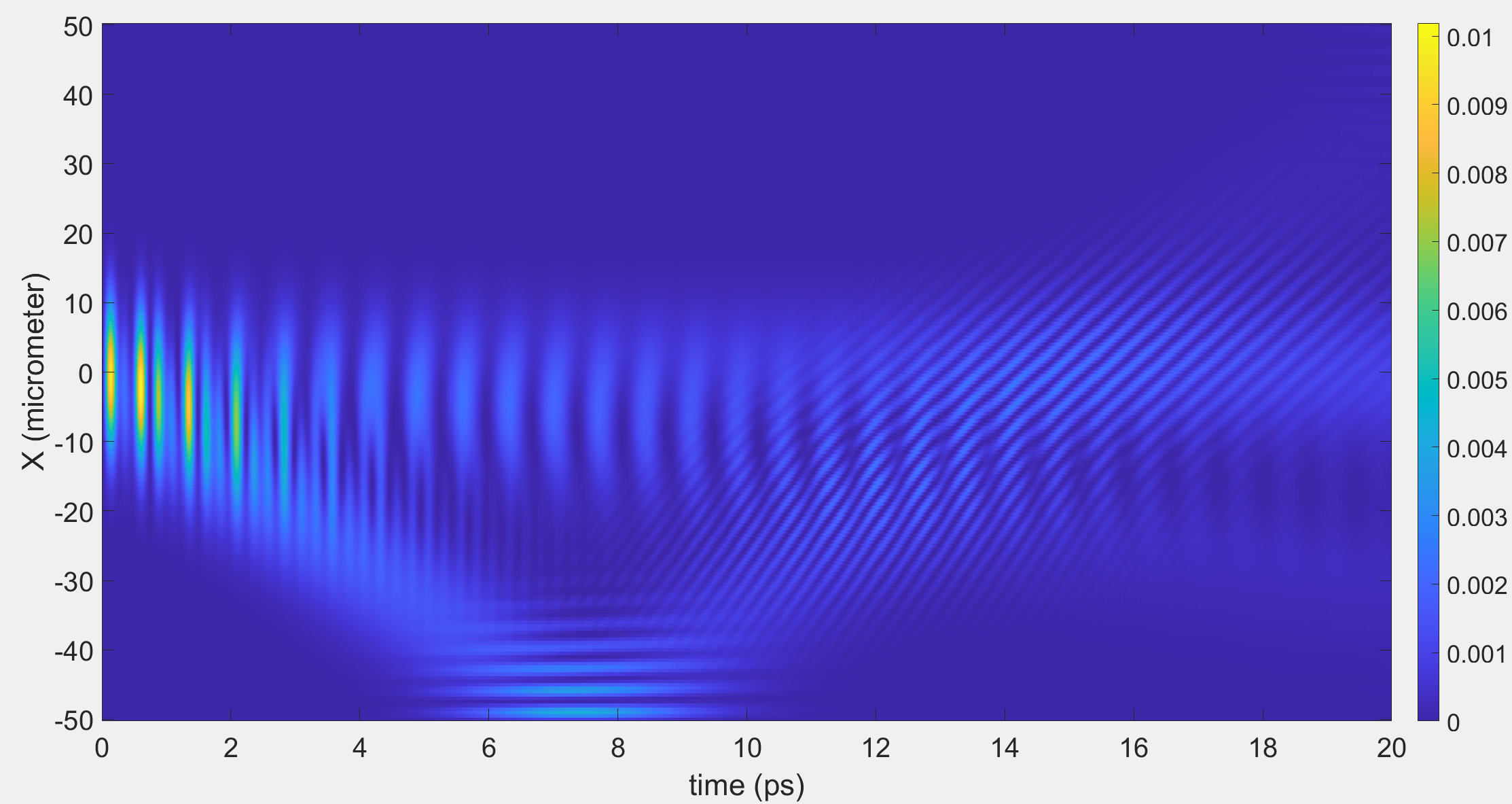}
		\caption{Photons}
		\label{psicB1_str_corr3_30sec}
	\end{subfigure}
	\hfill
	\begin{subfigure}{0.48\textwidth}
		\centering
		\includegraphics[width=0.8\columnwidth]{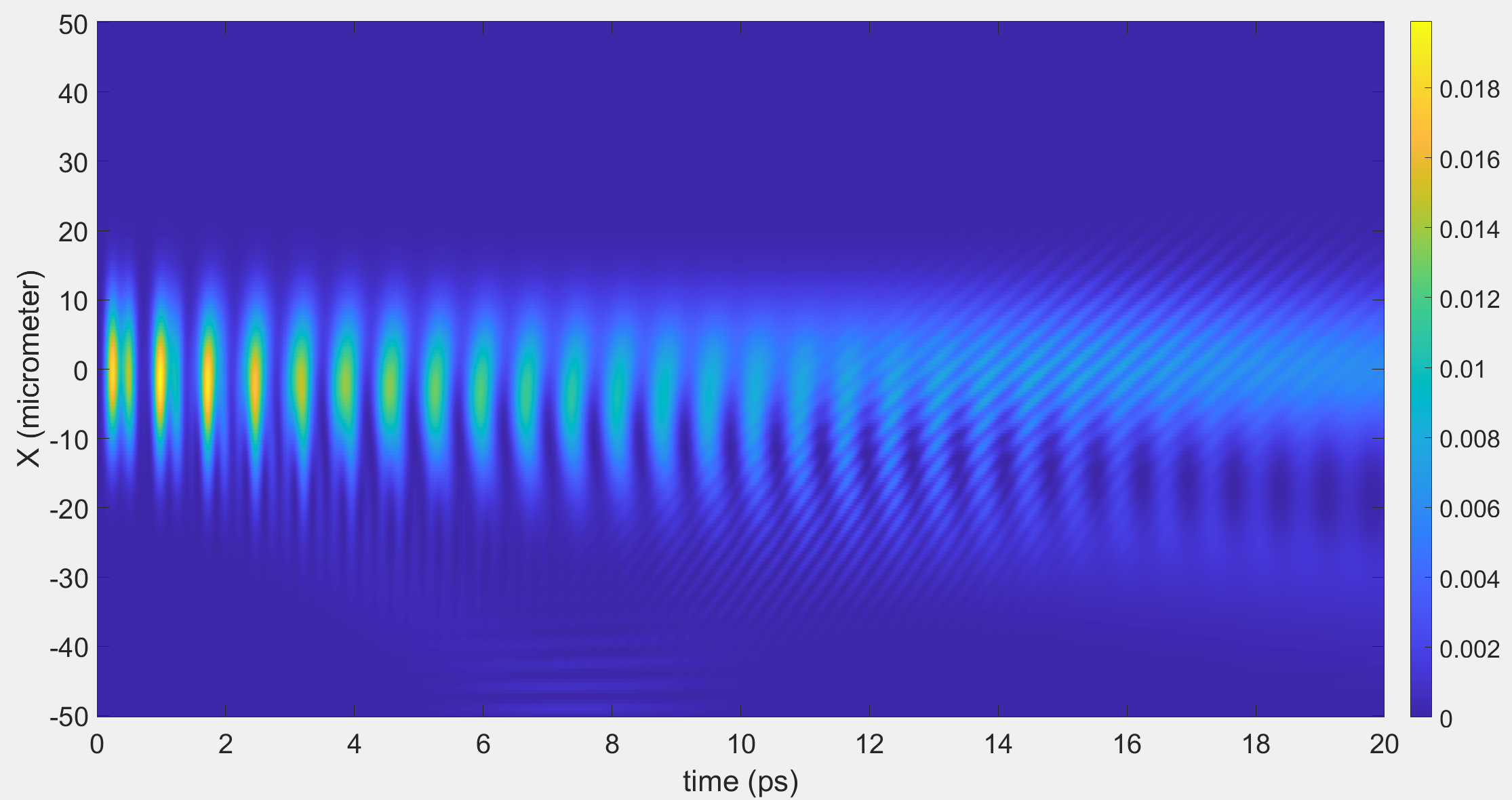}
		\caption{Excitons}
		\label{psixB1_str_corr3_30sec}
	\end{subfigure}
	\caption{Density of exciton-polaritons with the ratio $g/\Gamma_p = 100$ for coherent, near-resonant pumping for spin $\sigma = -1$.}
	\label{CNRP1B_str_corr3}
\centering
\end{figure}

The results including the effect of spin are shown in Figures \ref{CNRP1A_str_corr1}, \ref{CNRP1A_str_corr2}, \ref{CNRP1A_str_corr3}, \ref{CNRP1B_str_corr1}, \ref{CNRP1B_str_corr2}, and \ref{CNRP1B_str_corr3}. 
Theoretically, it is expected that the same spin interaction constant is much higher than the opposite spin one~\cite{snoke2023reanalysis} and experimentally the difference was found to be 10 times~\cite{cuevas2018first}. We thus consider the same spin interaction strength $g_1$ to be 10 times the opposite spin interaction strength $g_2$ in Equation~\ref{CNRP1eqspin}.
 
We observe that there is a certain periodicity in the density of both excitons and photons and there is a spread in their distribution in both time and space. This is true for both $ \sigma = +1 $ and $ \sigma = -1 $ spins. 

In the following, we analyze the trend for photons with $\sigma = +1$:

There is a negative shift in space of the probability density for photons with the peak value attained only in the first 4 clusters. It is also observed that as time passes, the density of photons gradually decreases. Between about 10 to 15 ps, the density of photons peaks a little between -40 to -50 $ \mu m $ in contrast to the without spin case. There is almost no difference between the plots corresponding to $ U/\Gamma = 1.132 $ (figure \ref{psicA1_str_corr1_30sec}) and $ U/\Gamma = 10 $ (figure \ref{psicA1_str_corr2_30sec}) for photons. The plot corresponding to $ U/\Gamma = 100 $ (figure \ref{psicA1_str_corr3_30sec}) shows a slight increase in probability density. Otherwise, the trend remains unchanged.

Observations corresponding to $ \sigma = -1 $ are identical to the above.

In the following, we analyze the trend for excitons with $\sigma = +1$:

We observe that there is a certain periodicity in the density of both excitons and photons and there is a spread in their distribution in both time and space. There is a general increase in probability density of excitons relative to photons. As time progresses, the probability density sees a general decrease for excitons with the peak value attained in the first cluster with it decreasing gradually. There is almost no difference between the plots corresponding to $ U/\Gamma = 1.132 $ (figure \ref{psixA1_str_corr1_30sec}), $ U/\Gamma = 10 $ (figure \ref{psixA1_str_corr2_30sec}), and $ U/\Gamma = 100 $ (figure \ref{psixA1_str_corr3_30sec}) for excitons (this is in slight contrast to the trend observed for photons). Beyond 8 ps, there are multiple probability density clusters visible at several spatial points across multiple times.

Observations corresponding to $ \sigma = -1 $ are identical to the above.

\subsection{Coherent, near-resonant pumping 2}
\label{CNRP2res}
\begin{figure}[htbp!]
	\centering
	\centering
\includegraphics[width=0.8\columnwidth]{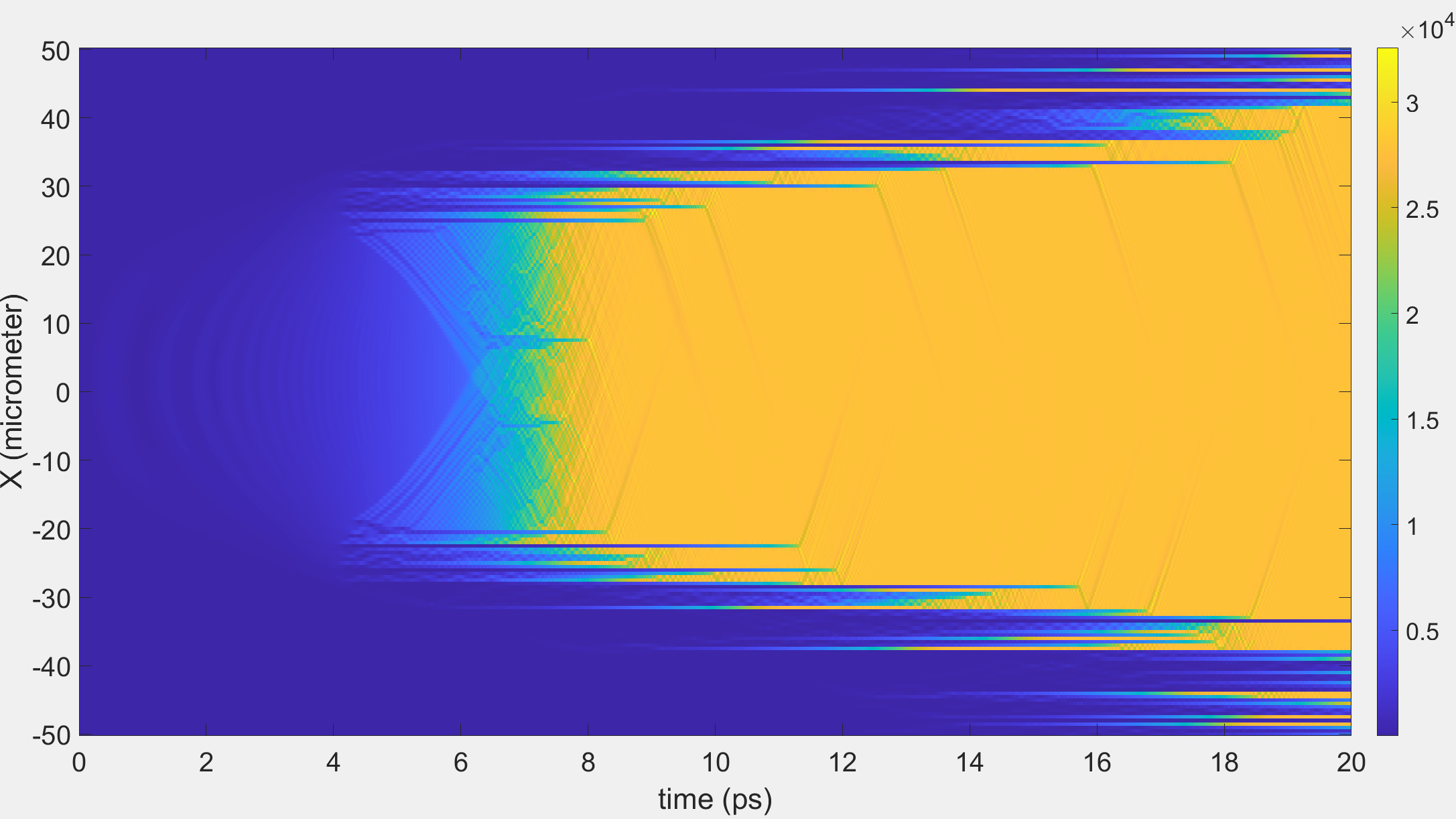}
	\caption{Polariton condensation corresponding to coherent, near-resonant pumping for 1D microwire microcavity}
	\label{DDGPE_psi_x_respump_20ps}
\centering
\end{figure}

We plot the results for polariton condensation corresponding to coherent, near-resonant pumping 2 given by Equation \ref{CNRP2eq}. The results corresponding to 1D microwire are shown in Figure~\ref{DDGPE_psi_x_respump_20ps}. The polariton condensation starts at about 4.2 ps and the number of polaritons in the condensate is of the order of $10^4$.

\begin{figure}[htbp!]
	\centering
	\begin{subfigure}{0.48\textwidth}
		\centering
		\centering
\includegraphics[width=0.6\columnwidth]{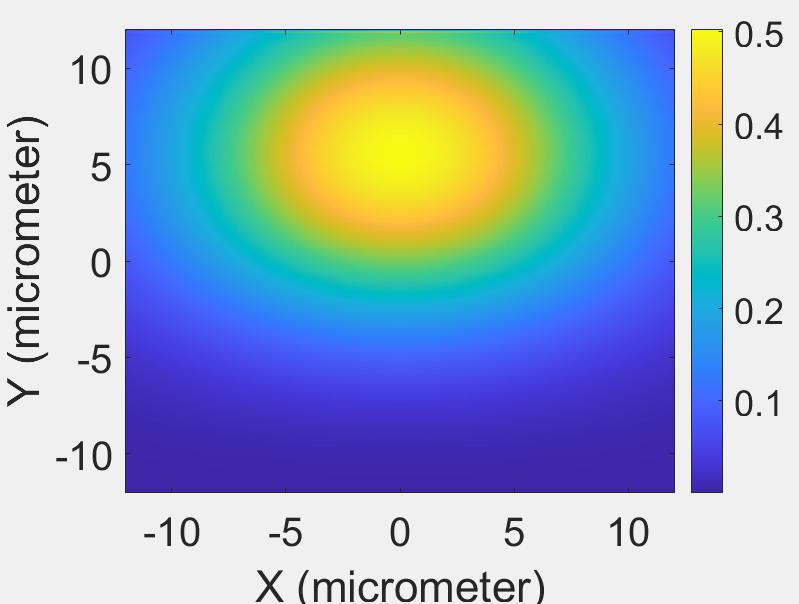}
		\caption{$t = 0$ s}
		\label{DDGPE_psi_x_2D_respump_0sec}
	\end{subfigure}
	\hfill
	\begin{subfigure}{0.48\textwidth}
		\centering
		\includegraphics[width=0.6\columnwidth]{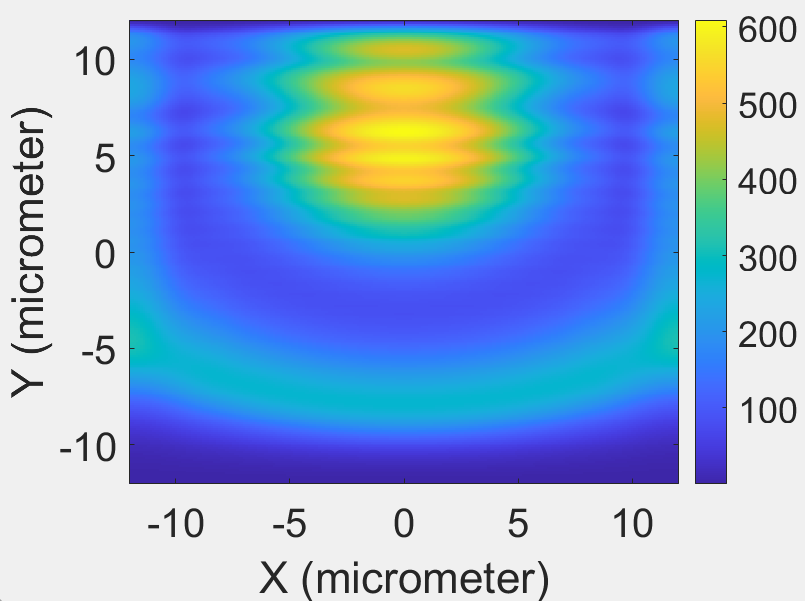}
		\caption{$t = 1$ ps}
		\label{DDGPE_psi_x_2D_respump_1ps}
	\end{subfigure}
	\vfill
	\begin{subfigure}{0.48\textwidth}
		\centering
		\includegraphics[width=0.6\columnwidth]{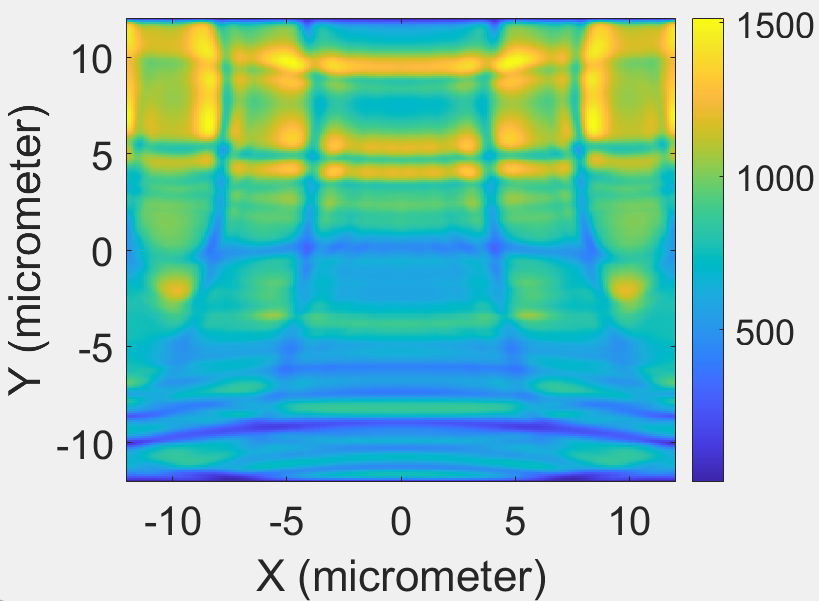}
		\caption{$t = 2.5$ ps}
		\label{DDGPE_psi_x_2D_respump_2.5ps}
	\end{subfigure}
	\hfill
	\begin{subfigure}{0.48\textwidth}
		\centering
		\includegraphics[width=0.6\columnwidth]{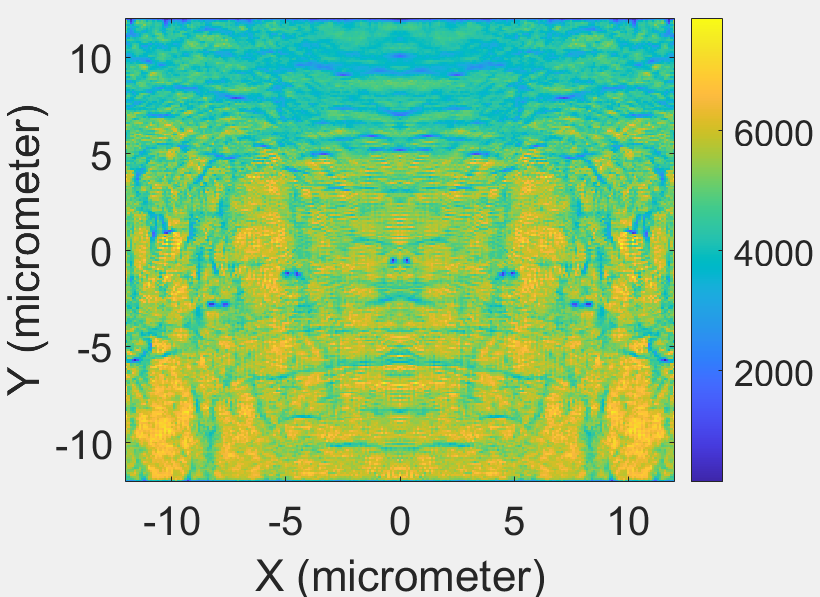}
		\caption{$t = 5$ ps}
		\label{DDGPE_psi_x_2D_respump_5ps}
	\end{subfigure}
	\vfill
	\begin{subfigure}{0.48\textwidth}
		\centering
		\includegraphics[width=0.6\columnwidth]{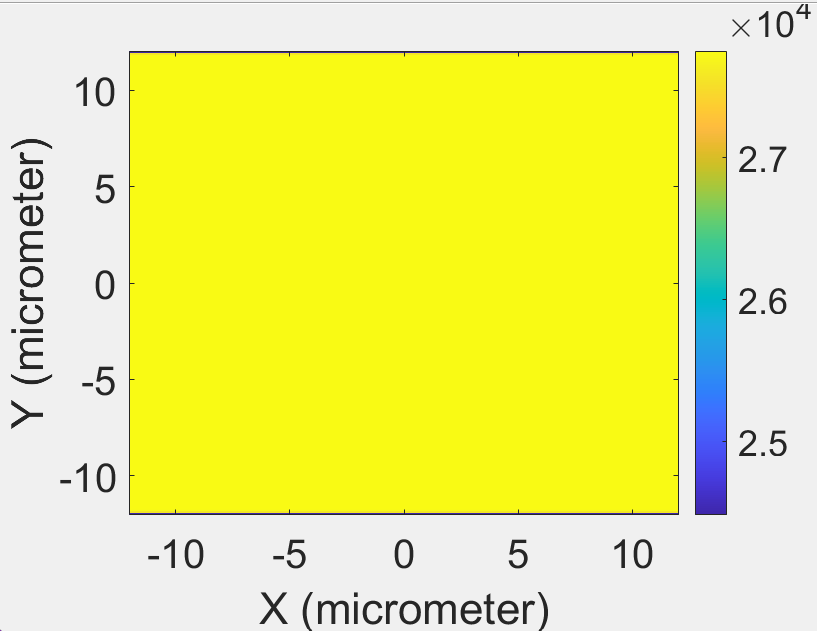}
		\caption{$t = 10$ ps}
		\label{DDGPE_psi_x_2D_respump_10ps}
	\end{subfigure}
	\caption{Density of condensate polaritons for two-dimensional microcavity with coherent, near-resonant pumping at different times with $ F_p = 0.5 $ meV~$(\mu\text{m})^{-1}$. The initial distribution is Gaussian in x and y.}
	\label{CNRP2_psi_x_2D}
\centering
\end{figure}

For the 2D case, for coherent, near-resonant pumping, the plots are given in Figure~\ref{CNRP2_psi_x_2D}. While the polariton condensation is initially (at 1 ps) symmetric about the X axis, it is not so about the Y axis. The polaritons seems to majorly condense along the positive side of the Y axis. The number of polaritons are between 100-600 at 1 ps, and gradually peak to the order of $ 10^4 $ at 10 ps. This sequence of figures provides a detailed snapshot of the polariton condensation numerical trend relative to 1D figure~\ref{DDGPE_psi_x_respump_20ps}.

\begin{figure}[htbp!]
	\centering
	\begin{subfigure}{0.48\textwidth}
		\centering
		\centering
\includegraphics[width=0.8\columnwidth]{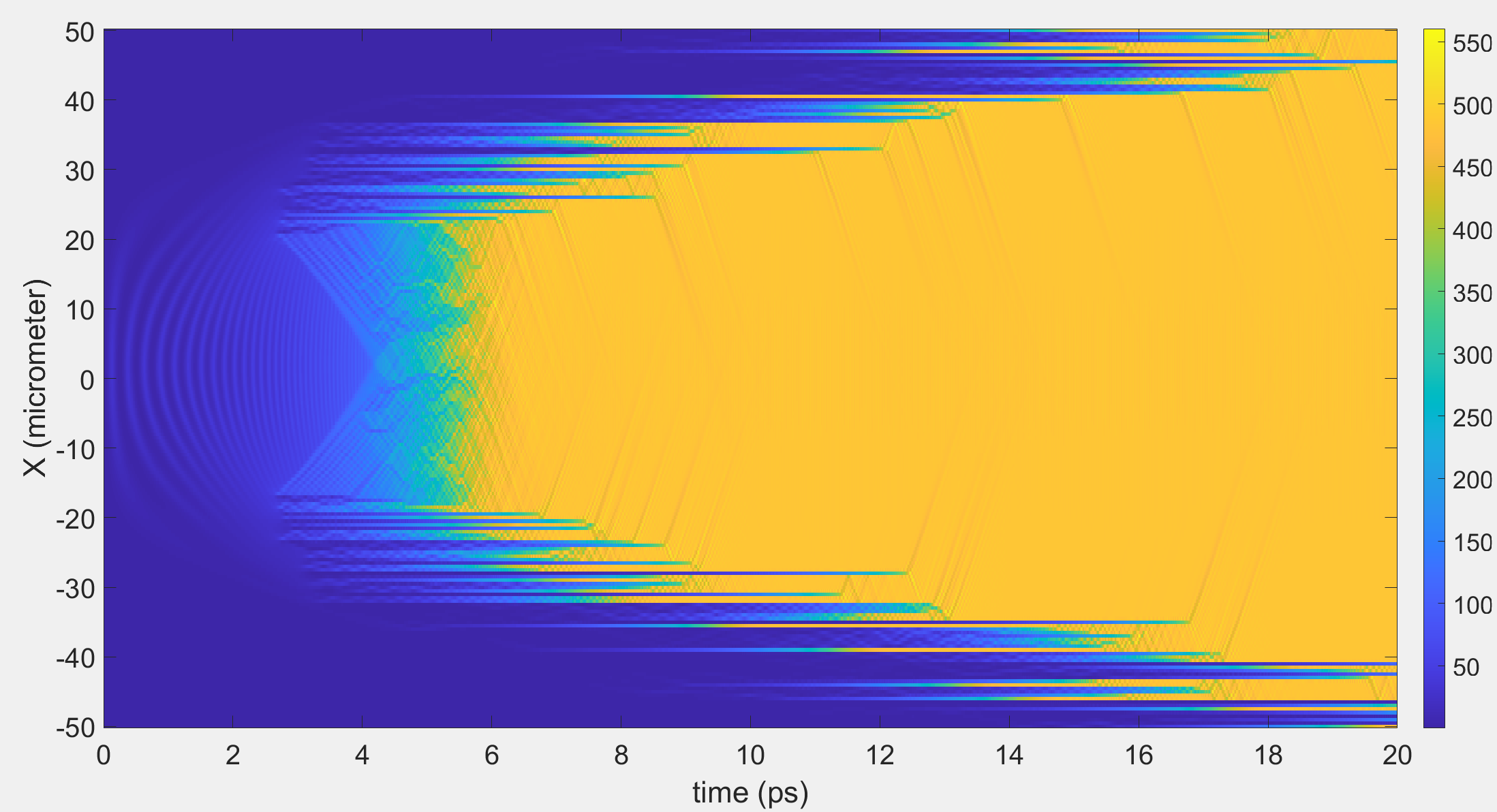}
		\caption{$F_p = 0.5$ $meV/\mu m^{1/2}$}
		\label{DDGPE_psi_x_respump_str_corr1_Fpdef_20ps}
	\end{subfigure}
	\hfill
	\begin{subfigure}{0.48\textwidth}
		\centering
		\includegraphics[width=0.8\columnwidth]{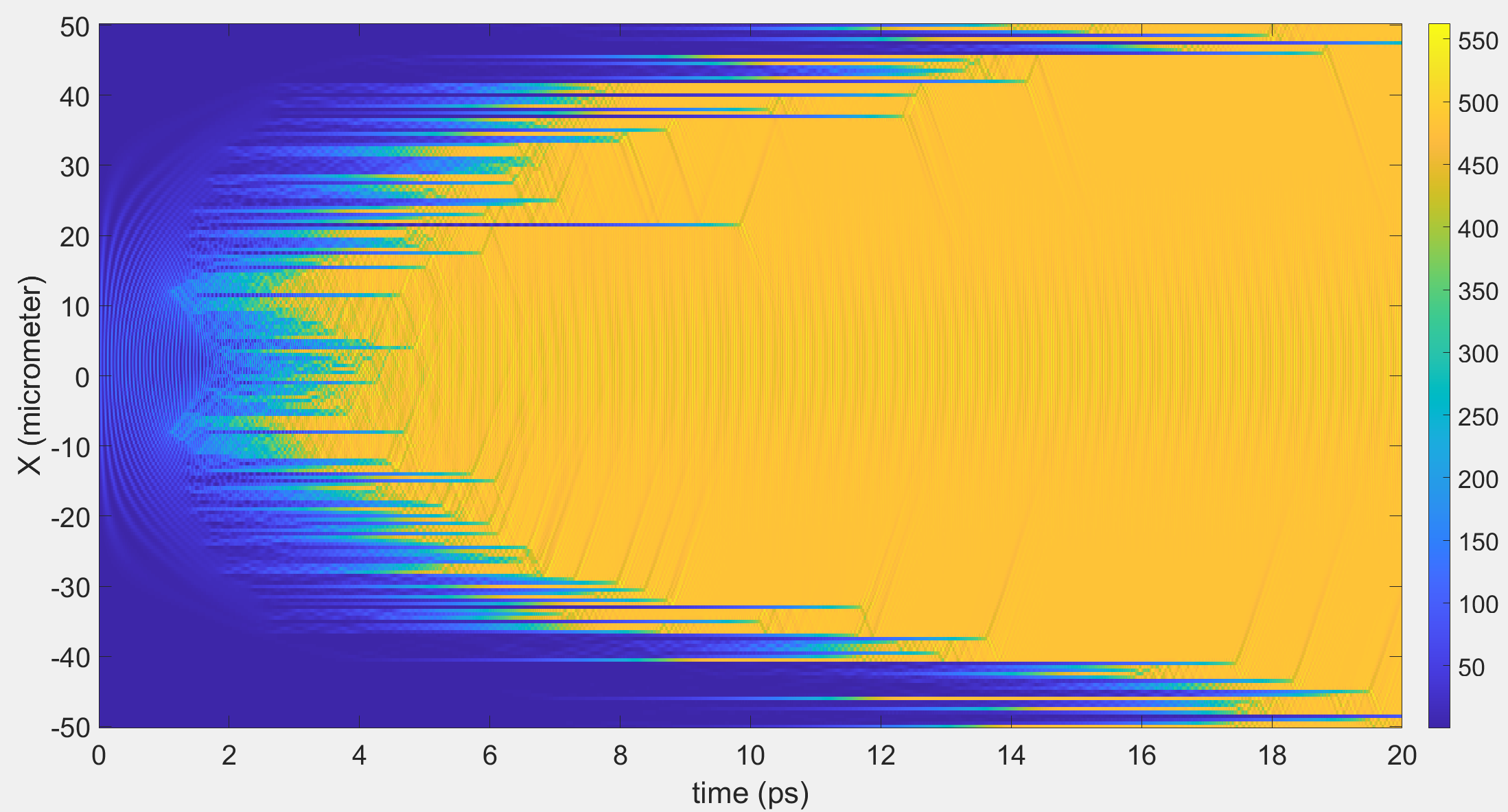}
		\caption{$F_p = 5$ $meV/\mu m^{1/2}$}
		\label{DDGPE_psi_x_respump_str_corr1_Fp19_20ps}
	\end{subfigure}
	\caption{Density of the condensate polaritons with the ratio $g/\gamma_c = 1.132$ for coherent, near-resonant pumping.}
	\label{CNRP2_psi_x_str_corr1}
\centering
\end{figure}

\begin{figure}[htbp!]
	\centering
	\begin{subfigure}{0.48\textwidth}
		\centering
		\centering
\includegraphics[width=0.8\columnwidth]{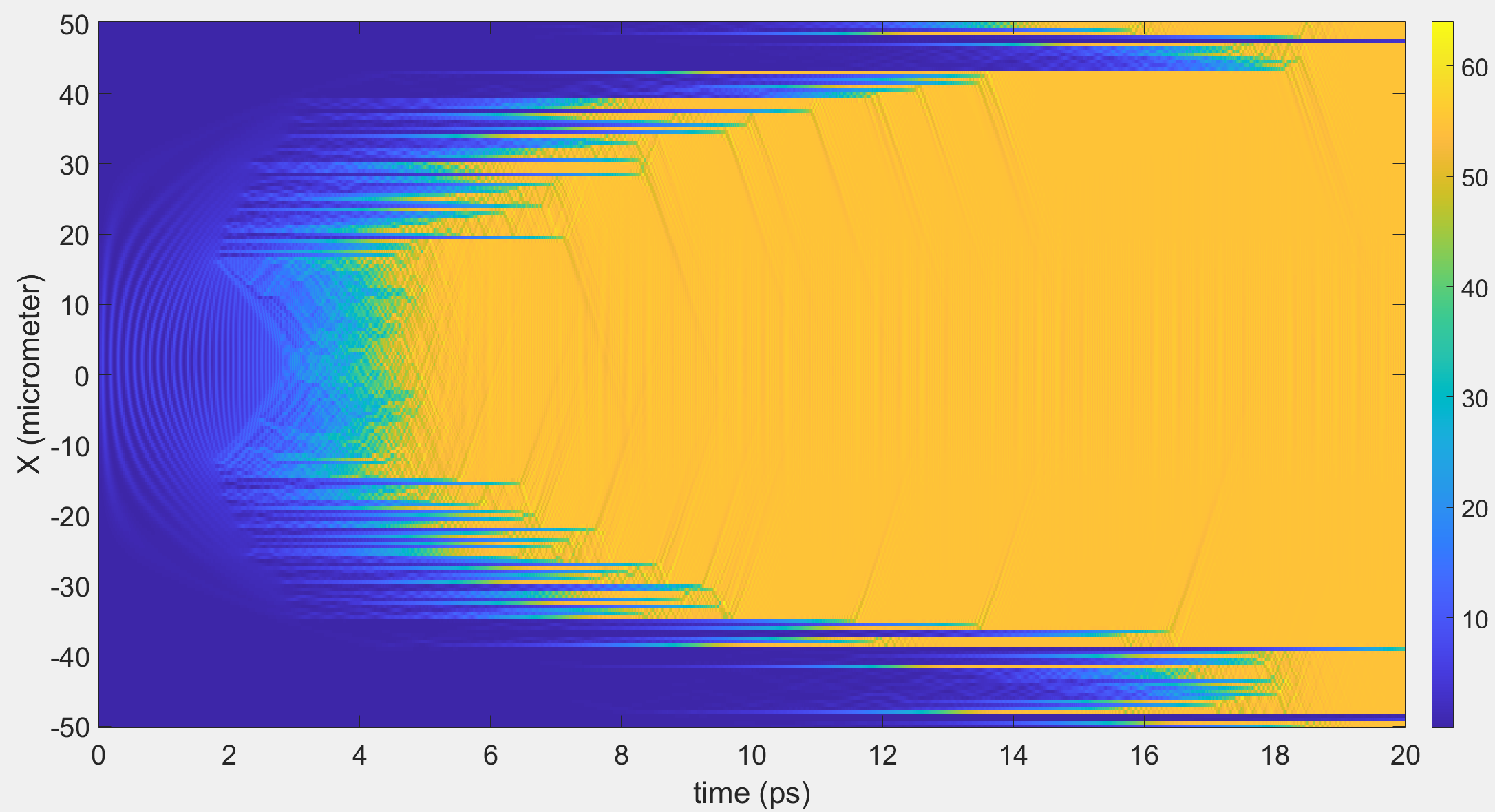}
		\caption{$F_p = 0.5$ $meV/\mu m^{1/2}$}
		\label{DDGPE_psi_x_respump_str_corr2_Fpdef_20ps}
	\end{subfigure}
	\hfill
	\begin{subfigure}{0.48\textwidth}
		\centering
		\includegraphics[width=0.8\columnwidth]{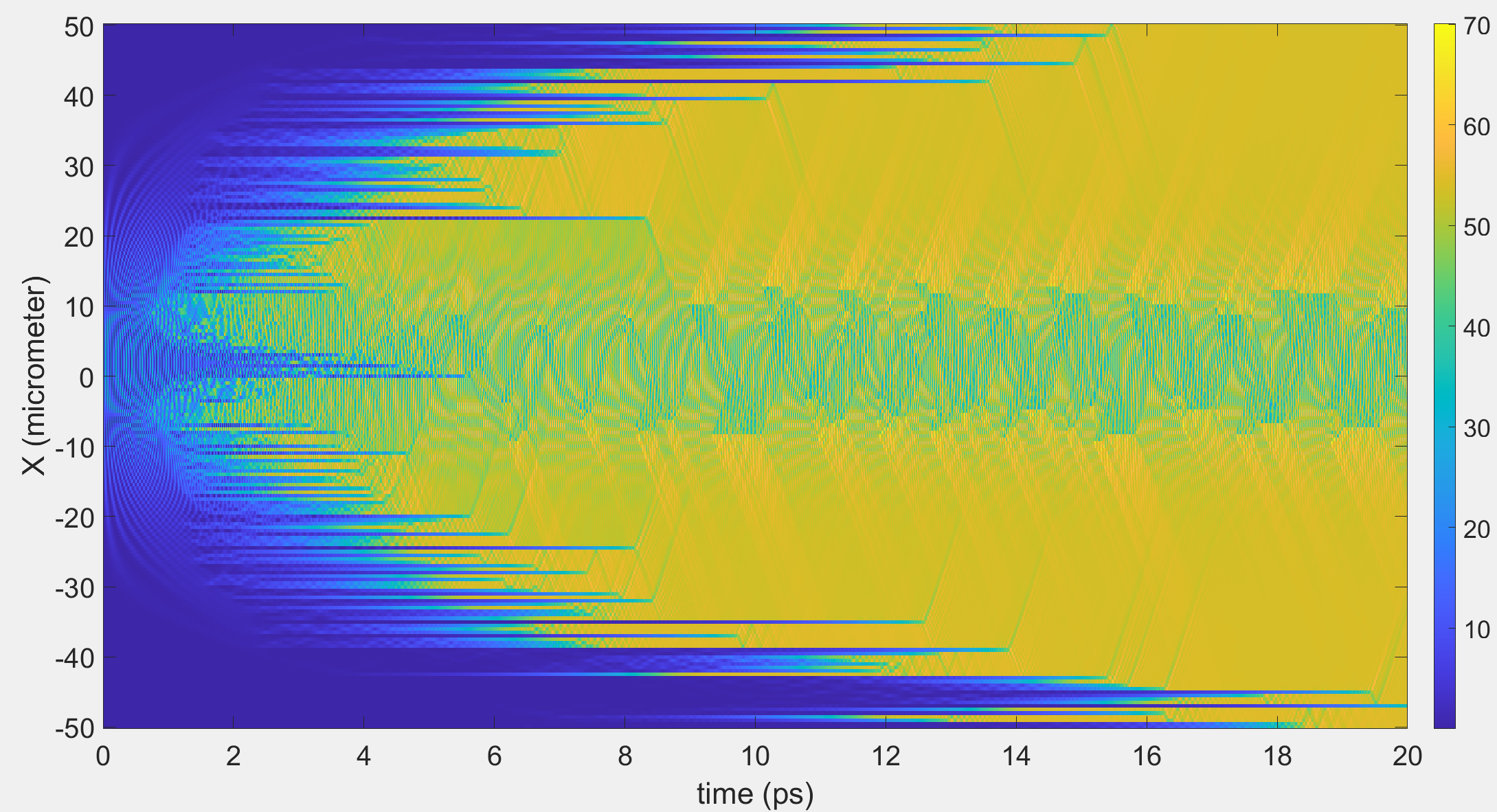}
		\caption{$F_p = 5$ $meV/\mu m^{1/2}$}
		\label{DDGPE_psi_x_respump_str_corr2_Fp19_20ps}
	\end{subfigure}
	\caption{Density of the condensate polaritons with the ratio $g/\gamma_c = 10$ for coherent, near-resonant pumping.}
	\label{CNRP2_psi_x_str_corr2}
\centering
\end{figure}

\begin{figure}[htbp!]
	\centering
	\begin{subfigure}{0.48\textwidth}
		\centering
		\centering
\includegraphics[width=0.8\columnwidth]{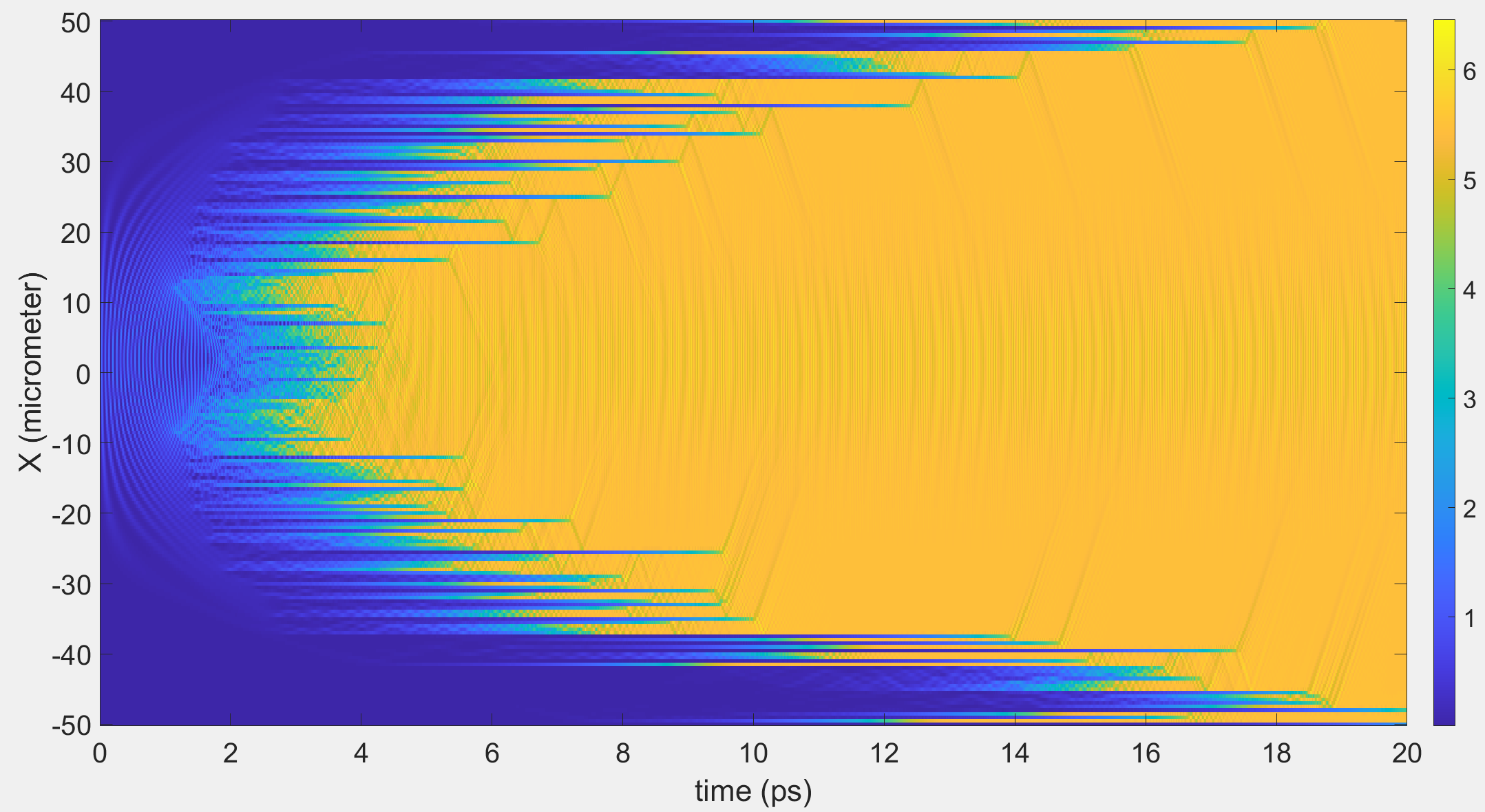}
		\caption{$F_p = 0.5$ $meV/\mu m^{1/2}$}
		\label{DDGPE_psi_x_respump_str_corr3_Fpdef_20ps}
	\end{subfigure}
	\hfill
	\begin{subfigure}{0.48\textwidth}
		\centering
		\includegraphics[width=0.8\columnwidth]{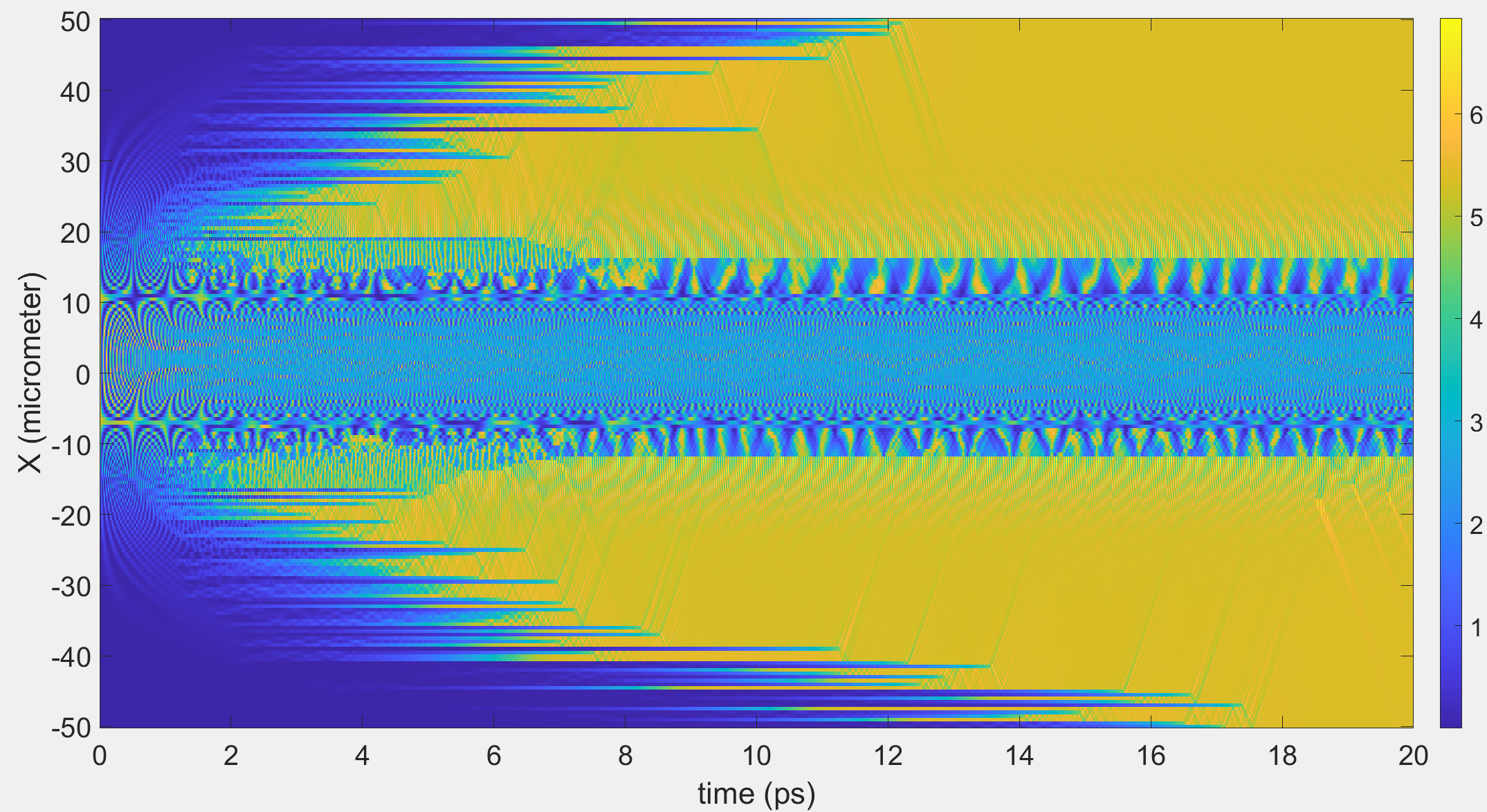}
		\caption{$F_p = 5$ $meV/\mu m^{1/2}$}
		\label{DDGPE_psi_x_respump_str_corr3_Fp19_20ps}
	\end{subfigure}
	\caption{Density of the condensate polaritons with the ratio $g/\gamma_c = 100$ for coherent, near-resonant pumping.}
	\label{CNRP2_psi_x_str_corr3}
\centering
\end{figure}

We first plot the condensation for coherent, near-resonant pumping case for the strongly correlated polariton regime for a 1D microwire microcavity for low and high pump amplitudes in Figures \ref{CNRP2_psi_x_str_corr1}, \ref{CNRP2_psi_x_str_corr2}, and \ref{CNRP2_psi_x_str_corr3}. It can be noticed that with increasing correlation ($g/\gamma_c$), at higher amplitude of the pump field ($F_p$), the polaritons are increasingly averse from occupying positions near the centre of the microcavity. Moreover, with higher correlation the number of polaritons added to the system reduces up to single digits.

\begin{figure}[htbp!]
	\centering
	\begin{subfigure}{0.48\textwidth}
		\centering
		\centering
\includegraphics[width=0.7\columnwidth]{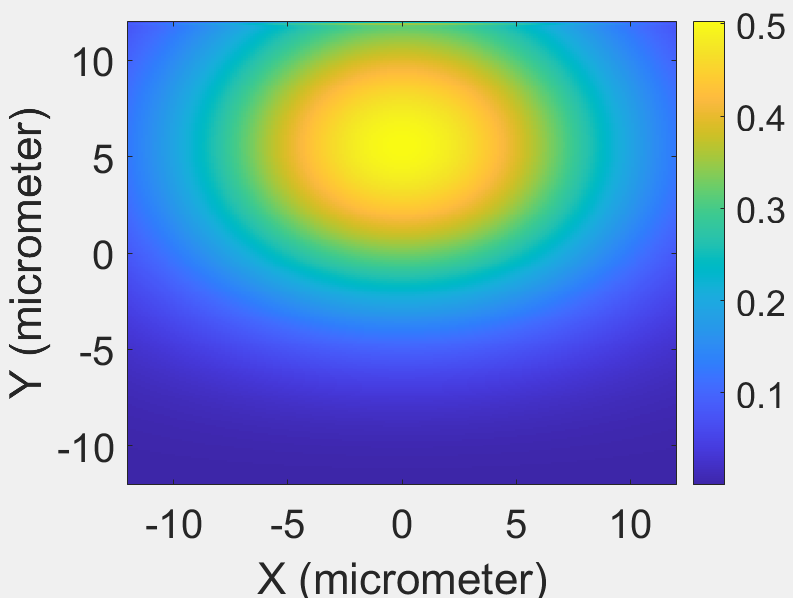}
		\caption{$t = 0$ s}
		\label{DDGPE_psi_x_2D_str_corr1_Fpdef_respump_0sec}
	\end{subfigure}
	\hfill
	\begin{subfigure}{0.48\textwidth}
		\centering
		\includegraphics[width=0.7\columnwidth]{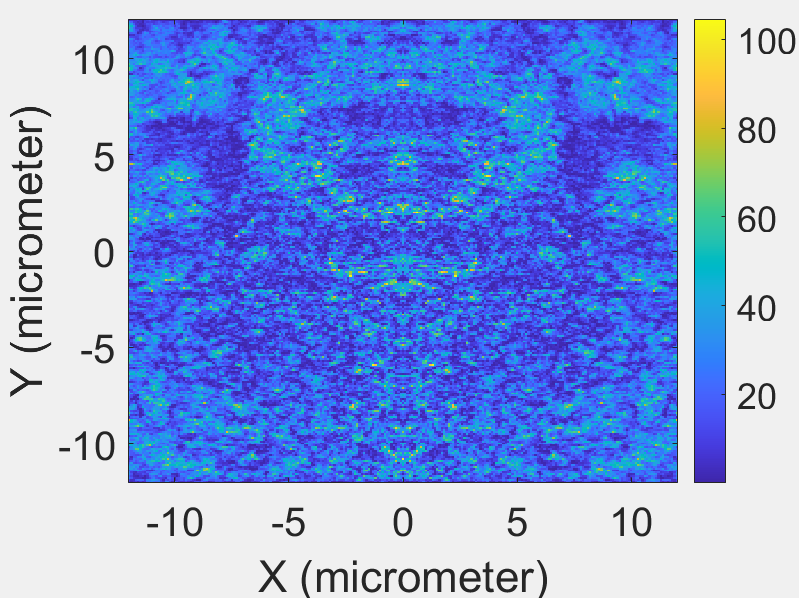}
		\caption{$t = 2.5$ ps}
		\label{DDGPE_psi_x_2D_str_corr1_Fpdef_respump_2.5ps}
	\end{subfigure}
	\vfill
	\begin{subfigure}{0.48\textwidth}
		\centering
		\includegraphics[width=0.7\columnwidth]{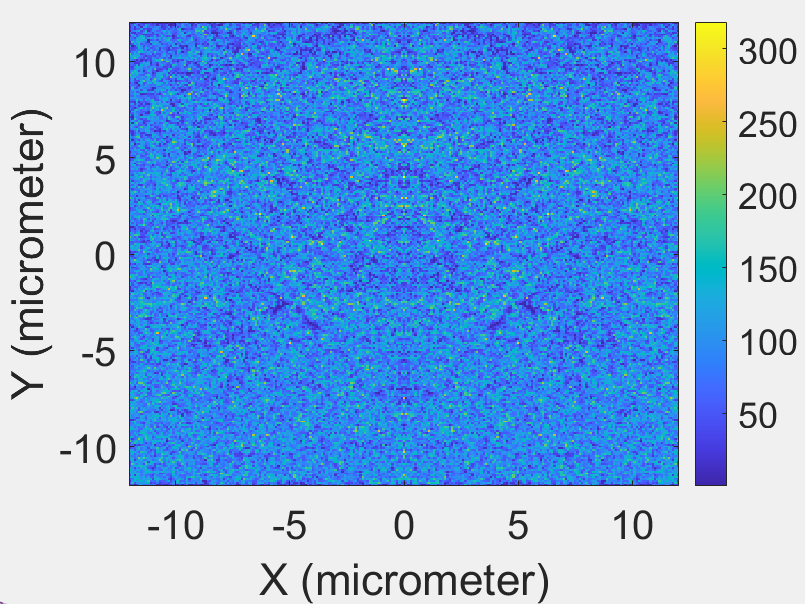}
		\caption{$t = 5$ ps}
		\label{DDGPE_psi_x_2D_str_corr1_Fpdef_respump_5ps}
	\end{subfigure}
	\hfill
	\begin{subfigure}{0.48\textwidth}
		\centering
		\includegraphics[width=0.7\columnwidth]{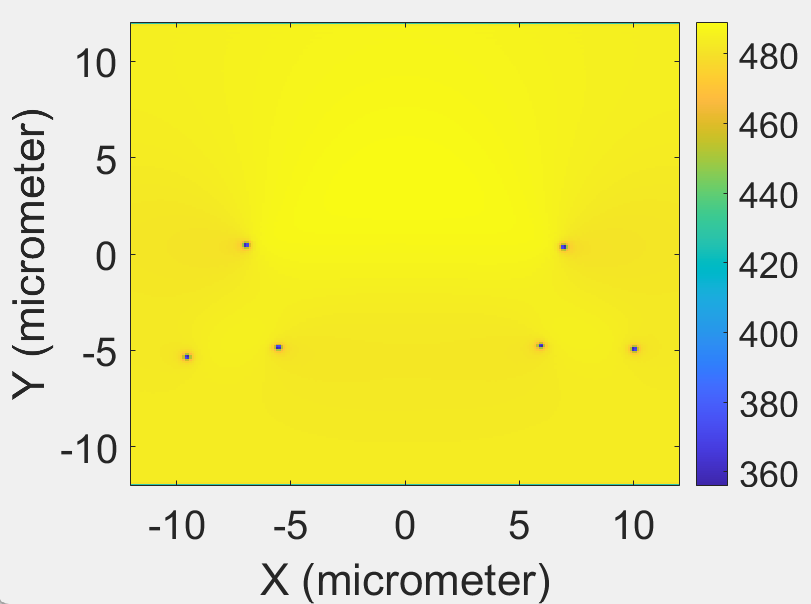}
		\caption{$t = 10$ ps}
		\label{DDGPE_psi_x_2D_str_corr1_Fpdef_respump_10ps}
	\end{subfigure}
	\caption{Density of condensate polaritons for two-dimensional microcavity with coherent, near-resonant pumping at different times in the strongly correlated polariton regime for the ratio $g/\gamma_c = 1.132$ with $ F_p = 0.5 $ meV~$(\mu\text{m})^{-1}$. The initial distribution is Gaussian in x and y.}
	\label{CNRP2_psi_x_2D_str_corr1_Fpdef}
\centering
\end{figure}

\begin{figure}[htbp!]
	\centering
	\begin{subfigure}{0.48\textwidth}
		\centering
		\centering
\includegraphics[width=0.7\columnwidth]{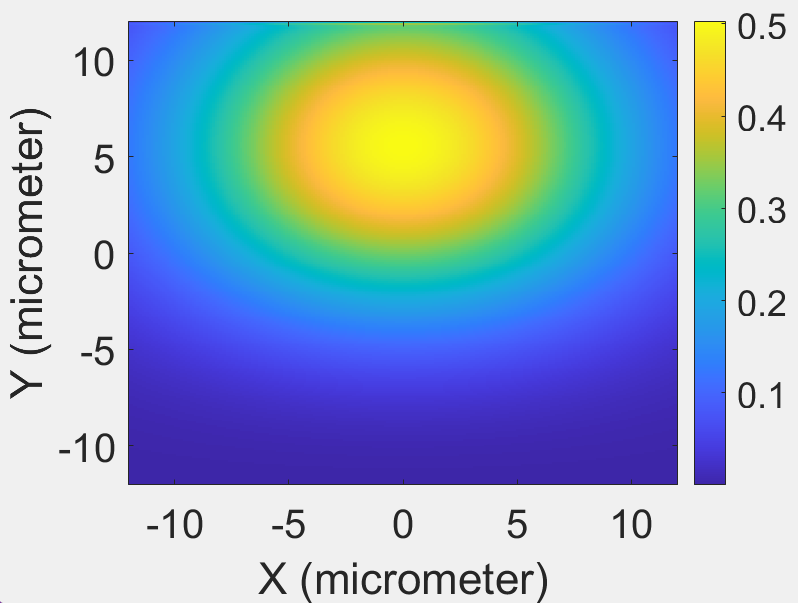}
		\caption{$t = 0$ s}
		\label{DDGPE_psi_x_2D_str_corr2_Fpdef_respump_0sec}
	\end{subfigure}
	\hfill
	\begin{subfigure}{0.48\textwidth}
		\centering
		\includegraphics[width=0.7\columnwidth]{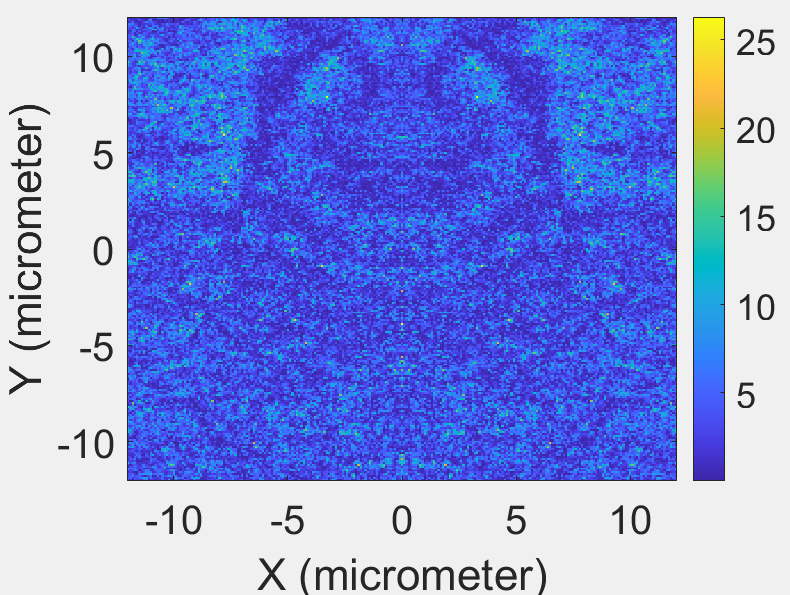}
		\caption{$t = 2.5$ ps}
		\label{DDGPE_psi_x_2D_str_corr2_Fpdef_respump_2.5ps}
	\end{subfigure}
	\vfill
	\begin{subfigure}{0.48\textwidth}
		\centering
		\includegraphics[width=0.7\columnwidth]{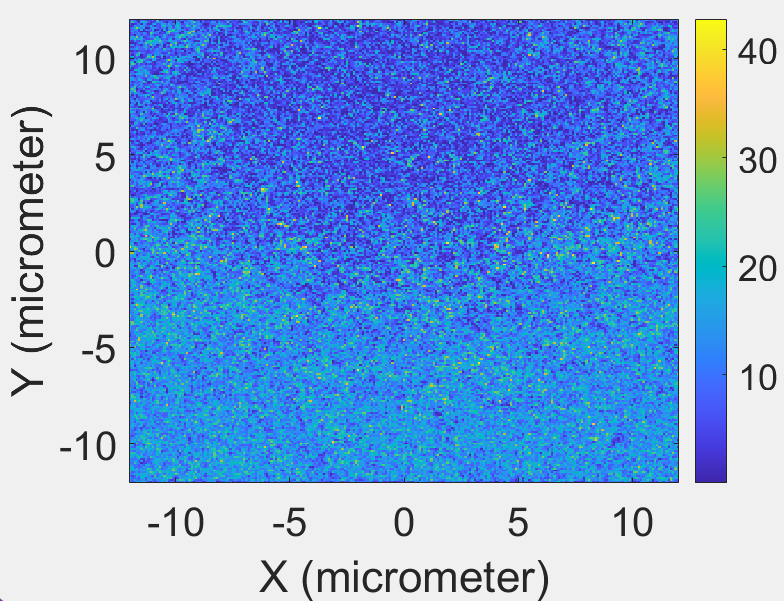}
		\caption{$t = 5$ ps}
		\label{DDGPE_psi_x_2D_str_corr2_Fpdef_respump_5ps}
	\end{subfigure}
	\hfill
	\begin{subfigure}{0.48\textwidth}
		\centering
		\includegraphics[width=0.7\columnwidth]{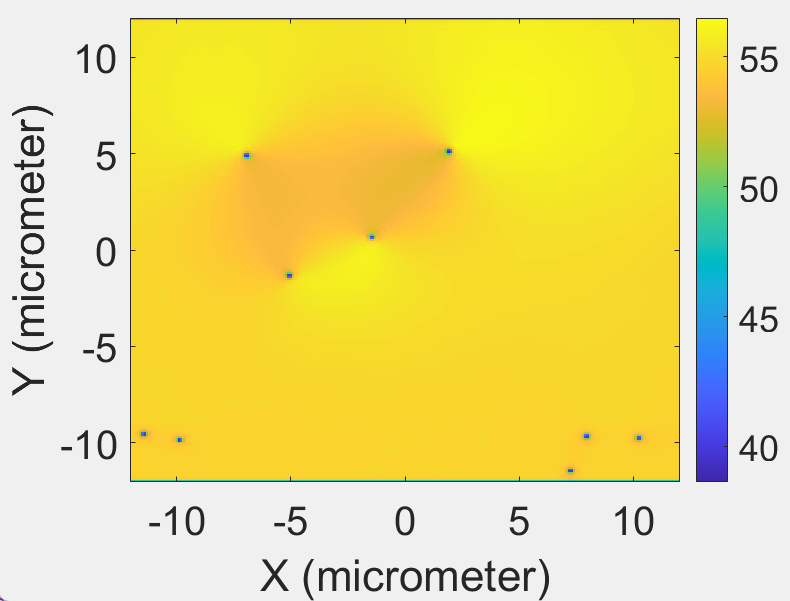}
		\caption{$t = 10$ ps}
		\label{DDGPE_psi_x_2D_str_corr2_Fpdef_respump_10ps}
	\end{subfigure}
	\caption{Density of condensate polaritons for two-dimensional microcavity with coherent, near-resonant pumping at different times in the strongly correlated polariton regime for the ratio $g/\gamma_c = 10$ with $ F_p = 0.5 $ meV~$(\mu\text{m})^{-1}$. The initial distribution is Gaussian in x and y.}
	\label{CNRP2_psi_x_2D_str_corr2_Fpdef}
\centering
\end{figure}

\begin{figure}[htbp!]
	\centering
	\begin{subfigure}{0.48\textwidth}
		\centering
		\centering
\includegraphics[width=0.7\columnwidth]{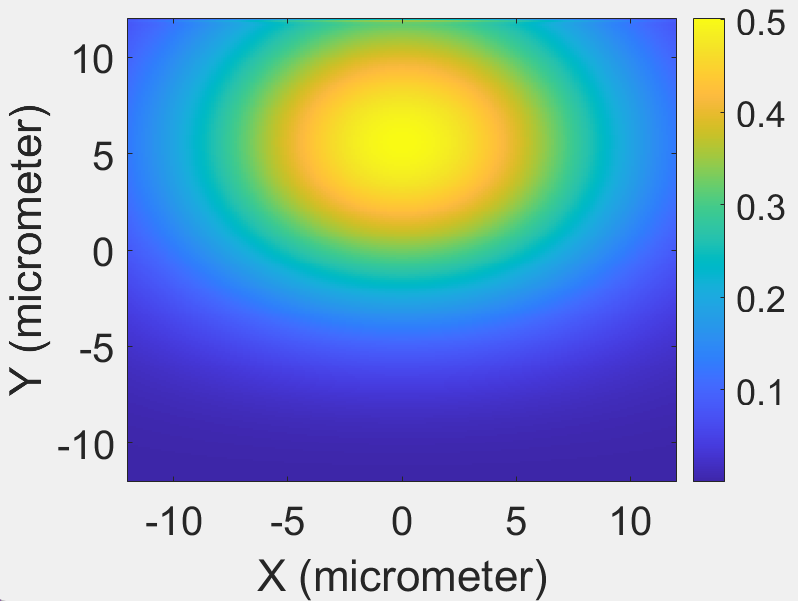}
		\caption{$t = 0$ s}
		\label{DDGPE_psi_x_2D_str_corr3_Fpdef_respump_0sec}
	\end{subfigure}
	\hfill
	\begin{subfigure}{0.48\textwidth}
		\centering
		\includegraphics[width=0.7\columnwidth]{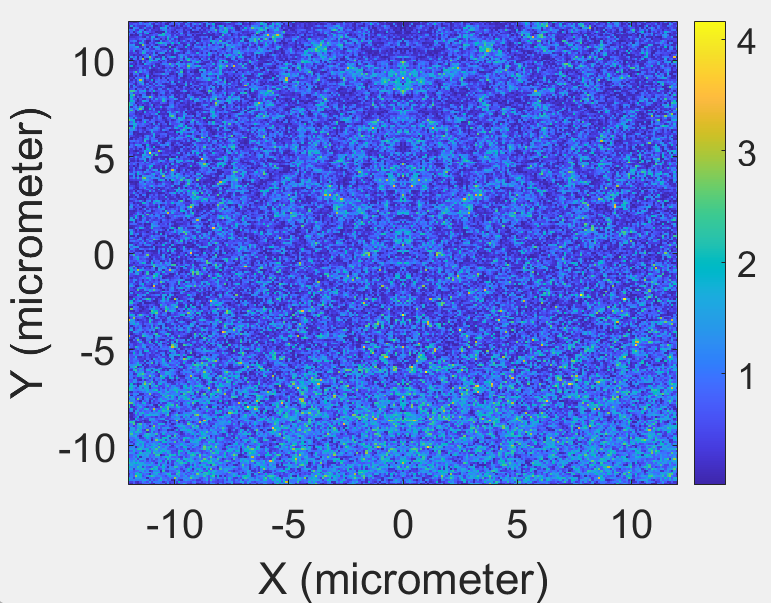}
		\caption{$t = 2.5$ ps}
		\label{DDGPE_psi_x_2D_str_corr3_Fpdef_respump_2.5ps}
	\end{subfigure}
	\vfill
	\begin{subfigure}{0.48\textwidth}
		\centering
		\includegraphics[width=0.7\columnwidth]{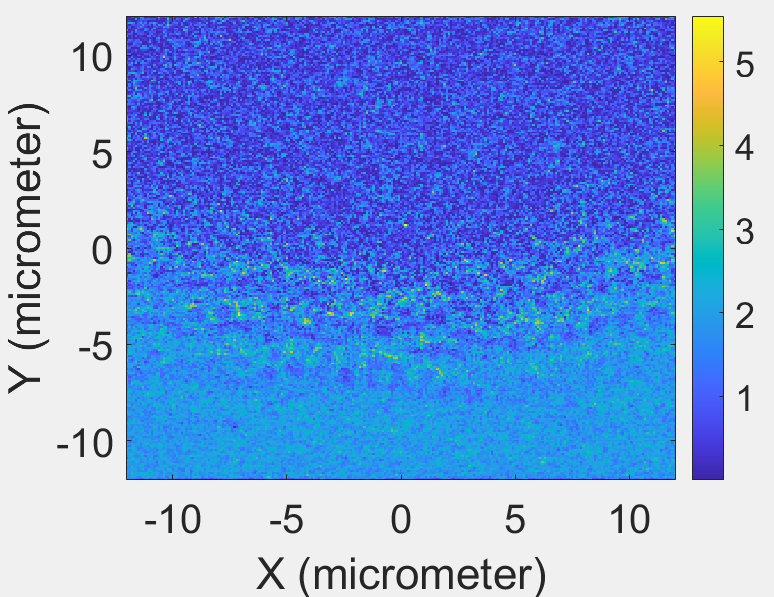}
		\caption{$t = 5$ ps}
		\label{DDGPE_psi_x_2D_str_corr3_Fpdef_respump_5ps}
	\end{subfigure}
	\hfill
	\begin{subfigure}{0.48\textwidth}
		\centering
		\includegraphics[width=0.7\columnwidth]{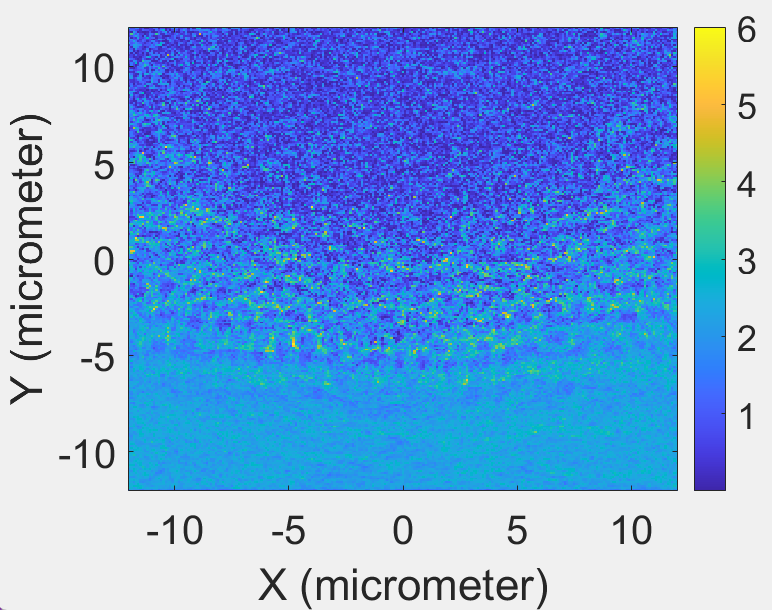}
		\caption{$t = 10$ ps}
		\label{DDGPE_psi_x_2D_str_corr3_Fpdef_respump_10ps}
	\end{subfigure}
	\caption{Density of condensate polaritons for two-dimensional microcavity with coherent, near-resonant pumping at different times in the strongly correlated polariton regime for the ratio $g/\gamma_c = 100$ with $ F_p = 0.5 $ meV~$(\mu\text{m})^{-1}$. The initial distribution is Gaussian in x and y.}
	\label{CNRP2_psi_x_2D_str_corr3_Fpdef}
\centering
\end{figure}

The results for 2D microcavity corresponding to coherent, near-resonant pumping in the strongly correlated polariton regime are shown in Figures \ref{CNRP2_psi_x_2D_str_corr1_Fpdef}, \ref{CNRP2_psi_x_2D_str_corr2_Fpdef}, and \ref{CNRP2_psi_x_2D_str_corr3_Fpdef}. As the correlation ratio is increased, we summarily observe that the peak number of polaritons added to the system decreases from about 480 ($ g/\gamma_c = 1.132 $), about 55 ($ g/\gamma_c = 10 $), to about 2 ($g/\gamma_c = 100$). It is observed that the approximate percentage change in the number of polaritons at time snapshots of 2.5 ps, 5 ps, and 10 ps for $ g/\gamma_c = 1.132 $ and $ g/\gamma_c = 10 $ is nearly the same, which is in contrast to the $ g/\gamma_c = 100 $ case, where the number of polaritons remains almost constant. We also observe certain islands of low polariton occupancy close to the saturation of the polariton number at about 10 ps in figures \ref{CNRP2_psi_x_2D_str_corr1_Fpdef} and \ref{CNRP2_psi_x_2D_str_corr2_Fpdef}. We also observe distinct patterns within the microcavity space at the same time points (see figures \ref{CNRP2_psi_x_2D_str_corr1_Fpdef}, \ref{CNRP2_psi_x_2D_str_corr2_Fpdef}, and \ref{CNRP2_psi_x_2D_str_corr3_Fpdef}). 

In experiments which hinted at the polariton blockade~\cite{delteil2019towards,munoz2019emergence}, the laser power used was 10 nW. We convert this to the amplitude of the pump field $F_p$ as follows: $P = IA $ and $I = (1/2)c\epsilon_0 F_p^2 $ $\implies $ $P/A = (c \epsilon_0/2) F_p^2$, which leads to $F_p = \frac{\sqrt{2P}}{\sqrt{c\epsilon_0 A}}$ for 2D microcavity and $F_p = \frac{\sqrt{2P}}{\sqrt{c\epsilon_0 L}}$ for 1D microwire microcavity. Using $P = 10 nW = 6.24 \cross 10^13 meV/sec$, $c = 3 \cross 10^8 m/s = 3 \cross 10^14 \mu m/sec$, $\epsilon_0 = 55.2635 \cross 10^3 \frac{e^2}{meV \mu m}$, $A = 576 \mu m^2$, $L = 100 \mu m$, we get $F_{p1D} = 0.2743642 \frac{meV}{\sqrt{\mu m}}$ and $F_{p1D} = 0.114318 \frac{meV}{\mu m}$.

Although we do not show those plots here, there is no appreciable difference in the population of polaritons for coherent, non-resonant pumping in the strongly correlated polariton regime when the amplitude of the pump field is reduced to $F_p = 0.05$ $meV/\mu m^{1/2}$, which is ten times less than the value used earlier. Thus, we have already used an even lower amplitude of the pump field for coherent, near-resonant pumping than that used for the experiments which hinted at the polariton blockade~\cite{delteil2019towards,munoz2019emergence} and found that it has no effect on the polariton dynamics.
\subsection{Homogeneous, incoherent, non-resonant pumping}	
\label{HINRPres}
\begin{figure}[htbp!]
	\centering
	\begin{subfigure}{0.48\textwidth}
		\centering
		\centering
\includegraphics[width=0.6\columnwidth]{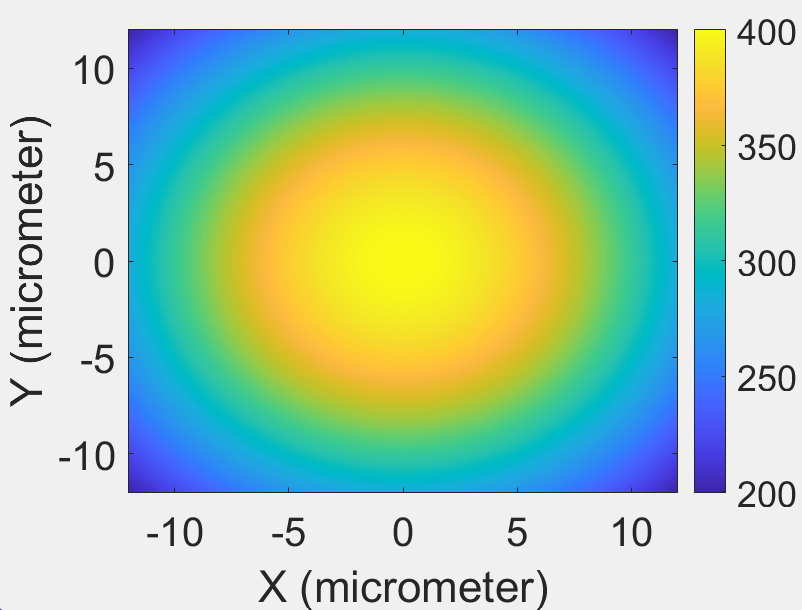}
		\caption{$t = 0$ s}
		\label{DDGPE_nR_x_2D_corr_0sec}
	\end{subfigure}
	\hfill
	\begin{subfigure}{0.48\textwidth}
		\centering
		\includegraphics[width=0.6\columnwidth]{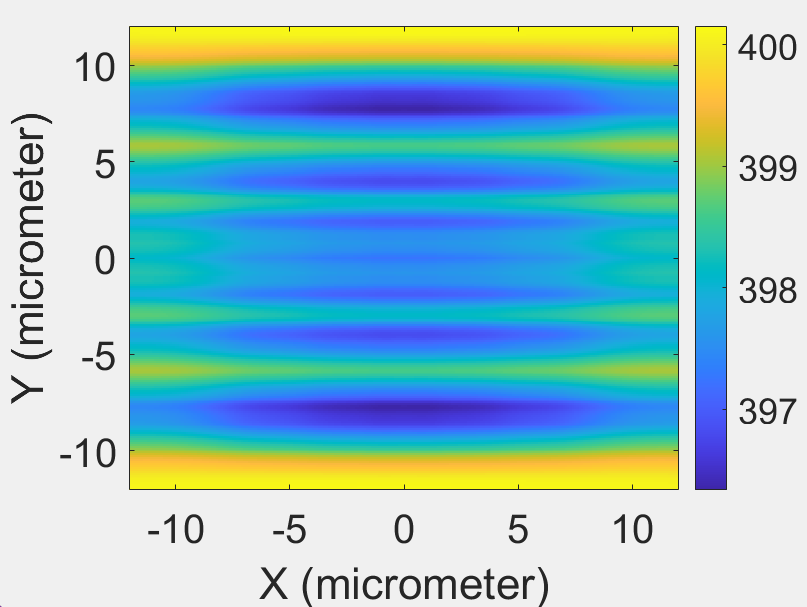}
		\caption{$t = 2.5$ ps}
		\label{DDGPE_nR_x_2D_corr_2.5ps}
	\end{subfigure}
	\vfill
	\begin{subfigure}{0.48\textwidth}
		\centering
		\includegraphics[width=0.6\columnwidth]{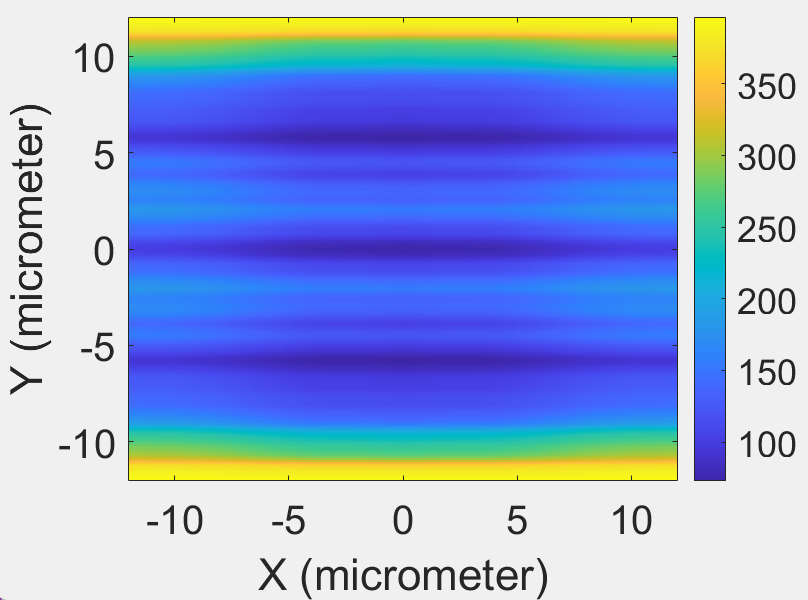}
		\caption{$t = 5$ ps}
		\label{DDGPE_nR_x_2D_corr_5ps}
	\end{subfigure}
	\hfill
	\begin{subfigure}{0.48\textwidth}
		\centering
		\includegraphics[width=0.6\columnwidth]{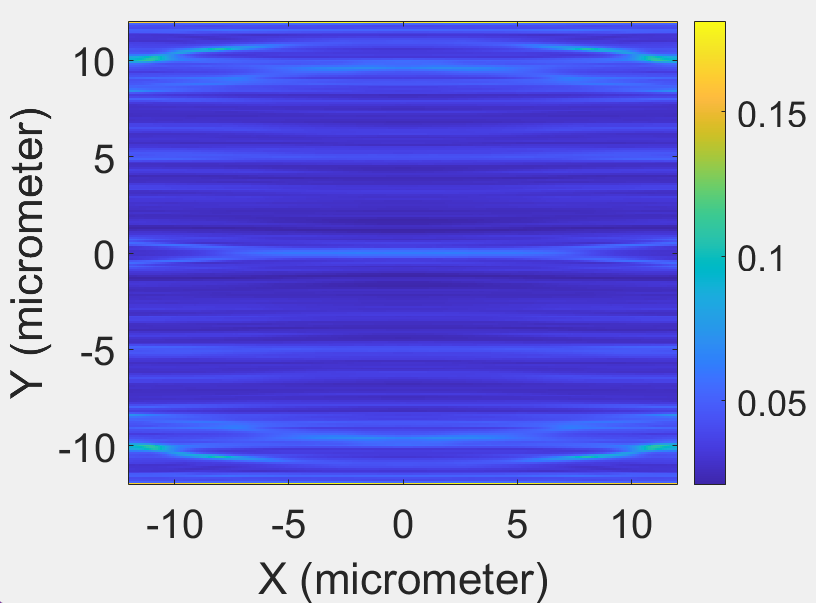}
		\caption{$t = 10$ ps}
		\label{DDGPE_nR_x_2D_corr_10ps}
	\end{subfigure}
	\vfill
	\begin{subfigure}{0.48\textwidth}
		\centering
		\includegraphics[width=0.6\columnwidth]{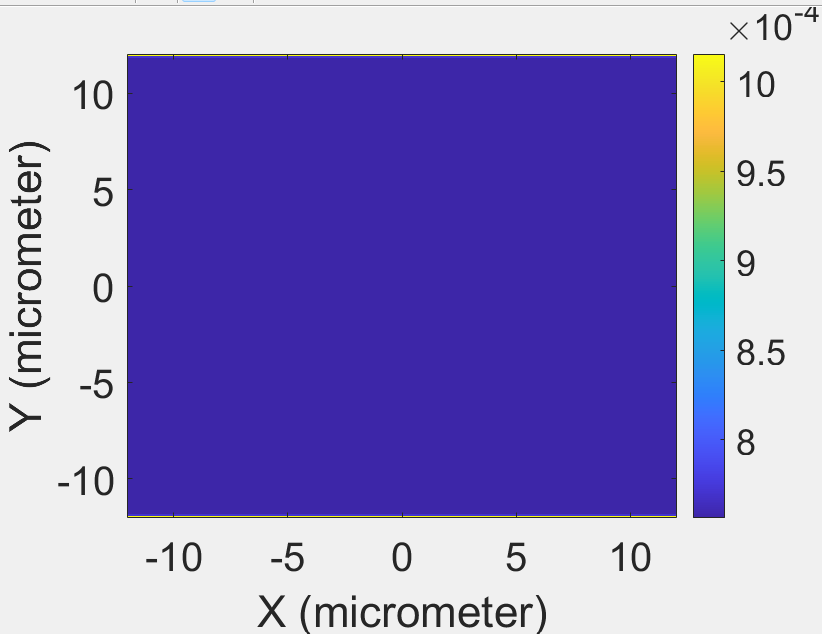}
		\caption{$t = 15$ ps}
		\label{DDGPE_nR_x_2D_corr_15ps}
	\end{subfigure}
	\caption{Density of reservoir polaritons for two-dimensional microcavity with homogeneous, incoherent, non-resonant pumping at different times. The initial distribution is Gaussian in x and y.}
	\label{HINRP_nR_x_2D}
\centering
\end{figure}

\begin{figure}[htbp!]
	\centering
	\begin{subfigure}{0.48\textwidth}
		\centering
		\centering
\includegraphics[width=0.6\columnwidth]{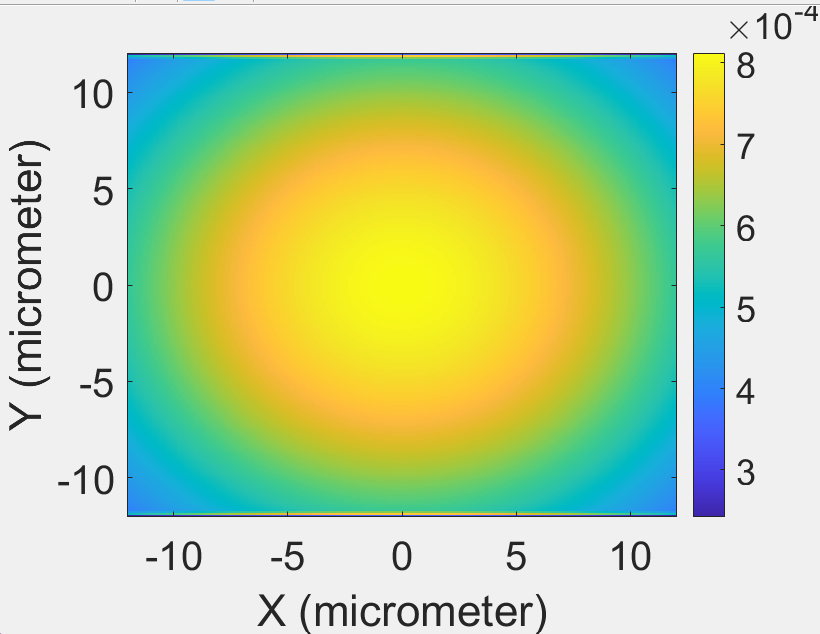}
		\caption{$t = 0$ s}
		\label{DDGPE_psi_x_2D_corr_0sec}
	\end{subfigure}
	\hfill
	\begin{subfigure}{0.48\textwidth}
		\centering
		\includegraphics[width=0.6\columnwidth]{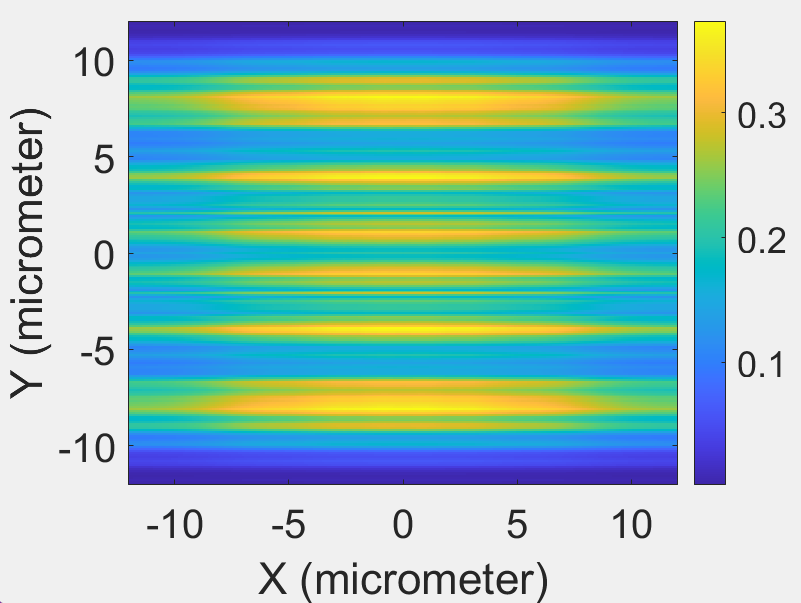}
		\caption{$t = 2.5$ ps}
		\label{DDGPE_psi_x_2D_corr_2.5ps}
	\end{subfigure}
	\vfill
	\begin{subfigure}{0.48\textwidth}
		\centering
		\includegraphics[width=0.6\columnwidth]{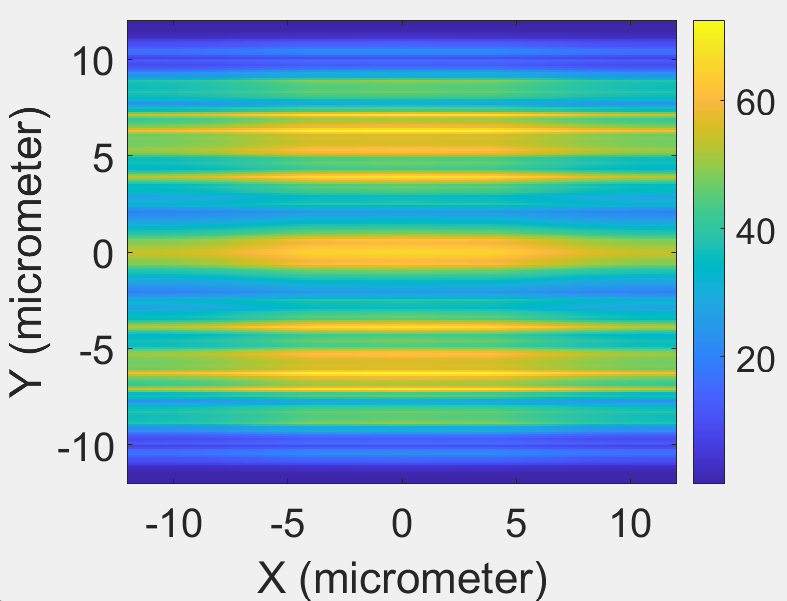}
		\caption{$t = 5$ ps}
		\label{DDGPE_psi_x_2D_corr_5ps}
	\end{subfigure}
	\hfill
	\begin{subfigure}{0.48\textwidth}
		\centering
		\includegraphics[width=0.6\columnwidth]{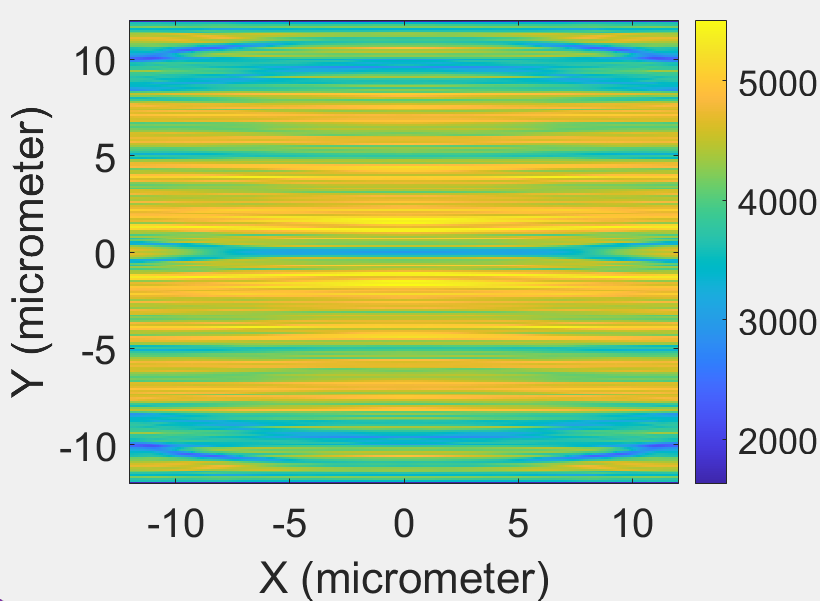}
		\caption{$t = 10$ ps}
		\label{DDGPE_psi_x_2D_corr_10ps}
	\end{subfigure}
	\vfill
	\begin{subfigure}{0.48\textwidth}
		\centering
		\includegraphics[width=0.6\columnwidth]{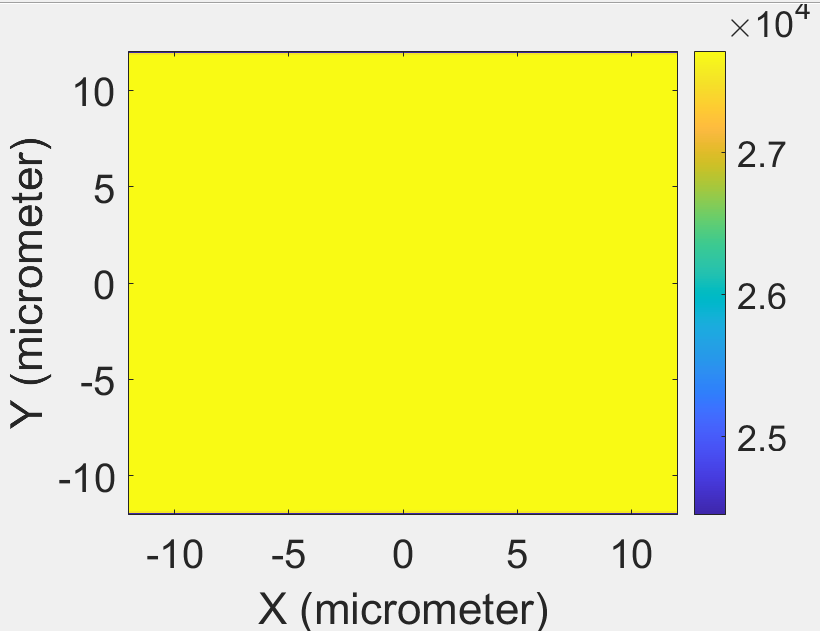}
		\caption{$t = 15$ ps}
		\label{DDGPE_psi_x_2D_corr_15ps}
	\end{subfigure}
	\caption{Density of condensate polaritons for two-dimensional microcavity with homogeneous, incoherent, non-resonant pumping at different times. The initial distribution is Gaussian in x and y.}
	\label{HINRP_psi_x_2D}
\centering
\end{figure}

We plot results for 2D microcavity in Figures \ref{HINRP_nR_x_2D} and \ref{HINRP_psi_x_2D}. From observing the distribution of density of condensate and reservoir polaritons at different times, it is evident that the two densities show complementary behaviour: in parts of the microcavity where the density of condensate polaritons is high, the density of reservoir polaritons is low. Additionally, with time, due to relaxation from top of the lower polariton branch (where reservoir polaritons are injected into the system) to the bottom of the lower polariton branch (where the condensate forms), the density of reservoir polaritons in the microcavity decreases with time while that of the condensate polaritons increases with time.

\begin{figure}[htbp!]
	\centering
	\begin{subfigure}{0.48\textwidth}
		\centering
		\centering
\includegraphics[width=0.8\columnwidth]{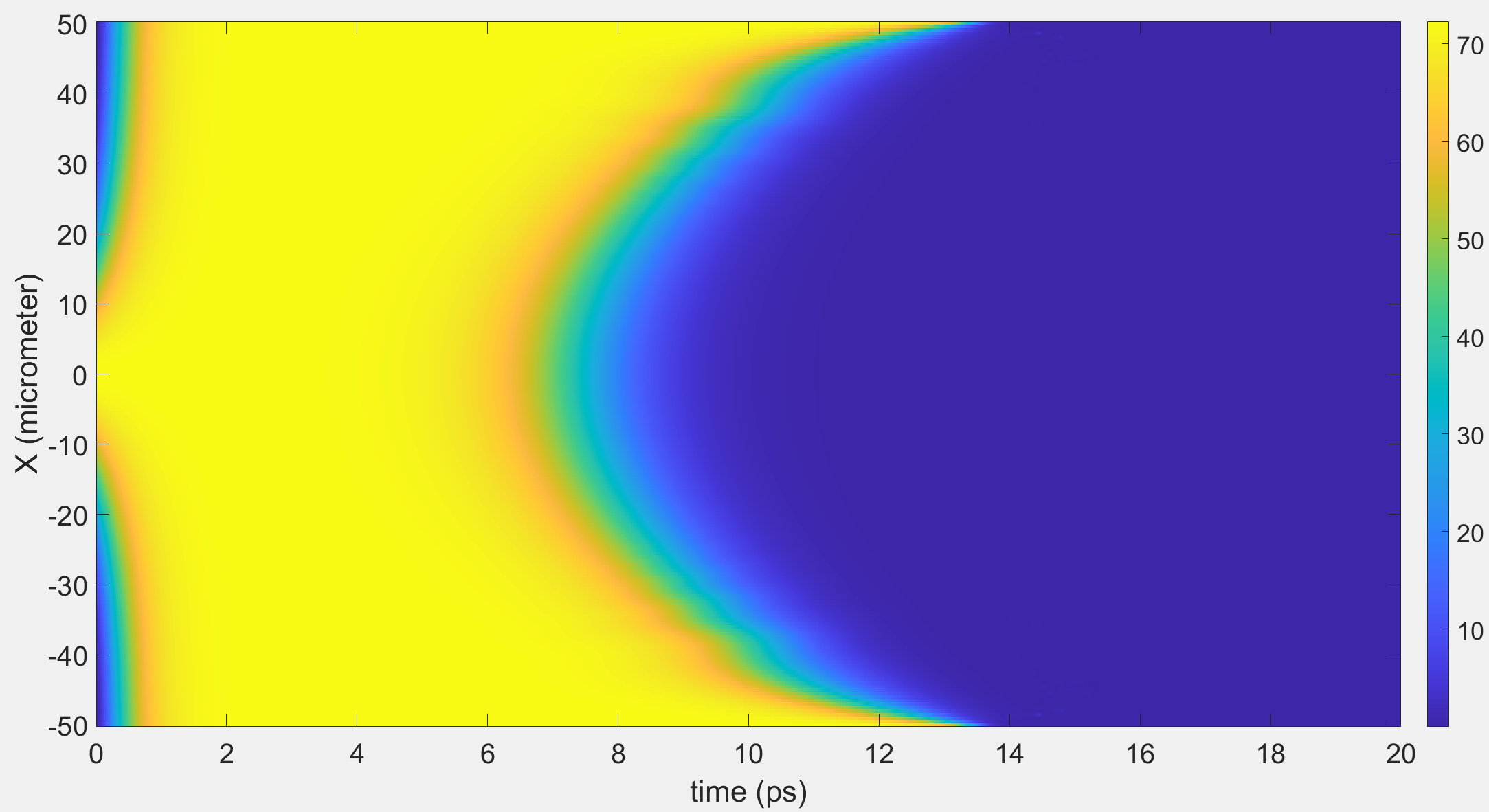}
		\caption{Reservoir polaritons}
		\label{DDGPE_nR_x_P2_20ps}
	\end{subfigure}
	\hfill
	\begin{subfigure}{0.48\textwidth}
		\centering
		\includegraphics[width=0.8\columnwidth]{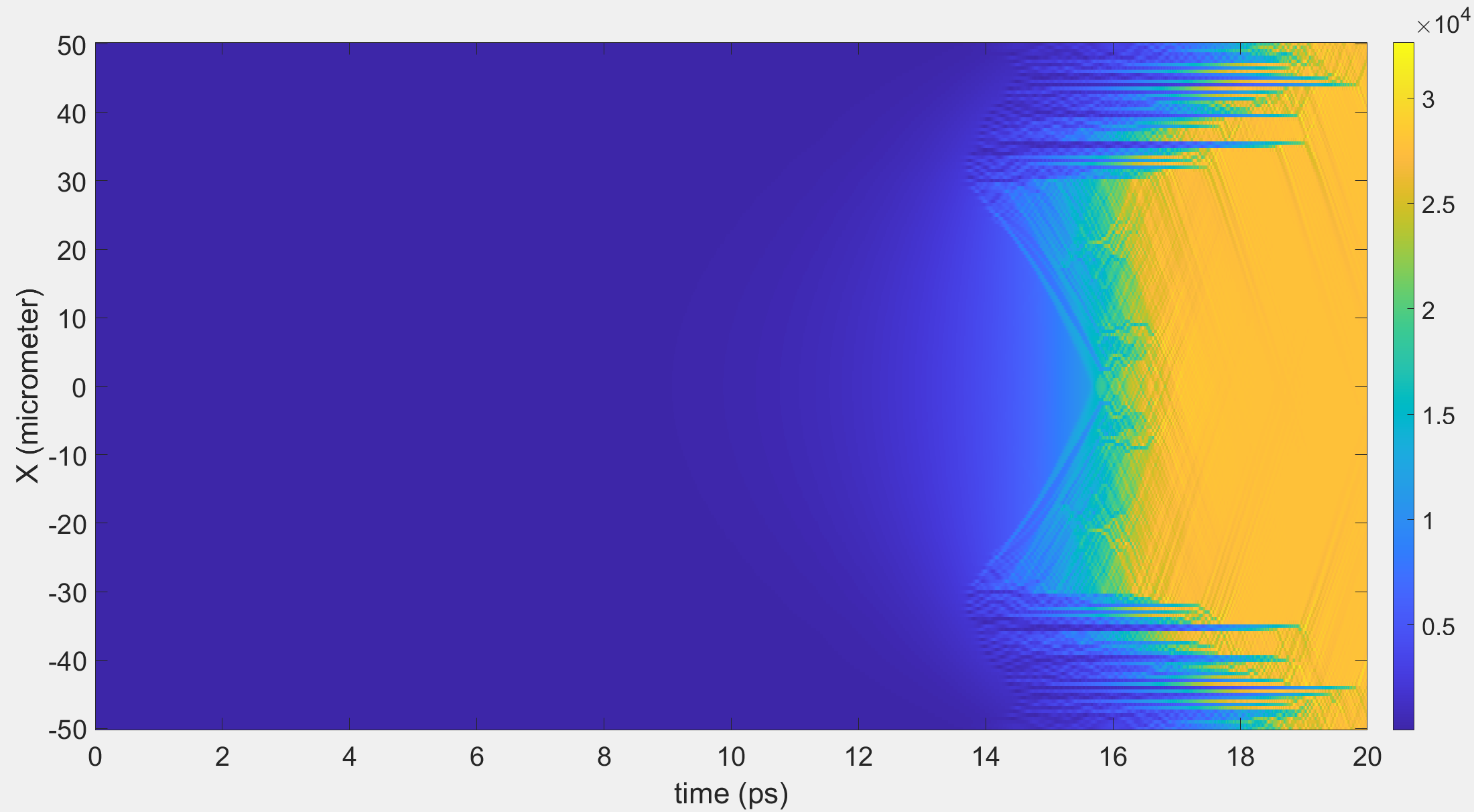}
		\caption{Condensate polaritons}
		\label{DDGPE_psi_x_P2_20ps}
	\end{subfigure}
	\caption{Density of the condensate polaritons with a pumping rate (pump power) of P = 25.835 $\mu m^{-1} ps^{-1}$ for the homogeneous, incoherent, nonresonant pumping.}
	\label{HINRP_x_P2}
\centering
\end{figure}

\begin{figure}[htbp!]
	\centering
	\begin{subfigure}{0.48\textwidth}
		\centering
		\centering
\includegraphics[width=0.8\columnwidth]{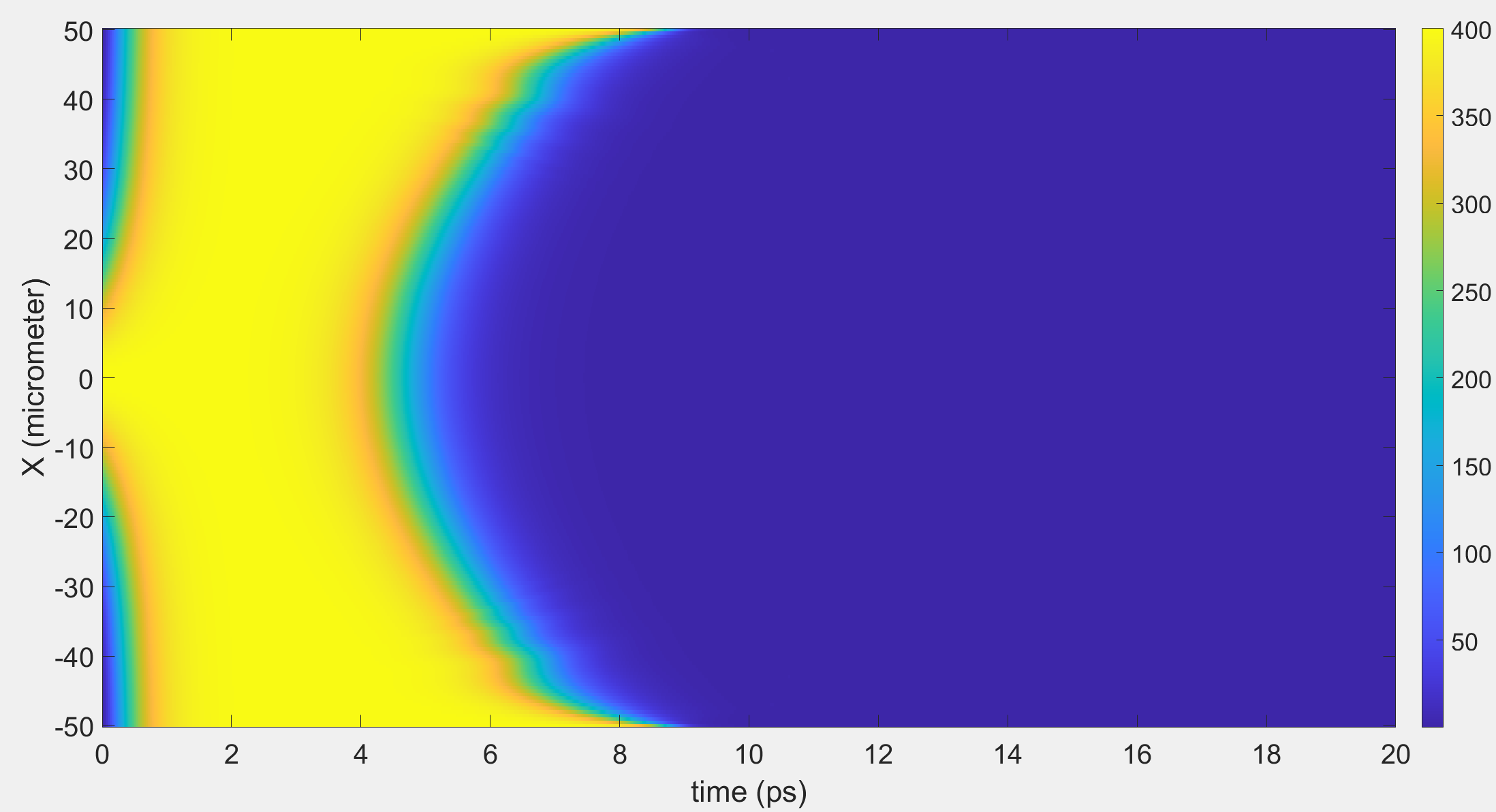}
		\caption{Reservoir polaritons}
		\label{DDGPE_nR_x_20ps}
	\end{subfigure}
	\hfill
	\begin{subfigure}{0.48\textwidth}
		\centering
		\includegraphics[width=0.8\columnwidth]{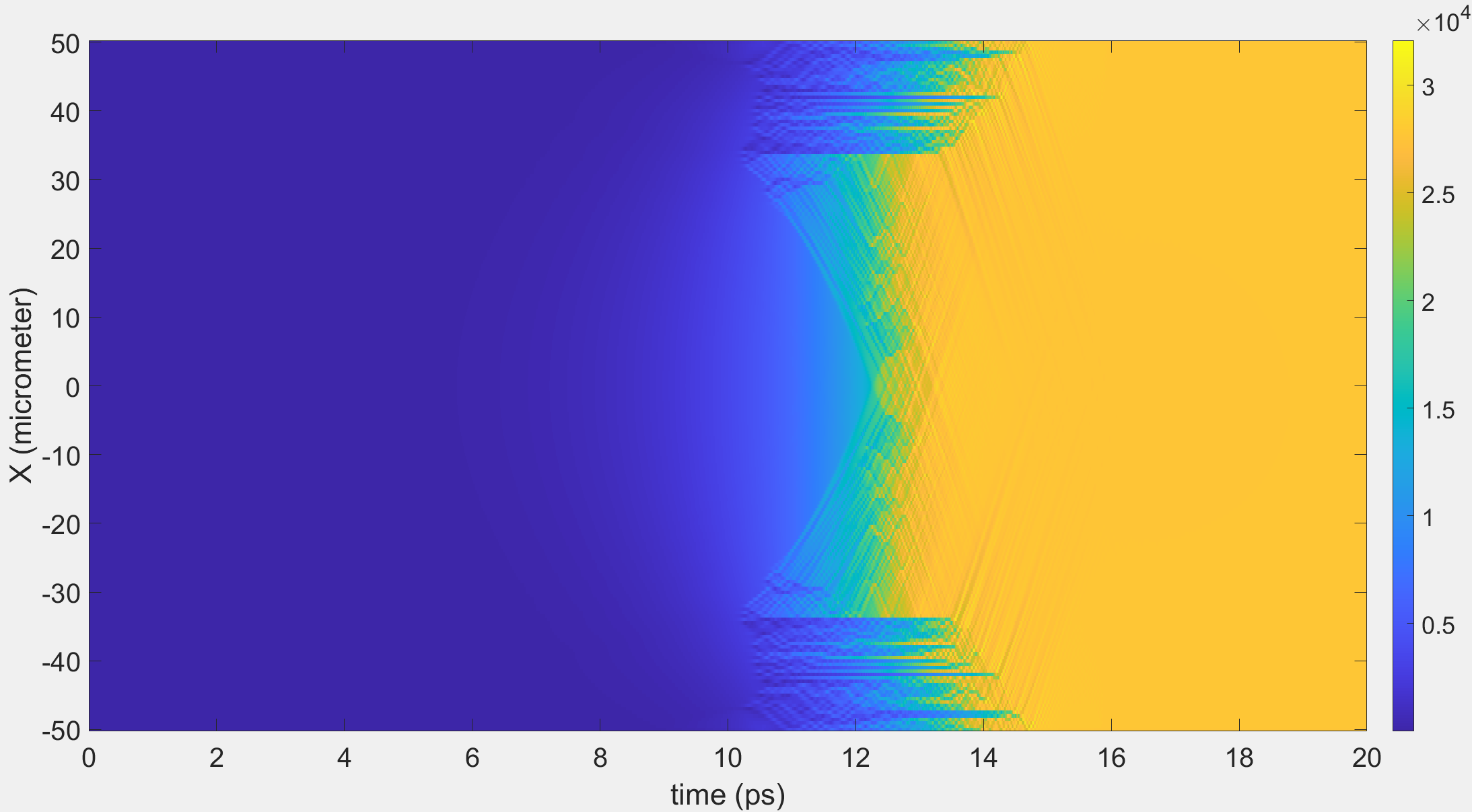}
		\caption{Condensate polaritons}
		\label{DDGPE_psi_x_20ps}
	\end{subfigure}
	\caption{Density of the condensate polaritons with a pumping rate (pump power) of P = 60.790 $\mu m^{-1} ps^{-1}$ for the homogeneous, incoherent, nonresonant pumping.}
	\label{HINRP_x_P1}
\centering
\end{figure}

\begin{figure}[htbp!]
	\centering
	\begin{subfigure}{0.48\textwidth}
		\centering
		\centering
\includegraphics[width=0.8\columnwidth]{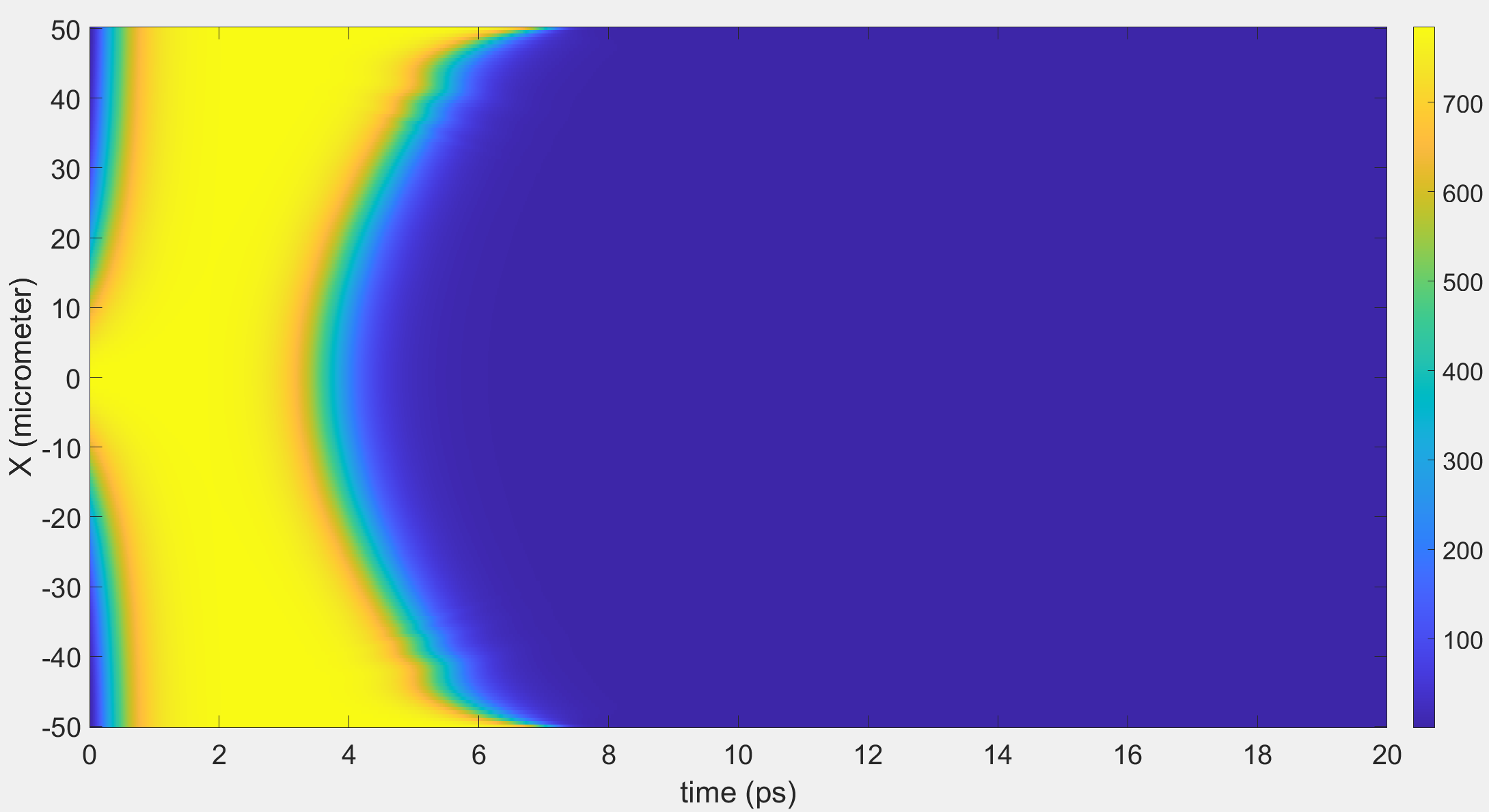}
		\caption{Reservoir polaritons}
		\label{DDGPE_nR_x_P3_20ps}
	\end{subfigure}
	\hfill
	\begin{subfigure}{0.48\textwidth}
		\centering
		\includegraphics[width=0.8\columnwidth]{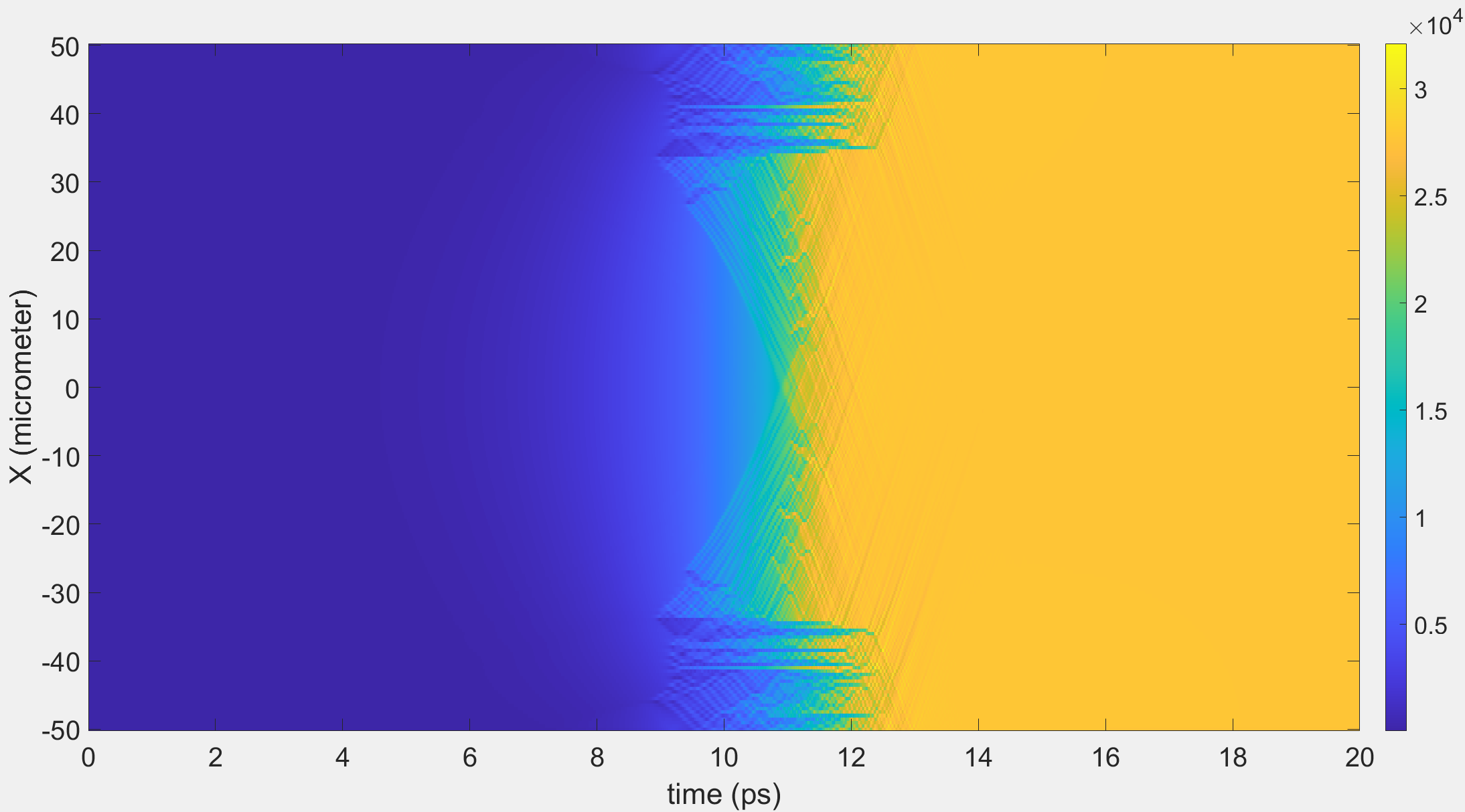}
		\caption{Condensate polaritons}
		\label{DDGPE_psi_x_P3_20ps}
	\end{subfigure}
	\caption{Density of the condensate polaritons with a pumping rate (pump power) of P = 85.106 $\mu m^{-1} ps^{-1}$ for the homogeneous, incoherent, nonresonant pumping.}
	\label{HINRP_x_P3}
\centering
\end{figure}

We plot results for the polariton condensation with homogeneous, non-resonant, incoherent pumping corresponding to three different pump powers~\cite{gargoubi2016polariton} for the 1D case in Figures \ref{HINRP_x_P2}, \ref{HINRP_x_P1}, and \ref{HINRP_x_P3}. We observe that the reservoir polariton life gets shorter as the power is increased with the corresponding increase in the life of the condensate polaritons. Also, the polariton number for the reservoir polaritons seems to increase with increasing power but there is no such visible change in the condensate polaritons. There is also complementarity between areas of the microcavity where condensate polaritons are observed and reservoir polaritons are observed and this evolves with time.

\begin{figure}[htbp!]
	\centering
	\begin{subfigure}{0.48\textwidth}
		\centering
		\centering
\includegraphics[width=0.8\columnwidth]{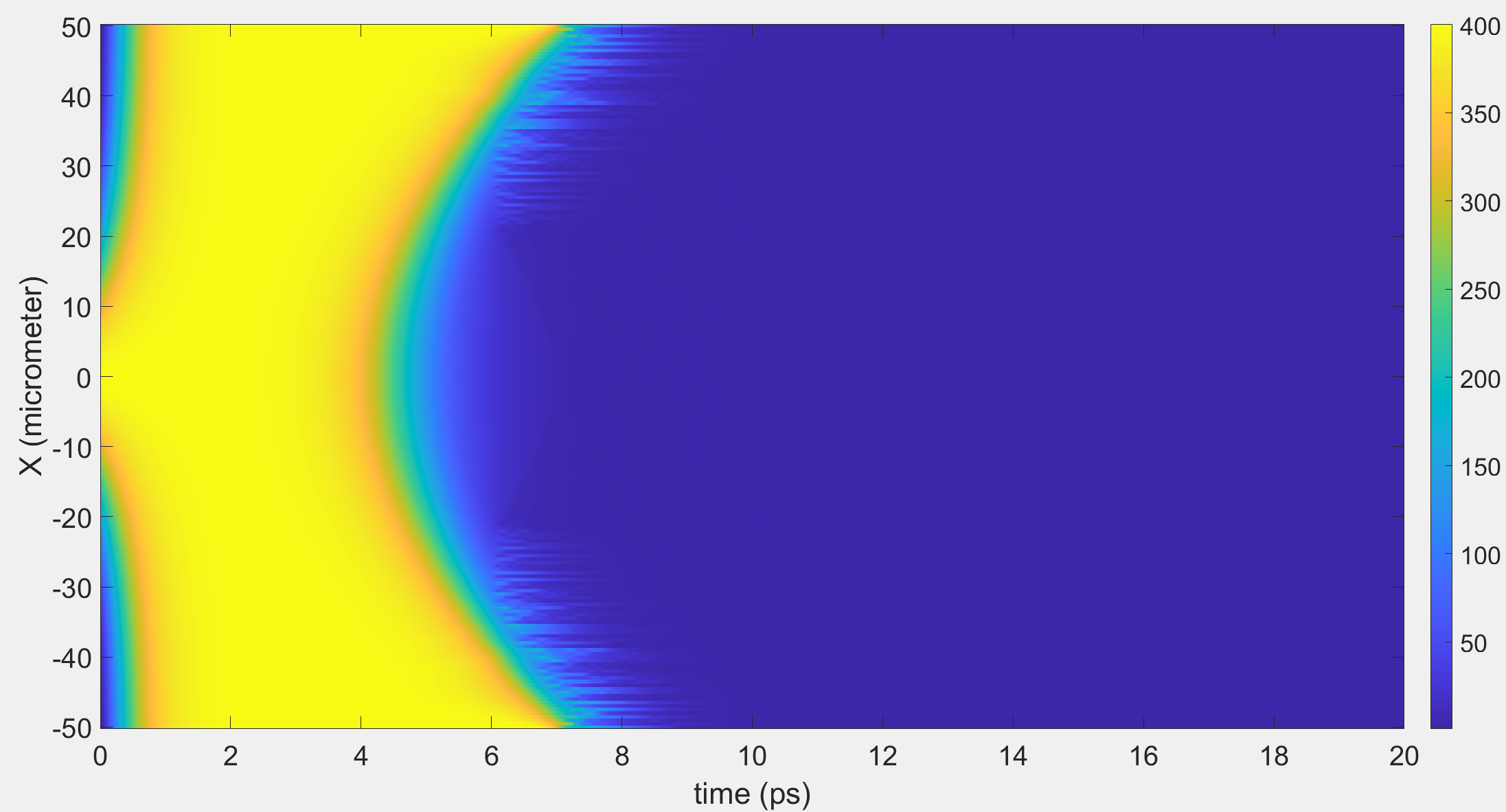}
		\caption{Reservoir polaritons}
		\label{DDGPE_nR_x_str_corr1_P1}
	\end{subfigure}
	\hfill
	\begin{subfigure}{0.48\textwidth}
		\centering
		\includegraphics[width=0.8\columnwidth]{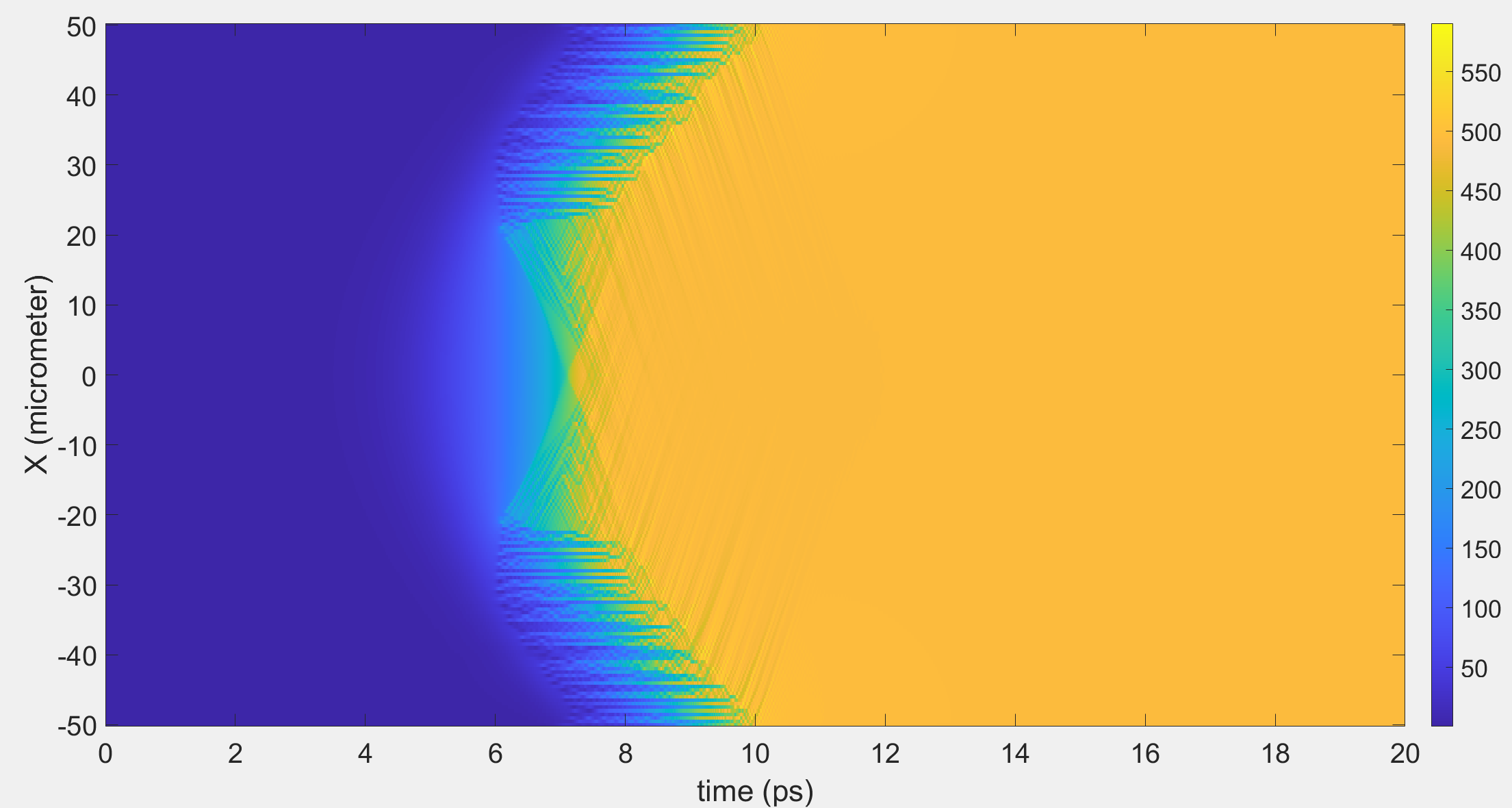}
		\caption{Condensate polaritons}
		\label{DDGPE_psi_x_str_corr1_P1}
	\end{subfigure}
	\caption{Density of the condensate polaritons with a pumping rate (pump power) of P = 60.790 $\mu m^{-1} ps^{-1}$ for the homogeneous, incoherent, nonresonant pumping in the strongly correlated polariton regime for $g/\gamma_c = 1.132$.}
	\label{HINRP_x_str_corr1_P1}
\centering
\end{figure}

\begin{figure}[htbp!]
	\centering
	\begin{subfigure}{0.48\textwidth}
		\centering
		\centering
\includegraphics[width=0.8\columnwidth]{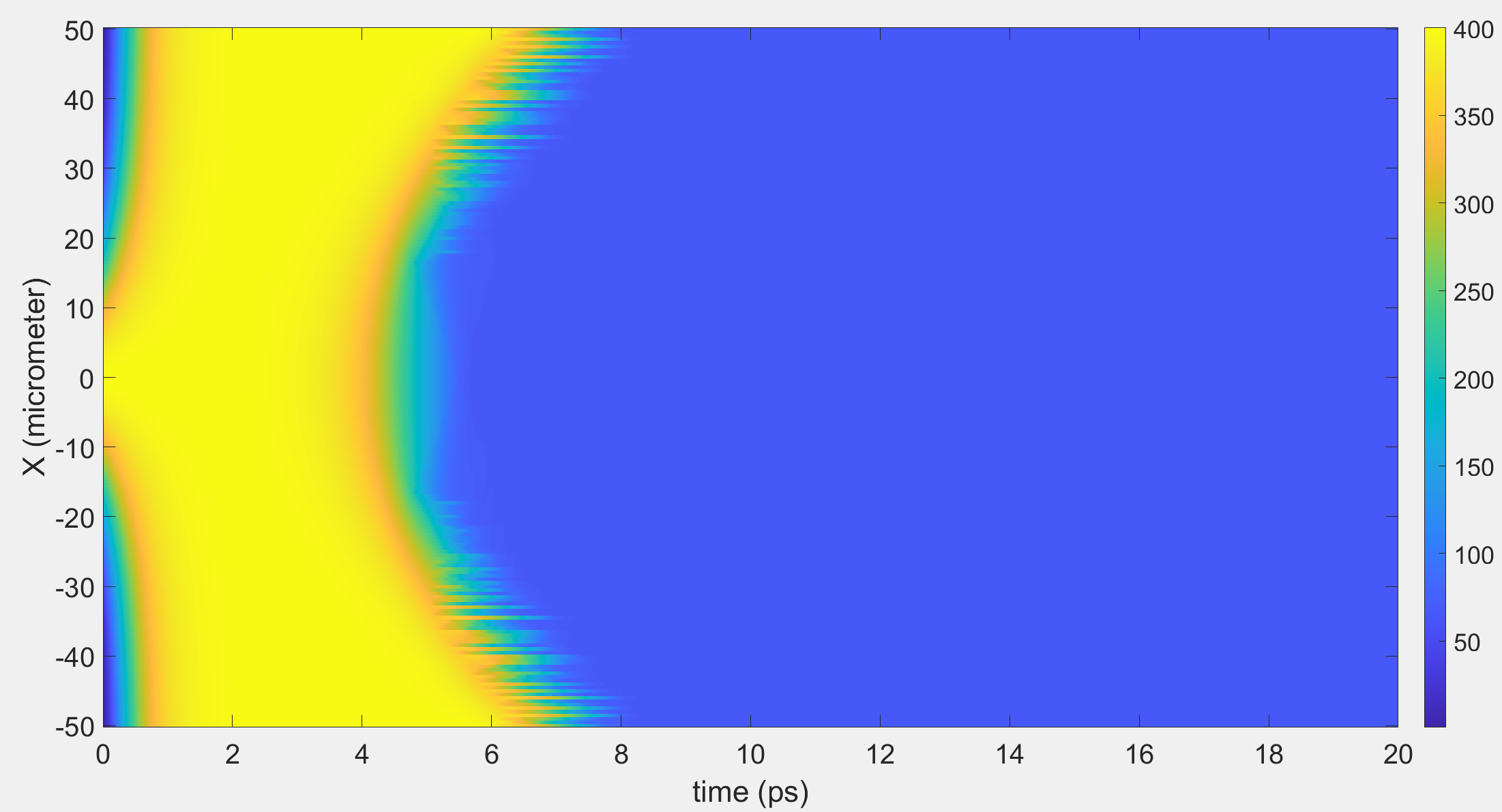}
		\caption{Reservoir polaritons}
		\label{DDGPE_nR_x_str_corr2_P1}
	\end{subfigure}
	\hfill
	\begin{subfigure}{0.48\textwidth}
		\centering
		\includegraphics[width=0.8\columnwidth]{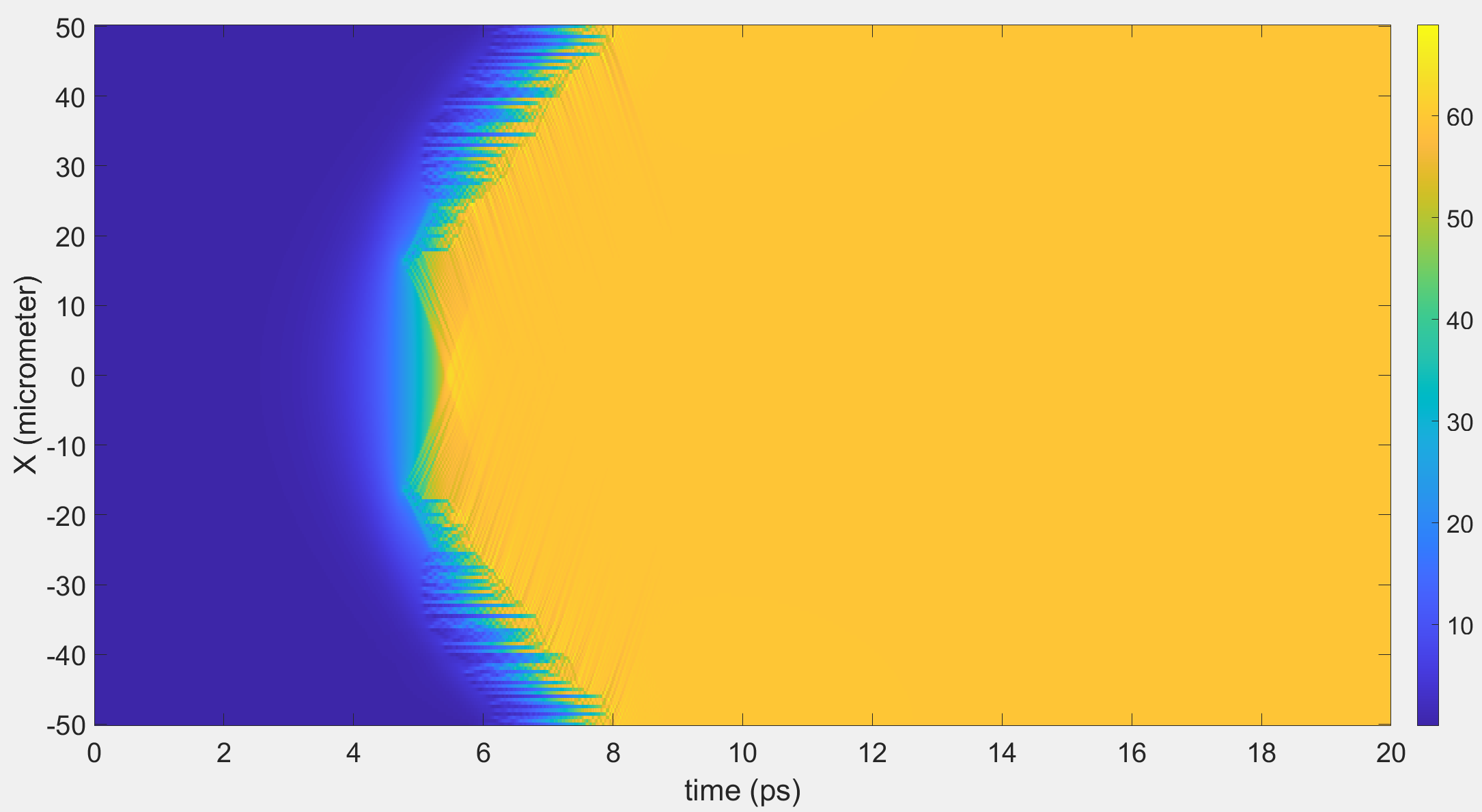}
		\caption{Condensate polaritons}
		\label{DDGPE_psi_x_str_corr2_P1}
	\end{subfigure}
	\caption{Density of the condensate polaritons with a pumping rate (pump power) of P = 60.790 $\mu m^{-1} ps^{-1}$ for the homogeneous, incoherent, nonresonant pumping in the strongly correlated polariton regime for $g/\gamma_c = 10$.}
	\label{HINRP_x_str_corr2_P1}
\centering
\end{figure}

\begin{figure}[htbp!]
	\centering
	\begin{subfigure}{0.48\textwidth}
		\centering
		\centering
\includegraphics[width=0.8\columnwidth]{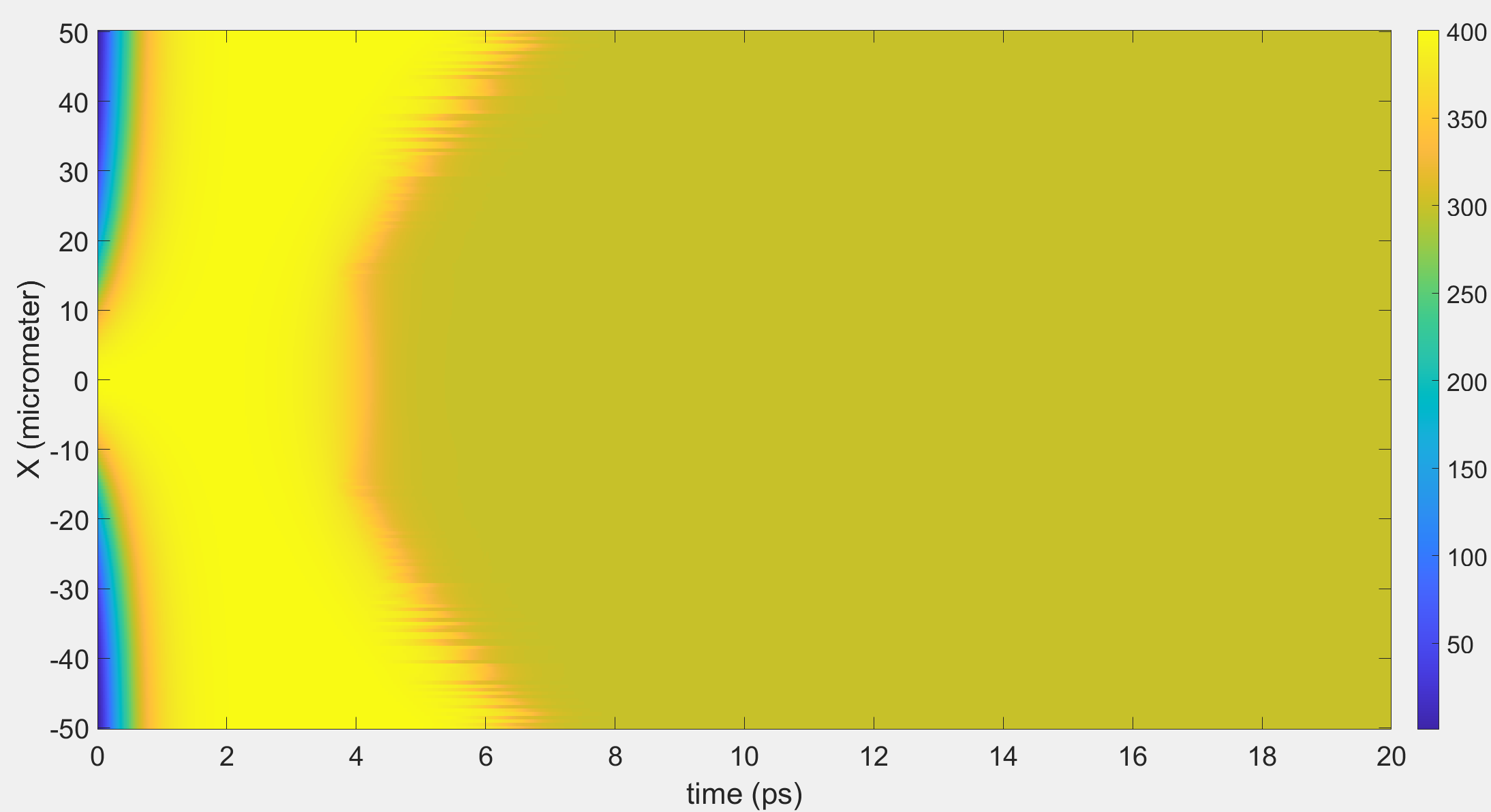}
		\caption{Reservoir polaritons}
		\label{DDGPE_nR_x_str_corr3_P1}
	\end{subfigure}
	\hfill
	\begin{subfigure}{0.48\textwidth}
		\centering
		\includegraphics[width=0.8\columnwidth]{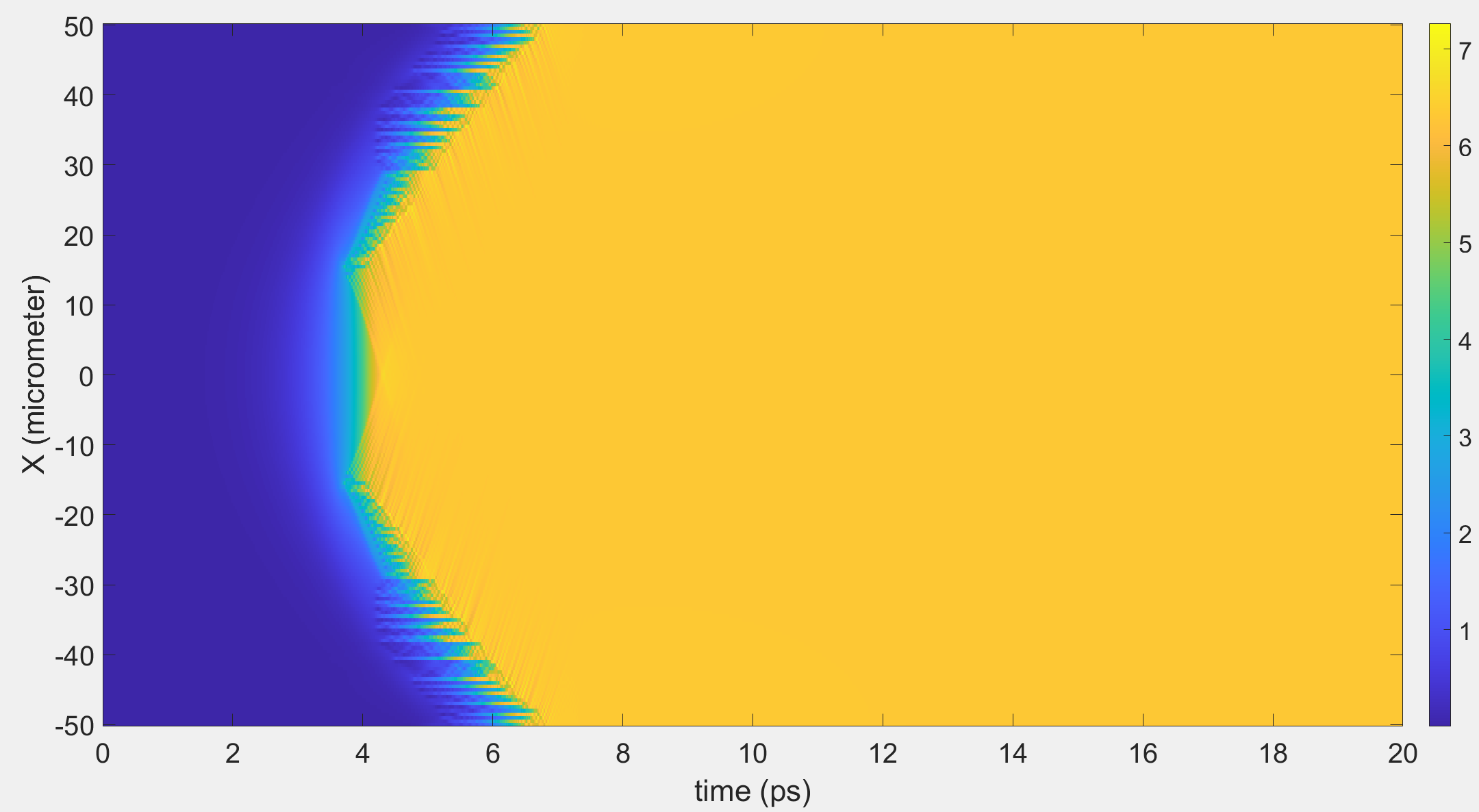}
		\caption{Condensate polaritons}
		\label{DDGPE_psi_x_str_corr3_P1}
	\end{subfigure}
	\caption{Density of the condensate polaritons with a pumping rate (pump power) of P = 60.790 $\mu m^{-1} ps^{-1}$ for the homogeneous, incoherent, nonresonant pumping in the strongly correlated polariton regime for $g/\gamma_c = 100$.}
	\label{HINRP_x_str_corr3_P1}
\centering
\end{figure}

Next, in Figures \ref{HINRP_x_str_corr1_P1}, \ref{HINRP_x_str_corr2_P1} and \ref{HINRP_x_str_corr3_P1}, we present the results for 1D microwire microcavity in the strongly correlated polariton regime for homogeneous, incoherent, non-resonant pumping. With increasing ratio $g/\gamma_c$, population of the reservoir polaritons remains almost constant while that of the condensate polaritons decreases sharply from a few hundreds to a few tens to single digits. While the complementary behavior noticed earlier continues, it can be noticed that the reservoir polaritons continue to stay dominant is parts of the microcavity where the condensate polaritons are supposed to be dominant in the low interaction strength regime. Condensate polaritons at bottom of the lower polariton branch prevent the reservoir polaritons from reaching there due to a polariton blockade stemming from strong interactions.

In experiments which hinted at the polariton blockade~\cite{delteil2019towards,munoz2019emergence}, the laser power used was 10 nW. The laser power can be converted to the pumping rate (which we call $P'$ to distinguish it from laser power $P$) as $P' = P/A$ for 2D and $P' = P/L$ for 1D. Using $A = 576 \mu m^2$, $L = 100 \mu m$, and $P = 10 nW = 62.42 \frac{meV}{ps}$, we get $P'_{2D} = 0.108368 \frac{meV}{\mu m^2 ps}$ and $P'_{1D} = 0.6242 \frac{meV}{\mu m ps}$.

\begin{figure}[htbp!]
	\centering
	\begin{subfigure}{0.48\textwidth}
		\centering
		\centering
\includegraphics[width=0.8\columnwidth]{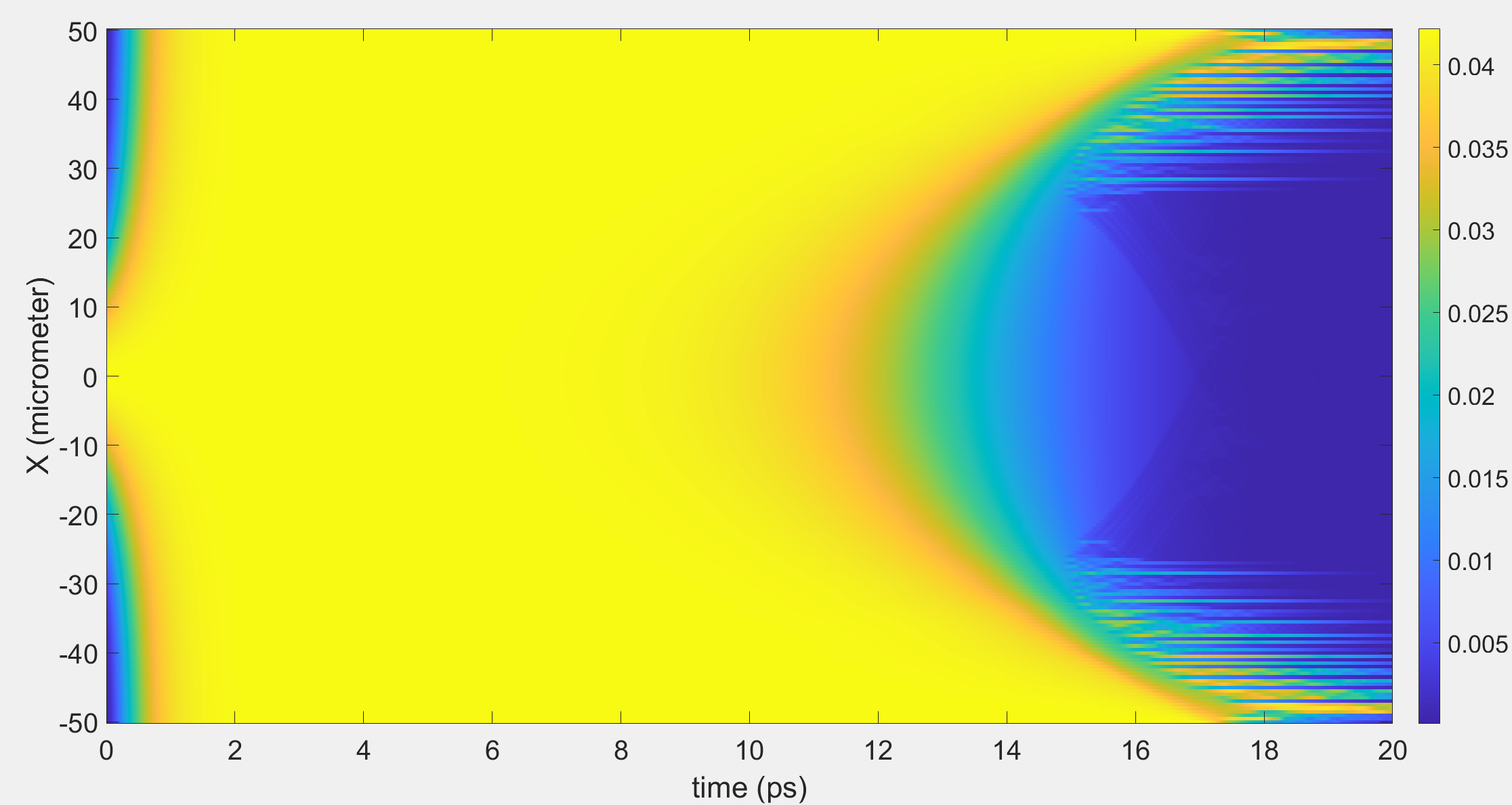}
		\caption{Reservoir polaritons}
		\label{DDGPE_nR_x_str_corr1_Plow}
	\end{subfigure}
	\hfill
	\begin{subfigure}{0.48\textwidth}
		\centering
		\includegraphics[width=0.8\columnwidth]{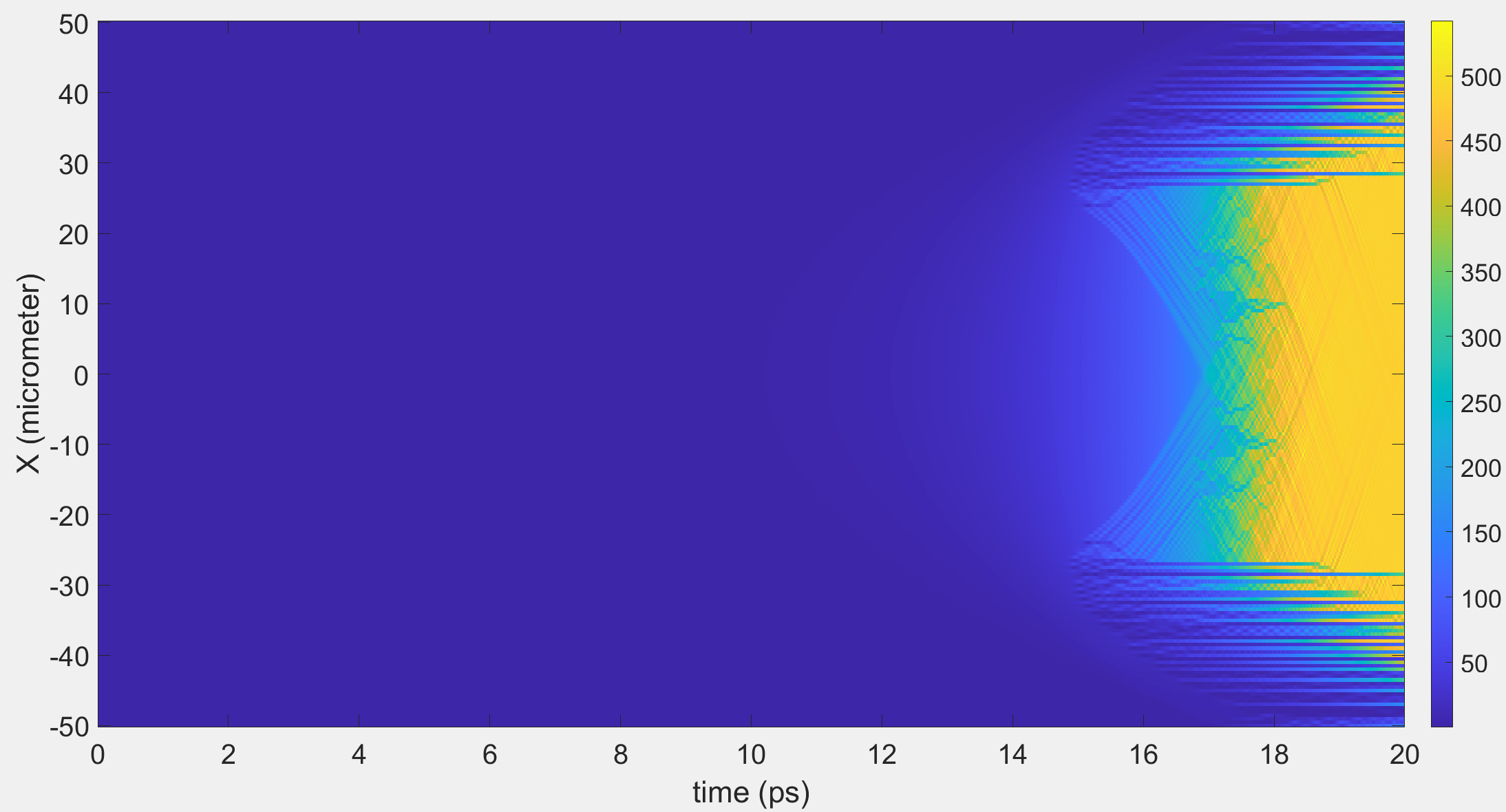}
		\caption{Condensate polaritons}
		\label{DDGPE_psi_x_str_corr1_Plow}
	\end{subfigure}
	\caption{Density of the condensate polaritons with a pumping rate (pump power) of P = 0.6242 $\mu m^{-1} ps^{-1}$ for the homogeneous, incoherent, nonresonant pumping in the strongly correlated polariton regime for $g/\gamma_c = 1.132$.}
	\label{HINRP_x_str_corr1_Plow}
\centering
\end{figure}

\begin{figure}[htbp!]
	\centering
	\begin{subfigure}{0.48\textwidth}
		\centering
		\centering
\includegraphics[width=0.8\columnwidth]{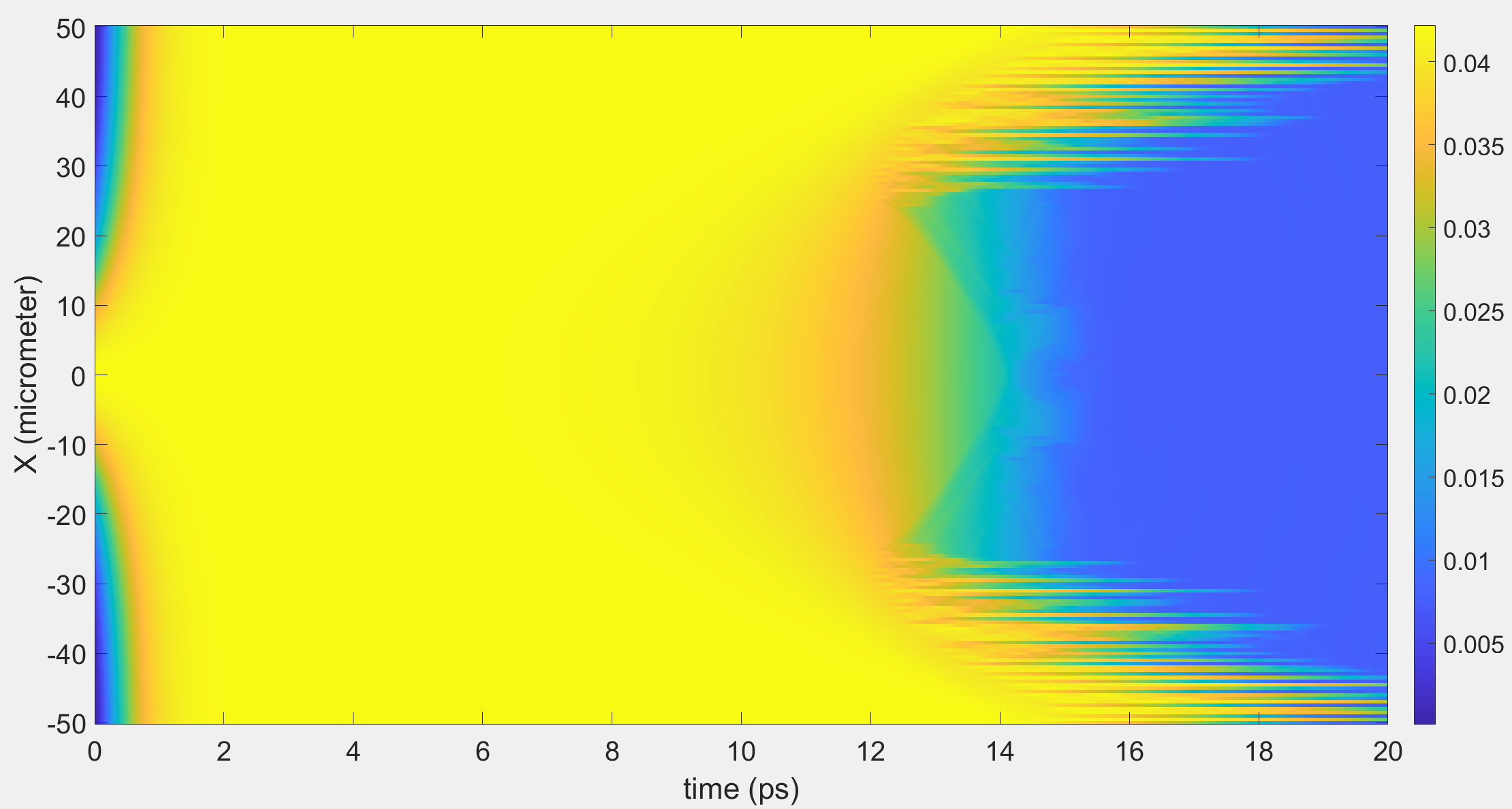}
		\caption{Reservoir polaritons}
		\label{DDGPE_nR_x_str_corr2_Plow}
	\end{subfigure}
	\hfill
	\begin{subfigure}{0.48\textwidth}
		\centering
		\includegraphics[width=0.8\columnwidth]{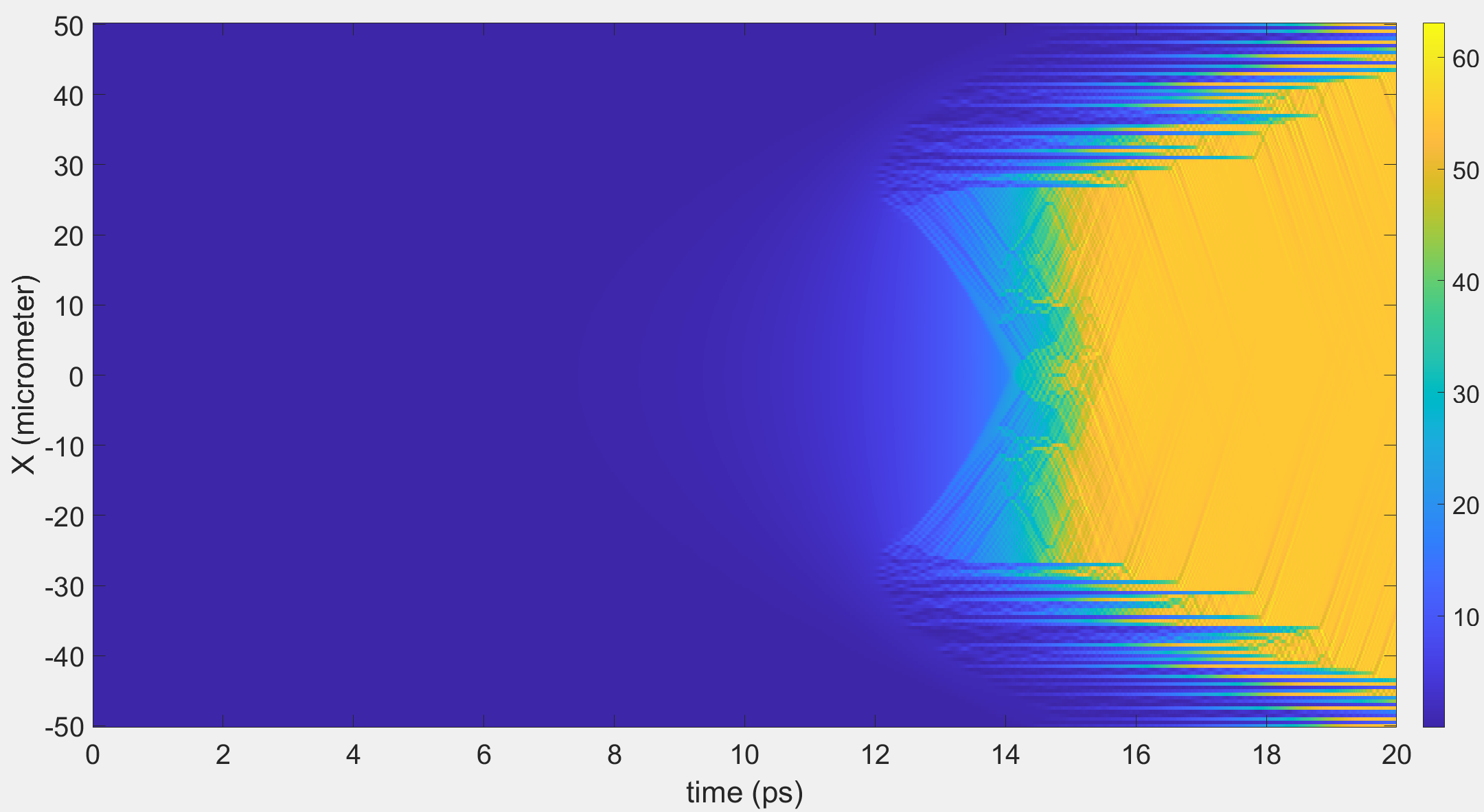}
		\caption{Condensate polaritons}
		\label{DDGPE_psi_x_str_corr2_Plow}
	\end{subfigure}
	\caption{Density of the condensate polaritons with a pumping rate (pump power) of P = 0.6242 $\mu m^{-1} ps^{-1}$ for the homogeneous, incoherent, nonresonant pumping in the strongly correlated polariton regime for $g/\gamma_c = 10$.}
	\label{HINRP_x_str_corr2_Plow}
\centering
\end{figure}

\begin{figure}[htbp!]
	\centering
	\begin{subfigure}{0.48\textwidth}
		\centering
		\centering
\includegraphics[width=0.8\columnwidth]{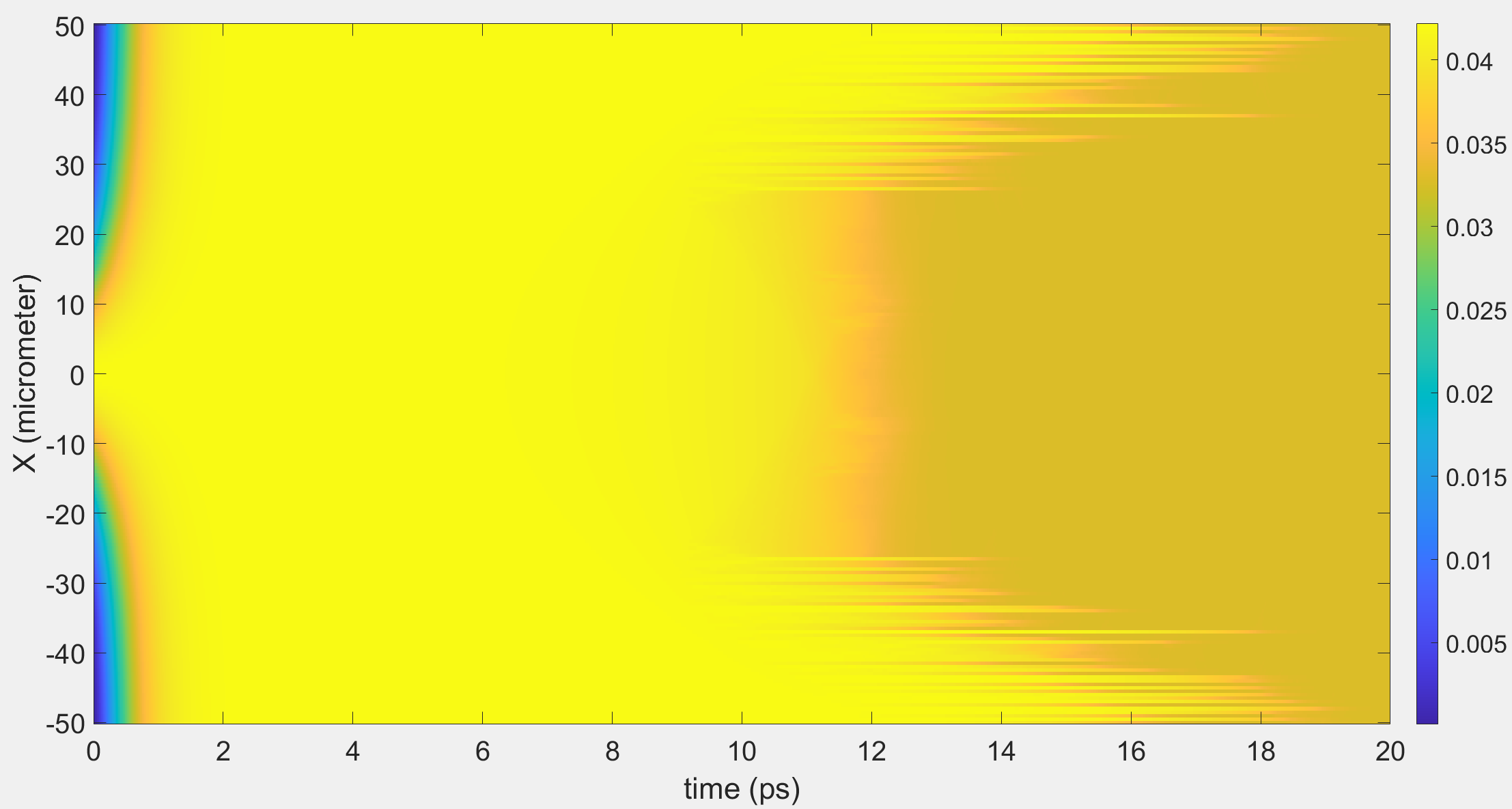}
		\caption{Reservoir polaritons}
		\label{DDGPE_nR_x_str_corr3_Plow}
	\end{subfigure}
	\hfill
	\begin{subfigure}{0.48\textwidth}
		\centering
		\includegraphics[width=0.8\columnwidth]{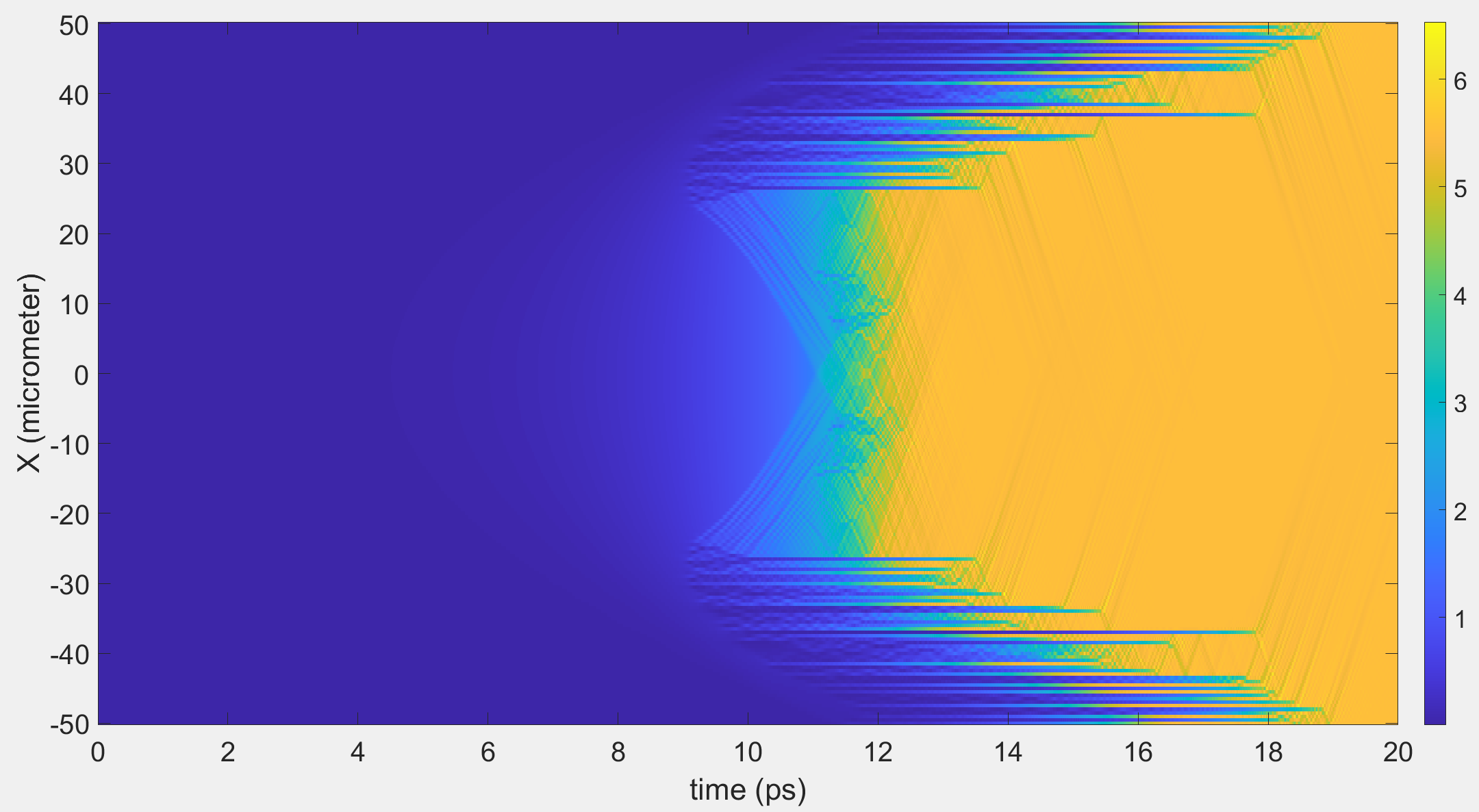}
		\caption{Condensate polaritons}
		\label{DDGPE_psi_x_str_corr3_Plow}
	\end{subfigure}
	\caption{Density of the condensate polaritons with a pumping rate (pump power) of P = 0.6242 $\mu m^{-1} ps^{-1}$ for the homogeneous, incoherent, nonresonant pumping in the strongly correlated polariton regime for $g/\gamma_c = 100$.}
	\label{HINRP_x_str_corr3_Plow}
\centering
\end{figure}

Below, we use the converted pumping rate -- as derived above -- for 1D microwire microcavity corresponding to homogeneous, incoherent, non-resonant pumping. The corresponding results are shown in Figures \ref{HINRP_x_str_corr1_Plow}, \ref{HINRP_x_str_corr2_Plow}, and \ref{HINRP_x_str_corr3_Plow}. We notice that with such very low pumping rate, it takes longer for the reservoir polaritons to get converted as condensate polaritons. Within the strongly correlated polariton regime, we can observe that this trend is sharper for lower correlation. Interestingly, the number of condensate polaritons in the strongly correlated regime remains almost the same and shows the same trend as with ordinary pump powers whereas the number of reservoir polaritons decreases sharply.

\section{Conclusion}
\label{conc}
We plotted the results in real space and time for 1D and 2D microcavities corresponding to all three equations which capture the dynamics of exciton-polaritons. We did the same with condition for strongly correlated polaritons included in the mean-field equations. While the results for equations corresponding to the exciton-polariton superfluid, that is coherent, near-resonant pumping of type 1 (Equation \ref{CNRP1eq}), are not particularly exciting as far as the numerical dynamics of excitons and polaritons is concerned, we find some promising results with the solution of Equations \ref{CNRP2eq} and \ref{HINRPeq} corresponding to coherent, near-resonant pumping of type 2 and homogeneous, incoherent, non-resonant pumping respectively. 

With the increase in the ratio of the polariton-polariton interaction strength and polariton dissipation rate, it is summarily observed that the number of condensate polaritons decrease sharply. Interestingly, this effect is agnostic to the change in laser pump power. There are also more detailed observations regarding the location inside the microcavity where polaritons condense as well as the contrasting dynamics of reservoir and condensate polaritons as done in the Results~\ref{res} section. By comparing our space-time plots for the reservoir and condensate polaritons, one can also calculate the exact time it takes for the reservoir polaritons to traverse along the potential landscape to reach the bottom of the lower polariton branch. Remarkably, a polariton blockade of the condensate polaritons can be achieved even with non-resonant pumping.

We note that a polariton-polariton interaction strength of 1.74 $\pm$ 0.46 meV $\mu m^2$ has been obtained, and a polariton lifetime of 300 ps has been achieved in a high quality microcavity~\cite{sun2017direct}. It is straightforward to see that ratio of the two values surpasses 100. While doubts have been raised about the experimentally obtained high value in contrast to the theoretical expectation~\cite{snoke2023reanalysis}, it is still interesting to investigate what exactly is operationally relevant for the observation of polariton blockade or entering the strongly correlated polariton regime. It is also noted that quantum gates such as the C-NOT can be implemented with high-fidelity when the ratio $ g/\Gamma_p $ gets closer to 100~\cite{ghosh2020quantum}.

\newpage
\section*{Acknowledgments}
The author acknowledges discussions with Lizy Abraham, Martin Aeschlimann, Omer Ali, Saeid Asgarnezhadzorgabad, Frank Bello, Daniel Clarke, Ortwin Hess, Zahra Jalali-Mola, Ghulam Murtaza, and Kavindu Sellahewa. This publication has emanated from research conducted with the financial support of Taighde Éireann – Research Ireland under Grant number 13/RC/2077\_P2 at CONNECT: the Research Ireland Centre for Future Networks.

\bibliography{EPCA}
\bibliographystyle{apsrev4-2}
	
\end{document}